\documentclass[structabstract]{aa}  
\usepackage{graphicx}
\usepackage{txfonts}
\usepackage{natbib}
\usepackage{breqn}
\usepackage{longtable,lscape}
\usepackage{hyperref} 
\usepackage{siunitx}  

\newcommand{\GBP}{\mbox{$G_{\rm BP}$}}
\newcommand{\GRP}{\mbox{$G_{\rm RP}$}}
\newcommand{\pmra}{\mbox{$\mu_{\alpha *}$}}
\newcommand{\pmdec}{\mbox{$\mu_{\delta}$}}
\newcommand{\mci}[1]{\multicolumn{1}{c}{#1}}
\newcommand{\mcii}[1]{\multicolumn{2}{c}{#1}}
\newcommand{\HeI}[1]{\mbox{He\,{\sc i}~$\lambda${#1}}}
\newcommand{\HeII}[1]{\mbox{He\,{\sc ii}~$\lambda${#1}}}
\newcommand{\CII}[1]{\mbox{C\,{\sc ii}~$\lambda${#1}}}
\newcommand{\CIIIt}[1]{\mbox{C\,{\sc iii}~$\lambda\lambda\lambda${#1}}}
\newcommand{\NIIId}[1]{\mbox{N\,{\sc iii}~$\lambda\lambda${#1}}}
\newcommand{\NIIIt}[1]{\mbox{N\,{\sc iii}~$\lambda\lambda\lambda${#1}}}
\newcommand{\NVd}[1]{\mbox{N\,{\sc v}~$\lambda\lambda${#1}}}
\newcommand{\SiIII}[1]{\mbox{Si\,{\sc iii}~$\lambda${#1}}}
\newcommand{\SiIV}[1]{\mbox{Si\,{\sc iv}~$\lambda${#1}}}
\newcommand{\MgII}[1]{\mbox{Mg\,{\sc ii}~$\lambda${#1}}}
\newcommand{\lili}{\mbox{LiLiMaRlin}}
\newcommand{\phm}{\phantom{--}}
\begin{document}

   \title{Lucky spectroscopy, an equivalent technique to Lucky Imaging:}
   \subtitle{II. Spatially-resolved intermediate-resolution blue-violet spectroscopy of \\
             19 close massive binaries using the William Herschel Telescope.}

   \author{J. Ma{\'\i}z Apell{\'a}niz\inst{1}
           \and
           R. H. Barb\'a\inst{2}
           \and
           C. Fari\~na\inst{3,4}
           \and
           A. Sota\inst{5}
           \and
           M. Pantaleoni Gonz\'alez\inst{1,6}
           \and
           G. Holgado\inst{1}
           \and
           I. Negueruela\inst{7}
           \and
           S. Sim\'on-D{\'\i}az\inst{3,8}
          }

   \institute{Centro de Astrobiolog{\'\i}a, CSIC-INTA. Campus ESAC. Camino bajo del castillo s/n. E-\num{28692} Vill. de la Ca\~nada. Madrid. Spain. \\
              \email{jmaiz@cab.inta-csic.es} 
         \and
              Departamento de Astronom{\'\i}a. Universidad de La Serena. Av. Cisternas 1200 Norte. La Serena. Chile. 
         \and
              Instituto de Astrof{\'\i}sica de Canarias. E-\num{38200} La Laguna, Tenerife. Spain. 
         \and
              Isaac Newton Group of Telescopes. Apartado de correos 321. E-\num{38700} Santa Cruz de La Palma, La Palma, Spain. 
         \and
              Instituto de Astrof{\'\i}sica de Andaluc{\'\i}a-CSIC. Glorieta de la Astronom\'{\i}a s/n. E-\num{18008} Granada. Spain. 
         \and
              Departamento de Astrof{\'\i}sica y F{\'\i}sica de la Atm\'osfera. Universidad Complutense de Madrid. E-\num{28040} Madrid. Spain. 
         \and
              Departamento de F{\'\i}sica. Ingenier{\'\i}a de Sistemas y Teor{\'\i}a de la Se\~nal. Escuela Polit\'ecnica Superior. Universidad de Alicante. Carretera San Vicente del Raspeig s/n. E-\num{03690} San Vicente del Raspeig. Alicante. Spain. 
         \and
              Departamento de Astrof{\'\i}sica. Universidad de La Laguna. E-\num{38205} La Laguna, Tenerife. Spain. 
             }

   \date{Submitted 21 Sep 2020; accepted 23 Nov 2020}

 
  \abstract
  {Many massive stars have nearby companions whose presence hamper their characterization through spectroscopy.}
  {We want to continue obtaining spatially resolved spectroscopy of close massive visual binaries to derive their spectral types.}
  {We have used the lucky spectroscopy technique to obtain a large number of short long-slit spectroscopic exposures of 19 close visual binaries under good seeing 
   conditions. We selected those with the best characteristics, extracted the spectra using multiple-profile fitting, and combined the results to derive spatially 
   separated spectra. The results are analyzed in combination with data from lucky imaging, regular intermediate-resolution single-order spectroscopy, and \'echelle
   high-resolution spectroscopy.}
  {The new application of lucky spectroscopy has allowed us (among other results) to [a] spatially disentangle for the first time two O stars (FN~CMa~B and 6~Cas~B) with 
   brighter BA supergiant companions; [b] determine that two B stars ($\alpha$~Sco~B and HD~\num{164492}~B) with close and more massive companions are fast rotators (in the
   second case solving a case of mistaken identity); [c] extend the technique to cases with extreme magnitude differences (the previous two cases plus CS~Cam~A,B), shorter 
   separations (HD~\num{193443}~A,B), and fainter primary magnitudes down to $B=11$ (HD~\num{219460}~A,B); [d] spatially disentangle the spectra of stars with companions
   as diverse as an A~supergiant (6~Cas~A), a WR~star (HD~\num{219460}~B~=~WR~157), and an M~supergiant ($\alpha$~Sco~A); [e] discover the unexpected identity of some 
   targets such as two previously unknown bright O stars (HD~\num{51756}~B and BD~$+$60~544) and a new member of the rare OC category (HD~8768~A); and [f] identify and classify
   (in some cases for the first time) which of the components of four visual binaries ($\sigma$~Ori, HD~\num{219460}, HD~\num{194649}, and HD~\num{191201}) is a double-lined 
   spectroscopic binary while for another seven systems (FN~CMa, $\sigma$~Sco, HD~\num{51756}, HD~\num{218195}, HD~\num{17520}, HD~\num{24431}, and HD~\num{164492}) we detect signs of  
   spectroscopic binarity using high-spectral-resolution spectroscopy. We also present a determination of the limits of the technique.
  }
  {}


   \maketitle
%

\section{Introduction}

$\,\!$ \indent Massive stars are born in company of others but their multiplicity is not always easy to detect or characterize. The main complication lies in the large range of
periods, from $\sim$1~d orbits to systems that get to complete just one or a few revolutions during the lifetimes of the stars. Therefore, for a complete account of massive-star
multiplicity one needs to use different techniques, as long-period systems are better studied by high-spatial-resolution techniques and short-period systems by
high-spectral-resolution techniques. Putting it in a slightly different manner, the first type are usually studied as ``visual binaries'' (or multiples) and the
second type as ``spectroscopic binaries'' (or multiples) \citep{Masoetal98} and in multiplicity studies it should be clear which type (one, the other, or both) we are referring to.
Otherwise, confusion is likely to arise and wrong results will be obtained.

\begin{table*}
 \caption{Multiple systems observed with lucky spectroscopy in this paper. PC and SC refer to the primary and secondary visual components, respectively, and $B_{\rm P}$ is 
         the $B$-band magnitude of the primary. For each observation we give the evening date, detector used (EEV for EEV12, QUC for QUCAM3), number of total exposures, 
         number of exposures used for lucky spectroscopy (single value for EEV12, range for QUCAM3), exposure time (for one exposure), separation $d$ (measured from
         the data), position angle used for the slit (which may be different from the one for the system), and magnitude difference in $B$ (measured from the data).}
\label{sample}
\begin{center}
\begin{tabular}{lcccrccccccrc}
\hline
Name            & PC    & SC & WDS ID        & $B_{\rm P}$ & Even. date & Det. & $n_{\rm exp}$ & $n_{\rm used}$ & $t_{\rm exp.}$ & $d$       & PA      & $\Delta B$ \\
                &       &    &               &             & (YYMMDD)   &      &               &                & (s)            & (\arcsec) & (\degr) & (mag)      \\
\hline
FN CMa          & A     & B  & J07067$-$1118 &  5.8        & 190220     & EEV  &  120          &  60            & 1.00           & 0.57      & 109     & 1.2        \\
                &       &    &               &             & 190917     & EEV  &  100          &  15            & 1.00           & 0.57      & 109     & 1.2        \\
6 Cas           & A     & B  & J23488$+$6213 &  6.3        & 180729     & EEV  &  143          &  65            & 1.00           & 1.50      & 194     & 2.3        \\
$\alpha$ Sco    & A     & B  & J16294$-$2626 &  2.8        & 120708     & EEV  &   15          &  15            & 4.33           & 2.57      & 297     & 0.0        \\
                &       &    &               &             & 180730     & EEV  &  100          &  23            & 1.00           & 2.70      & 273     & 3.6        \\
$\sigma$ Sco    & Aa    & Ab & J16212$-$2536 &  3.2        & 180730     & EEV  &  100          &  13            & 1.00           & 0.47      & 242     & 2.1        \\
CS Cam          & A     & B  & J03291$+$5956 &  4.6        & 190917     & EEV  &  100          &  66            & 1.00           & 2.30      & 159     & 4.2        \\
$\sigma$ Ori    & Aa,Ab & B  & J05387$-$0236 &  4.4        & 170907     & EEV  &  100          &   7            & 1.00           & 0.26      &  84     & 1.2        \\
                &       &    &               &             & 181203     & EEV  &  100          &  12            & 1.00           & 0.25      &  67     & 1.2        \\
                &       &    &               &             & 190917     & EEV  &  100          &  10            & 1.00           & 0.26      &  84     & 1.2        \\
HD 5005         & A     & B  & J00528$+$5638 &  8.6        & 181219     & QUC  & 1004          & 100-402        & 1.00           & 1.53      &  82     & 1.6        \\ 
                &       &    &               &             & 190906     & EEV  &  100          &  58            & 1.00           & 1.53      &  82     & 1.6        \\
HD \num{51756}  & A     & B  & J06585$-$0301 &  7.8        & 181219     & QUC  & 1004          &  26-51         & 1.00           & 0.63      & 100     & 0.3        \\ 
HD \num{218195} & A     & B  & J23052$+$5815 &  8.7        & 181220     & QUC  &  502          & 216-365        & 1.00           & 0.94      &  81     & 2.7        \\ 
HD \num{219460} & A     & B  & J23152$+$6027 & 11.0        & 181220     & QUC  &  502          & 423-499        & 1.00           & 1.38      & 127     & 0.3        \\ 
HD 8768         & A     & B  & J01281$+$6317 &  9.3        & 181220     & QUC  &  502          & 138-247        & 1.00           & 0.66      &  28     & 2.8        \\ 
BD $+$60 544    & A     & B  & J02411$+$6108 & 10.4        & 181220     & QUC  &  502          & 403-502        & 1.00           & 1.91      & 170     & 0.8        \\ 
HD \num{17520}  & A     & B  & J02512$+$6023 & 10.3        & 181220     & QUC  &  502          & 262-279        & 1.00           & 0.32      & 298     & 0.7        \\ 
HD \num{24431}  & A     & B  & J03556$+$5238 &  7.2        & 181220     & QUC  &  502          & 209-273        & 1.00           & 0.72      & 177     & 2.9        \\ 
HD \num{164492} & A     & B  & J18024$-$2302 &  7.6        & 190914     & EEV  &  100          &  71            & 1.00           & 6.19      &  20     & 3.6        \\
HD \num{168021} & A     & B  & J18187$-$1837 &  7.4        & 190421     & EEV  &  100          &  33            & 1.00           & 0.48      & 137     & 1.0        \\
HD \num{193443} & A     & B  & J20189$+$3817 &  8.3        & 190731     & EEV  &  100          &  13            & 1.00           & 0.14      & 258     & 0.3        \\
HD \num{194649} & A     & B  & J20254$+$4014 & 10.6        & 190914     & EEV  &  100          &  95            & 1.00           & 0.30      & 214     & 0.7        \\
HD \num{191201} & A     & B  & J20074$+$3543 &  7.8        & 190914     & EEV  &  100          &  47            & 1.00           & 1.00      &  84     & 1.8        \\
\hline
\end{tabular}
\end{center}
\end{table*}

{The first golden era of spectroscopic surveys of OB stars lasted from the 1950s to the 1970s. \citet{Morgetal55,Hilt56,Lesh68,Hiltetal69,Walb72,Walb73a,Garretal77} are probably
the best exponent of the results of that period. Interest in such surveys waned in the following two decades (with some exceptions) but regained strength in the 2000s with new technology 
and telescopes that significantly increased observing efficiency and with the need to address scientific issues such as the ubiquitous multiplicity already mentioned.
As part of that renewed interest, for some time now} 
we have been conducting different types of surveys that have the ultimate goal of characterizing massive stars and, more specifically, their multiplicity.
With the Galactic O-Star Spectroscopic Survey (GOSSS, \citealt{Maizetal11}) we are identifying and producing spectral classifications for O stars (and other massive stars) in the
Galactic neighborhood with intermediate-resolution blue-violet spectroscopy. GOSSS has resulted in three major papers: \citealt{Sotaetal11a} (GOSSS~I), \citealt{Sotaetal14}
(GOSSS~II), and \citealt{Maizetal16} (GOSSS~III), with a fourth paper in the series to be submitted soon. We have also collected a library of high-resolution spectroscopy
of massive stars (\lili, \citealt{Maizetal19a}) from public and private observations. In the northern hemisphere we are using it to study the spectroscopic multiplicity of O stars
(MONOS project) of which the first paper appeared as \citealt{Maizetal19b} (MONOS~I) and a second paper will be submitted soon. In a parallel effort, GOSSS is also serving as a 
source of information for IACOB \citep{SimDetal15b}, a long-term project aimed at extending further the empirical characterization of Galactic O (and B) stars using quantitative 
high-resolution spectroscopy (see, e.g. \citealt{SimDetal14,SimDetal17,Holgetal18,Holgetal20}). As for high-spatial-resolution observations, we have
observed a large number of systems with lucky imaging using AstraLux (\citealt{Maiz10a} and MONOS~I).

\begin{table*}
\caption{Spectral classifications. GOS/GBS/GAS/GWS/GLS stand for Galactic O/B/A/WR/Late Star, respectively. 
The information in this table is also available in electronic form at the GOSC web site (\url{http://gosc.cab.inta-csic.es}).} 
\label{GOSSS_spty}
\begin{center}
\begin{tabular}{lccclllll}
\hline
Name                   & GOSSS ID               & R.A. (J2000) & Decl. (J2000)  & ST    & LC     & Qual.      & Second.    & Notes               \\
\hline
FN CMa A               & GBS 224.71$-$01.79\_01 & 07:06:40.767 & $-$11:17:38.46 & B0.7  & Ib     & \ldots     & \ldots     & SB1 in \lili        \\
FN CMa B               & GOS 224.71$-$01.79\_02 & 07:06:40.803 & $-$11:17:38.71 & O6    & V      & ((f))z     & \ldots     & \ldots              \\
6 Cas A                & GAS 115.71$+$00.22\_02 & 23:48:50.166 & $+$62:12:52.22 & A3    & Ia     & \ldots     & \ldots     & \ldots              \\
6 Cas B                & GOS 115.71$+$00.22\_01 & 23:48:50.108 & $+$62:12:50.75 & O9.5  & II     & \ldots     & \ldots     & \ldots              \\
$\alpha$ Sco A         & GLS 351.95$+$15.06\_01 & 16:29:24.461 & $-$26:25:55.21 & M1.5  & Iab    & \ldots     & \ldots     & \ldots              \\
$\alpha$ Sco B         & GBS 351.95$+$15.06\_02 & 16:29:24.275 & $-$26:25:55.04 & B2    & V      & n          & \ldots     & \ldots              \\
$\sigma$ Sco Aa        & GBS 351.31$+$17.00\_01 & 16:21:11.313 & $-$25:35:34.09 & B1    & III    & \ldots     & \ldots     & SB2 in \lili        \\
$\sigma$ Sco Ab        & GBS 351.31$+$17.00\_02 & 16:21:11.282 & $-$25:35:34.30 & B1:   & V      & \ldots     & \ldots     & \ldots              \\
CS Cam A               & GBS 141.50$+$02.88\_01 & 03:29:04.136 & $+$59:56:25.20 & B9    & Ia     & \ldots     & \ldots     & \ldots              \\
CS Cam B               & GBS 141.50$+$02.88\_02 & 03:29:04.252 & $+$59:56:22.88 & B2    & III    & \ldots     & \ldots     & \ldots              \\
$\sigma$ Ori Aa,Ab     & GOS 206.82$-$17.34\_01 & 05:38:44.765 & $-$02:36:00.25 & O9.5  & V      & \ldots     & B0.2 V     & In LS~I + MONOS~I   \\ 
$\sigma$ Ori B         & GBS 206.82$-$17.34\_02 & 05:38:44.782 & $-$02:36:00.27 & B0.2  & V      & (n)        & \ldots     & In LS~I + MONOS~I   \\ 
HD 5005 A              & GOS 123.12$-$06.24\_01 & 00:52:49.206 & $+$56:37:39.49 & O4.5  & V      & ((fc))z    & \ldots     & In GOSSS~I          \\
HD 5005 B              & GBS 123.12$-$06.24\_03 & 00:52:49.390 & $+$56:37:39.71 & B0    & V      & \ldots     & \ldots     & In GOSSS~I          \\
HD \num{51756} A       & GBS 216.42$+$00.18\_01 & 06:58:28.180 & $-$03:01:25.44 & B0    & IV     & \ldots     & \ldots     & SB1? in \lili       \\
HD \num{51756} B       & GOS 216.42$+$00.18\_02 & 06:58:28.221 & $-$03:01:25.57 & O9.7  & IV     & (n)        & \ldots     & \ldots              \\
HD \num{218195} A      & GOS 109.32$-$01.79\_01 & 23:05:12.928 & $+$58:14:29.34 & O8.5  & III    & Nstr       & \ldots     & In GOSSS~I~+~II     \\
HD \num{218195} B      & GBS 109.32$-$01.79\_02 & 23:05:13.043 & $+$58:14:29.49 & B1.5  & V      & \ldots     & \ldots     & SB2? in \lili       \\
HD \num{219460} A      & GBS 111.33$-$00.24\_01 & 23:15:12.394 & $+$60:27:01.84 & B0.7  & II     & \ldots     & B1: III:   & \ldots              \\
WR 157                 & GWS 111.33$-$00.24\_02 & 23:15:12.542 & $+$60:27:01.00 & WN5   & \ldots & \ldots     & \ldots     & \ldots              \\
HD 8768 A              & GOS 127.03$+$00.70\_01 & 01:28:03.153 & $+$63:16:57.20 & OC9.2 & II     & \ldots     & \ldots     & \ldots              \\
HD 8768 B              & GBS 127.03$+$00.70\_02 & 01:28:03.198 & $+$63:16:57.78 & B0.7: & V      & \ldots     & \ldots     & \ldots              \\
BD $+$60 544 A         & GOS 135.78$+$01.03\_01 & 02:41:08.194 & $+$61:08:07.02 & O9.5  & IV     & \ldots     & \ldots     & \ldots              \\
BD $+$60 544 B         & GBS 135.78$+$01.03\_02 & 02:41:08.244 & $+$61:08:05.14 & B1    & V      & \ldots     & \ldots     & \ldots              \\
HD \num{17520} A       & GOS 137.22$+$00.88\_01 & 02:51:14.434 & $+$60:23:09.97 & O9.2  & V      & e          & \ldots     & In GOSSS~III        \\
                       &                        &              &                &       &        &            &            & SB1 in \lili        \\
HD \num{17520} B       & GOS 137.22$+$00.88\_02 & 02:51:14.396 & $+$60:23:10.12 & O8    & V      & z          & \ldots     & In GOSSS~III        \\
HD \num{24431} A       & GOS 148.84$-$00.71\_01 & 03:55:38.420 & $+$52:38:28.75 & O9    & III    & \ldots     & \ldots     & In GOSSS~I          \\
                       &                        &              &                &       &        &            &            & SB1 in \lili        \\
HD \num{24431} B       & GBS 148.84$-$00.71\_02 & 03:55:38.424 & $+$52:38:28.03 & B1.5  & V      & \ldots     & \ldots     & \ldots              \\
HD \num{164492} A      & GOS 007.00$-$00.25\_01 & 18:02:23.553 & $-$23:01:51.06 & O7.5  & V      & z          & \ldots     & In GOSSS~II         \\
HD \num{164492} B      & GBS 007.00$-$00.25\_03 & 18:02:23.701 & $-$23:01:45.14 & B2    & V      & nn         & \ldots     & \ldots              \\
HD \num{164492} C      & GBS 007.00$-$00.25\_02 & 18:02:23.132 & $-$23:02:00.17 & B1    & V      & \ldots     & \ldots     & SB3 in \lili        \\
HD \num{168021} A      & GBS 012.70$-$01.47\_01 & 18:18:43.260 & $-$18:37:10.81 & B0    & Ia     & \ldots     & \ldots     & \ldots              \\
HD \num{168021} B      & GBS 012.70$-$01.47\_02 & 18:18:43.281 & $-$18:37:11.18 & B0.2  & II     & \ldots     & \ldots     & \ldots              \\
HD \num{168021} C      & GBS 012.71$-$01.47\_01 & 18:18:44.204 & $-$18:36:59.94 & B0.5  & II     & \ldots     & \ldots     & \ldots              \\
HD \num{193443} A      & GOS 076.15$+$01.28\_01 & 20:18:51.707 & $+$38:16:46.50 & O8.5  & III:   & ((f))      & \ldots     & In MONOS~I          \\
                       &                        &              &                & O8.5  & III    & ((f))      & \ldots     & Sp. cl. from STIS~I \\
HD \num{193443} B      & GOS 076.15$+$01.28\_02 & 20:18:51.697 & $+$38:16:46.48 & O9.2: & IV:    & \ldots     & \ldots     & In MONOS~I          \\
                       &                        &              &                & O9.2  & IV     & \ldots     & \ldots     & Sp. cl. from STIS~I \\
HD \num{194649} A      & GOS 078.46$+$01.35\_01 & 20:25:22.124 & $+$40:13:01.07 & O5.5  & V      & ((f))      & O9.5 V     & In MONOS~I          \\
HD \num{194649} B      & GOS 078.46$+$01.35\_02 & 20:25:22.109 & $+$40:13:00.82 & O7    & V      & z          & \ldots     & In MONOS~I          \\
HD \num{191201} A      & GOS 072.75$+$01.78\_01 & 20:07:23.684 & $+$35:43:05.91 & O9.5  & III    & \ldots     & B0 V       & In MONOS~I          \\
HD \num{191201} B      & GOS 072.75$+$01.78\_02 & 20:07:23.766 & $+$35:43:06.01 & O9.7  & V      & \ldots     & \ldots     & In MONOS~I          \\
\hline
\end{tabular}
\end{center}
\end{table*}

The surveys in the previous paragraph use either spectroscopic or high-spatial-resolution techniques but do not combine both. Therefore, in some cases such as hierarchical systems
composed of a short-period spectroscopic binary and a close visual companion, it may not be clear how the three stars involved are related to each other or even what the precise
spectral types of each may be (for an example, $\sigma$~Ori, see \citealt{SimDetal15a} and section~\ref{sec_sigma_Ori}). 
Combining both types of techniques (i.e. spectroscopy with high spatial resolution) addresses this issue. We have done this on the one hand using STIS
spectroscopy with HST (\citealt{MaizBarb20}, from now on STIS~I) or applying the novel technique of lucky spectroscopy (\citealt{Maizetal18a}, from now on LS~I), which we developed
in LS~I. Lucky spectroscopy is the extension of lucky imaging \citep{Lawetal06} to spectroscopy: we obtain a large number of short long-slit spectroscopic exposures under good-seeing conditions, 
select those with the best characteristics, and combine them to derive spatially resolved spectra of two or more closely separated point sources aligned with the slit. 

In this paper we continue our analysis of massive close binaries with lucky spectroscopy that we started in LS~I. In the next section we describe the main data in the paper, our new
lucky spectroscopy observations for 19 close visual multiple systems, and their relationship with the previous data. We also describe there the complementary data we use to analyze 
those systems in the form of lucky imaging and other spectroscopy. In section 3 we present the results for our sample. We end
up with a summary of our results and a brief presentation of our future plans.

\section{Data and methods}

\subsection{Lucky spectroscopy}

$\,\!$ \indent  In LS~I we presented lucky spectroscopy results for five bright (primaries with $B$ magnitudes of 2-4) close visual binaries whose primaries have spectral subtypes between
O7~and~B0.7. The original data were obtained in September~2017 (though several tests had been previously performed, see below for an example) with the ISIS spectrograph at the William Herschel 
Telescope (WHT) and the success of the technique prompted us to attempt the separation of additional systems in 2018 and 2019. We had also obtained time for further observations in 2020 but 
those did not finally take place due to scheduling changes at WHT. Besides the obvious scientific interest of studying more objects for the knowledge of the members of the larger sample, there
were several avenues we were interested in exploring:

\begin{itemize}
 \item Repeatability of the results, including new epochs for systems that contain spectroscopic binaries that may be observed at a more favorable orbital phase.
 \item Coverage of the separation-magnitude difference ($d-\Delta m$) plane to test the limitations of the technique and comparison with other techniques (STIS~I).
 \item Exploration of systems with spectral types significantly different to the relatively narrow scope of LS~I. 
 \item Extension of the technique to systems with fainter primaries.
\end{itemize}

\begin{table*} 
 \caption{Measurements for visual pairs using our AstraLux lucky images. The evening date, Heliocentric Julian Date (HJD), separation ($d$), position angle ($\theta$), and magnitude difference is given in each case.
         Six different filters were used: Johnson $V$, H$\alpha$, SDSS $i$ and $z$, $zn$ (a narrow filter with a central wavelength similar to that of $z$), and $Y$, in order of increasing central wavelength.}
\centerline{\small
\addtolength{\tabcolsep}{-4pt}
\begin{tabular}{lcr@{.}lrrcccccc}
\hline
Pair               & Even. date & \multicolumn{2}{c}{HJD$-$\num{2400000}} & \mci{$d$}                 & \multicolumn{1}{c}{$\theta$} & $\Delta V$    & $\Delta$H$\alpha$ &  $\Delta i$       &  $\Delta z$       & $\Delta zn$   & $\Delta Y$    \\
                   & (YYMMDD)   & \multicolumn{2}{c}{(d)}                 & \mci{($^{\prime\prime}$)} & \multicolumn{1}{c}{(deg)}    & (mag)         & (mag)             &  (mag)            &  (mag)            & (mag)         & (mag)         \\
\hline
FN CMa A,B         & 121002     & $\;\;\;$56\,203&70                      &  0.577$\pm$0.005          & 115.58$\pm$0.11              & \ldots        & \ldots            &     \ldots        &     \ldots        & 1.66$\pm$0.02 & \ldots        \\ 
                   & 181126     & $\;\;\;$58\,449&63                      &  0.581$\pm$0.005          & 115.39$\pm$0.11              & \ldots        & \ldots            &     \ldots        &     \ldots        & 1.63$\pm$0.02 & \ldots        \\ 
6 Cas A,B          & 121002     & $\;\;\;$56\,203&52                      &  1.497$\pm$0.003          & 195.69$\pm$0.04              & 2.45$\pm$0.01 & \ldots            &     \ldots        & \phm3.05$\pm$0.01 & \ldots        & \ldots        \\ 
                   & 130918     & $\;\;\;$56\,554&63                      &  1.496$\pm$0.003          & 195.81$\pm$0.06              & \ldots        & \ldots            & \phm2.78$\pm$0.02 &     \ldots        & \ldots        & \ldots        \\ 
$\alpha$ Sco A,B   & 190615     & $\;\;\;$58\,650&44                      &  2.743$\pm$0.021          & 276.96$\pm$0.23              & \ldots        & 6.09$\pm$0.07     &     \ldots        &     \ldots        & 7.45$\pm$0.11 & 8.01$\pm$0.10 \\ 
$\sigma$ Sco Aa,Ab & 190616     & $\;\;\;$58\,651&40                      &  0.423$\pm$0.003          & 217.17$\pm$0.22              & \ldots        & \ldots            &     \ldots        &     \ldots        & 2.10$\pm$0.09 & 2.25$\pm$0.07 \\ 
CS Cam A,B         & 181227     & $\;\;\;$58\,480&41                      &  2.316$\pm$0.002          & 162.12$\pm$0.05              & \ldots        & \ldots            &     \ldots        &     \ldots        & 4.62$\pm$0.03 & 4.66$\pm$0.08 \\ 
HD 5005 A,B        & 071112     & $\;\;\;$54\,417&44                      &  1.526$\pm$0.002          &  82.12$\pm$0.04              & \ldots        & \ldots            &     \ldots        & \phm1.59$\pm$0.01 & \ldots        & \ldots        \\ 
                   & 130918     & $\;\;\;$56\,554&64                      &  1.526$\pm$0.002          &  82.25$\pm$0.04              & \ldots        & \ldots            & \phm1.60$\pm$0.01 &     \ldots        & \ldots        & \ldots        \\ 
                   & 181127     & $\;\;\;$58\,450&36                      &  1.526$\pm$0.002          &  82.19$\pm$0.04              & \ldots        & \ldots            & \phm1.57$\pm$0.01 & \phm1.59$\pm$0.01 & \ldots        & \ldots        \\ 
HD 51\,756 A,B     & 121013     & $\;\;\;$56\,204&69                      &  0.705$\pm$0.001          & 102.94$\pm$0.04              & \ldots        & \ldots            &     \ldots        & \phm0.29$\pm$0.01 & \ldots        & \ldots        \\ 
                   & 130919     & $\;\;\;$56\,555&69                      &  0.706$\pm$0.001          & 102.83$\pm$0.04              & \ldots        & \ldots            & \phm0.32$\pm$0.01 &     \ldots        & \ldots        & \ldots        \\ 
HD 218\,195 A,B    & 110913     & $\;\;\;$55\,818&48                      &  0.944$\pm$0.001          &  80.01$\pm$0.07              & \ldots        & \ldots            &     \ldots        & \phm2.66$\pm$0.04 & \ldots        & \ldots        \\ 
                   & 130918     & $\;\;\;$56\,554&52                      &  0.943$\pm$0.001          &  80.01$\pm$0.08              & \ldots        & \ldots            & \phm2.69$\pm$0.03 &     \ldots        & \ldots        & \ldots        \\ 
HD 219\,460 A,B    & 110913     & $\;\;\;$55\,818&45                      &  1.371$\pm$0.001          & 128.00$\pm$0.04              & \ldots        & \ldots            &     \ldots        &  $-$0.12$\pm$0.01 & \ldots        & \ldots        \\ 
                   & 130918     & $\;\;\;$56\,554&48                      &  1.370$\pm$0.001          & 128.08$\pm$0.04              & \ldots        & \ldots            &  $-$0.04$\pm$0.01 &     \ldots        & \ldots        & \ldots        \\ 
HD 8768 A,B        & 081021     & $\;\;\;$54\,761&63                      &  0.662$\pm$0.002          &  27.97$\pm$0.34              & \ldots        & \ldots            &     \ldots        & \phm2.62$\pm$0.09 & \ldots        & \ldots        \\ 
                   & 110913     & $\;\;\;$55\,818&50                      &  0.658$\pm$0.002          &  28.42$\pm$0.18              & \ldots        & \ldots            &     \ldots        & \phm2.73$\pm$0.04 & \ldots        & \ldots        \\ 
                   & 130918     & $\;\;\;$56\,554&65                      &  0.657$\pm$0.004          &  28.57$\pm$0.15              & \ldots        & \ldots            & \phm2.86$\pm$0.14 &     \ldots        & \ldots        & \ldots        \\ 
BD $+$60 544 A,B   & 180918     & $\;\;\;$58\,380&59                      &  1.948$\pm$0.001          & 168.51$\pm$0.04              & \ldots        & \ldots            &     \ldots        & \phm0.76$\pm$0.02 & \ldots        & \ldots        \\ 
HD 17\,520 A,B     & 080118     & $\;\;\;$54\,484&38                      &  0.317$\pm$0.001          & 298.72$\pm$0.07              & \ldots        & \ldots            &     \ldots        & \phm0.69$\pm$0.01 & \ldots        & \ldots        \\ 
                   & 130918     & $\;\;\;$56\,554&66                      &  0.320$\pm$0.001          & 299.05$\pm$0.08              & \ldots        & \ldots            & \phm0.71$\pm$0.02 &     \ldots        & \ldots        & \ldots        \\ 
HD 24\,431 A,B     & 080118     & $\;\;\;$54\,484&41                      &  0.734$\pm$0.001          & 177.18$\pm$0.15              & \ldots        & \ldots            &     \ldots        & \phm2.83$\pm$0.09 & \ldots        & \ldots        \\ 
                   & 110913     & $\;\;\;$55\,818&59                      &  0.737$\pm$0.004          & 176.65$\pm$0.11              & \ldots        & \ldots            &     \ldots        & \phm3.10$\pm$0.17 & \ldots        & \ldots        \\ 
HD 164\,492 A,B    & 130919     & $\;\;\;$56\,555&30                      &  6.261$\pm$0.001          &  20.37$\pm$0.04              & \ldots        & \ldots            &     \ldots        & \phm3.01$\pm$0.05 & \ldots        & \ldots        \\ 
HD 164\,492 A,C    & 130919     & $\;\;\;$56\,555&30                      & 10.707$\pm$0.008          & 211.93$\pm$0.04              & \ldots        & \ldots            &     \ldots        & \phm1.35$\pm$0.05 & \ldots        & \ldots        \\ 
HD 164\,492 A,H    & 130919     & $\;\;\;$56\,555&30                      &  1.489$\pm$0.019          & 345.03$\pm$0.10              & \ldots        & \ldots            &     \ldots        & \phm5.21$\pm$0.19 & \ldots        & \ldots        \\ 
\hline
\end{tabular}
\addtolength{\tabcolsep}{4pt}
}
\label{AstraLuxdata}
\end{table*}

In this paper we present our results for 19 close visual binaries observed with lucky spectroscopy 
(one from the previous sample and 18 new ones, Table~\ref{sample}), which were selected with those 
points in mind. The setup and technique used are the same as in LS~I with one difference. In that first paper we used the standard CCD for the blue arm of ISIS, an EEV12 array of 2048$\times$4096
13.5~$\mu$m~pixels. Here we used that CCD as the default setup but we also tried the alternative QUCAM3 CCD for the faint range of our sample. QUCAM3 is an electron-multiplying CCD with
1024$\times$1024 13~$\mu$m~pixels with a fast mode that has very little readout noise (that we did not use) and a standard slow mode in which it operates as a normal CCD (the one we used). The big 
advantage of QUCAM3 with respect to EEV12 is that it is a full-transfer device that allows for a quick transfer (20 milliseconds) of the accumulated charge (the image) into a temporary on-chip storage 
to be efectivelly read in 0.66 s (for the detector setup used for these data) while the detector remains actively acquiring data.
(see \url{http://www.ing.iac.es/Astronomy/instruments/isis/L3spectroscopy_v4.html}). Therefore, for one-second exposures the dead time is just 
$\sim$2\% while for EEV12 the corresponding dead time is $\sim$14~s, introducing a factor of nearly 15 in the efficiency in favor of QUCAM3. A second, less important, advantage of QUCAM3 is a slightly better 
pixel scale (0\farcs19 pixels vs. 0\farcs20 pixels). The main disadvantage of QUCAM3 arises from its smaller number of pixels: with EEV12 we can cover the 3900-5500~\AA\ range with a single grating 
position while with QUCAM3 to cover just the 3900-5100~\AA\ range we need three grating positions. 
Those characteristics translate into different observing sequences for the two setups. For EEV12 we have to:

\begin{itemize}
 \item Acquire the target and place the slit in the correct position angle (typically 4~m).
 \item Obtain 100 exposures of 1.0~s each in a single grating position (25~m).
 \item Do an arc (1~m).
\end{itemize}

That yields 30 minutes/object and 100~s of total exposure time, of which we typically end up using 10-50~s. For QUCAM3 we have to:

\begin{itemize}
 \item Acquire the target and place the slit in the correct position angle (typically 4~m).
 \item Obtain 500 exposures of 1.0~s each in the first grating position (8.5~m).
 \item Do an arc and change to the second grating position (1.5~m).
 \item Obtain 500 exposures of 1.0~s each in the second grating position (8.5~m).
 \item Do an arc and change to the third grating position (1.5~m).
 \item Obtain 500 exposures of 1.0~s each in the third grating position (8.5~m).
 \item Do an arc and change to the first grating position (1.5~m).
\end{itemize}

That yields 34 minutes/object, a similar amount to the previous one, but with 500~s of total exposure time as opposed to 100~s (QUCAM3 has a slightly lower quantum efficiency than EEV12 but that
effect is small). Therefore, one would expect QUCAM3 to be a more appropriate setup for faint stars, where the S/N per frame is low. For bright
stars, the advantage is lost because the S/N per frame is high and EEV12 has the advantages of being a simpler setup (less chances of making mistakes or having variable quality issues as a 
function of wavelength); covering a larger wavelength range; being easier to schedule as it is the default detector for the instrument; and, for the brightest objects such as some of the ones in 
LS~I, the possibility of having exposure times of 0.1~s for the very bright stars (avoiding saturation), which is not allowed in QUCAM3 as the minimum exposure time is set by the readout time of the 
image. We test the hypothesis about faint stars below with one example.

\begin{figure*}
\centerline{\includegraphics[width=\linewidth]{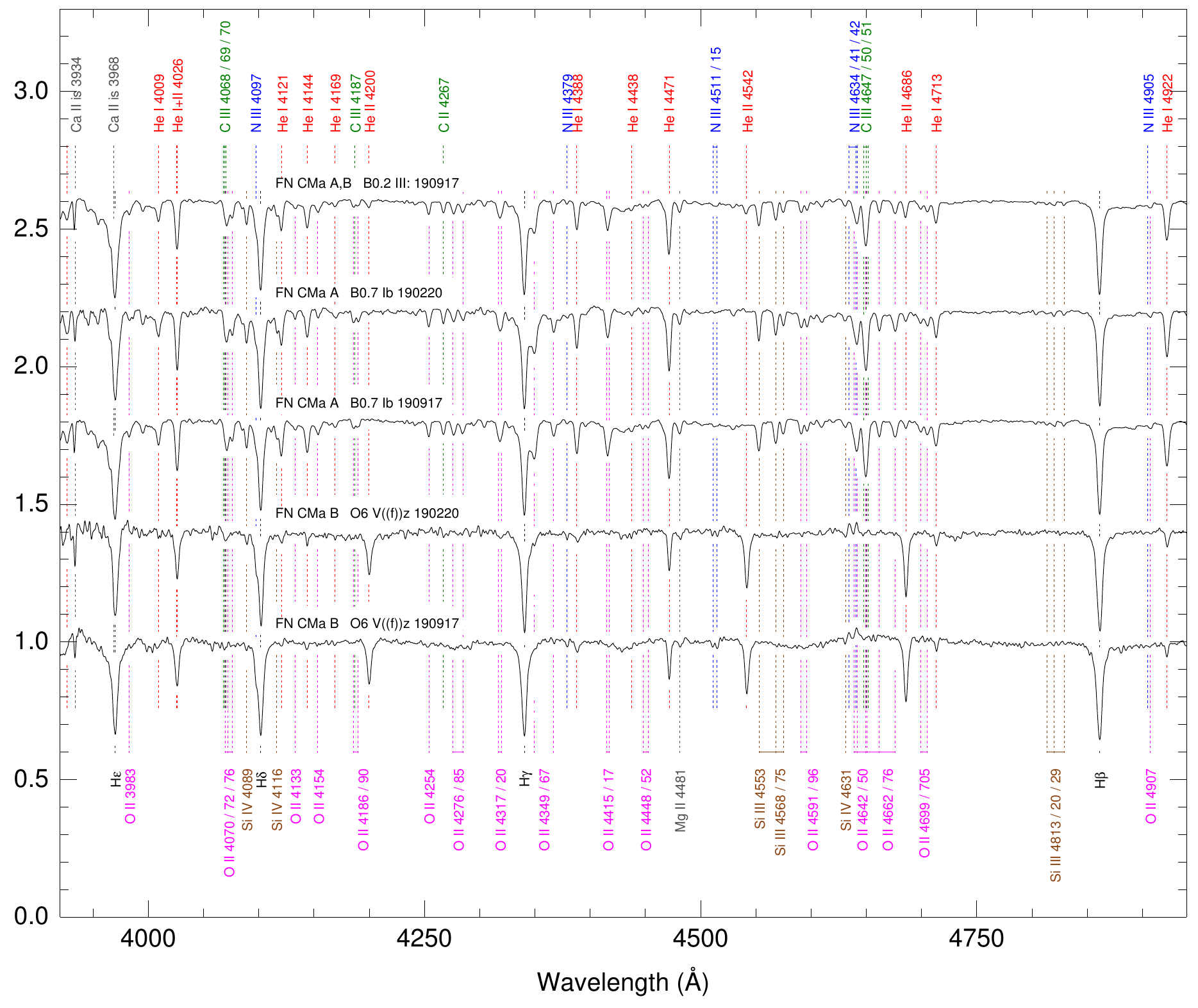}}
\caption{Rectified spectrograms for FN~CMa at the GOSSS spectral resolution of $R\sim 2500$ and on the stellar reference frame.
For each spectrogram, the name, spectral type, and evening date (YYMMDD) are shown. The top spectrogram is the weighted combination of the two
components for one of the epochs. Main atomic and ISM lines are indicated.}
\label{GOSSS_FN_CMa}
\end{figure*}

\subsection{Complementary data}

$\,\!$ \indent For some of our systems we presented lucky imaging results obtained with AstraLux at the 2.2~m CAHA telescope in \citet{Maiz10a} and MONOS~I. Here we present additional 
separations, position angles, and magnitude differences in Table~\ref{AstraLuxdata}. The references above describe the details of the setup and reduction, which have not changed since
the publication of MONOS~I. In this paper we also refer to the information listed in the Washington Double Star (WDS) catalog \citep{Masoetal01} and use astrometric results from 
{\it Gaia}~DR2.

Besides the new lucky spectroscopy data, we also include in this paper results from regular long-slit or IFU spectroscopy from GOSSS. The purpose is to compare the two methods and
to study an additional component in one of the systems. For another system, we also use STIS@HST data already presented in STIS~I as a comparison. Finally, we also use ground-based high-resolution
spectroscopy from \lili\ for some systems.

\section{Results}

$\,\!$ \indent We present our results in Table~\ref{GOSSS_spty}. We discuss each system individually below, comparing the new information with the previously available data. 
{As with our previous results, the spectral classifications and the spectra themselves will be made available at the GOSC web site (\url{http://gosc.cab.inta-csic.es}).} 

%
%
%
%

\subsection{FN~CMa~A,B, a bright O-type system}

$\,\!$ \indent This fifth magnitude system, also known as HD~\num{53974}~A,B, was classified as B0.5~IV by \citet{Morgetal55} and Simbad lists different 
classifications ranging in spectral subtypes from B0 to B3. It is a visual binary with $d$~=~0\farcs6 and $\Delta m$~=~1.21~mag listed in the 
WDS catalog. \citet{Rivietal11b} used spectral disentangling to obtain that the A visual component is a B0~III star and the B component is an O6~V((f)) background object, 
as they deemed the magnitude difference as too large to place two such stars at the same distance. FN CMa appears as a single source with no parallax in {\it Gaia}~DR2.
To estimate its distance we assume that it is the same as component C in the system. FN~CMa~C 
(Gaia~DR2~\num{3046209991397371392}) has a parallax of 0.6844$\pm$0.0509~mas and a RUWE of 0.98. For its distance we use the Bayesian prior described 
by \citet{Maiz01a,Maiz05c} with the updated Galactic (young) disk parameters from \citet{Maizetal08a}, apply a parallax zero point of 0.040$\pm$0.010~mas
(Ma{\'\i}z Apell\'aniz et al., accepted in A\&A), and obtain $1.48^{+0.12}_{-0.11}$~kpc.

We attempted to separate the two components of the system in three occasions and were successful in two of them (Fig.~\ref{GOSSS_FN_CMa}). The primary is indeed
an early B star but of later spectral subtype (B0.7) and higher luminosity class (Ib) than claimed by \citet{Rivietal11b}. The secondary is indeed an O6 dwarf but with
z suffix, thus likely underluminous for its spectral classification \citep{Ariaetal16}. These differences with the \citet{Rivietal11b} classifications explain the 
$\Delta m$ between the two components and allow them to be at the same distance, a more likely arrangement than a chance alignment between two such bright stars.
Note how the two components are cleanly separated, with no sign of \HeII{4542} in the primary or of Si\,{\sc iii} or O\,{\sc ii} lines in the secondary and with 
radically different (but expected for their spectral types) behaviors in the 4630-4650~\AA\ region. No appreciable differences are
seen between the two epochs (but see below for the high-spectral-resolution results). The combined spectrum (top one in Fig.~\ref{GOSSS_FN_CMa}) can be classified
as B0.2~III: but some peculiarities are seen in it that point towards a composite nature, such as the inability to consistently fit the Balmer profiles and the Si 
lines at the same time. This is a good example of the advantage of lucky spectroscopy (spatial disentangling) over spectral disentangling, as the latter can introduce 
artifacts that lead to erroneous spectral characterizations. In this particular case, we suspect that in the \citet{Rivietal11b} disentangling some \HeII{4542} may have 
remained in the spectrum of the primary (but as they did not publish the disentangled spectra we cannot be certain).

\begin{figure*}
\centerline{\includegraphics[width=\linewidth]{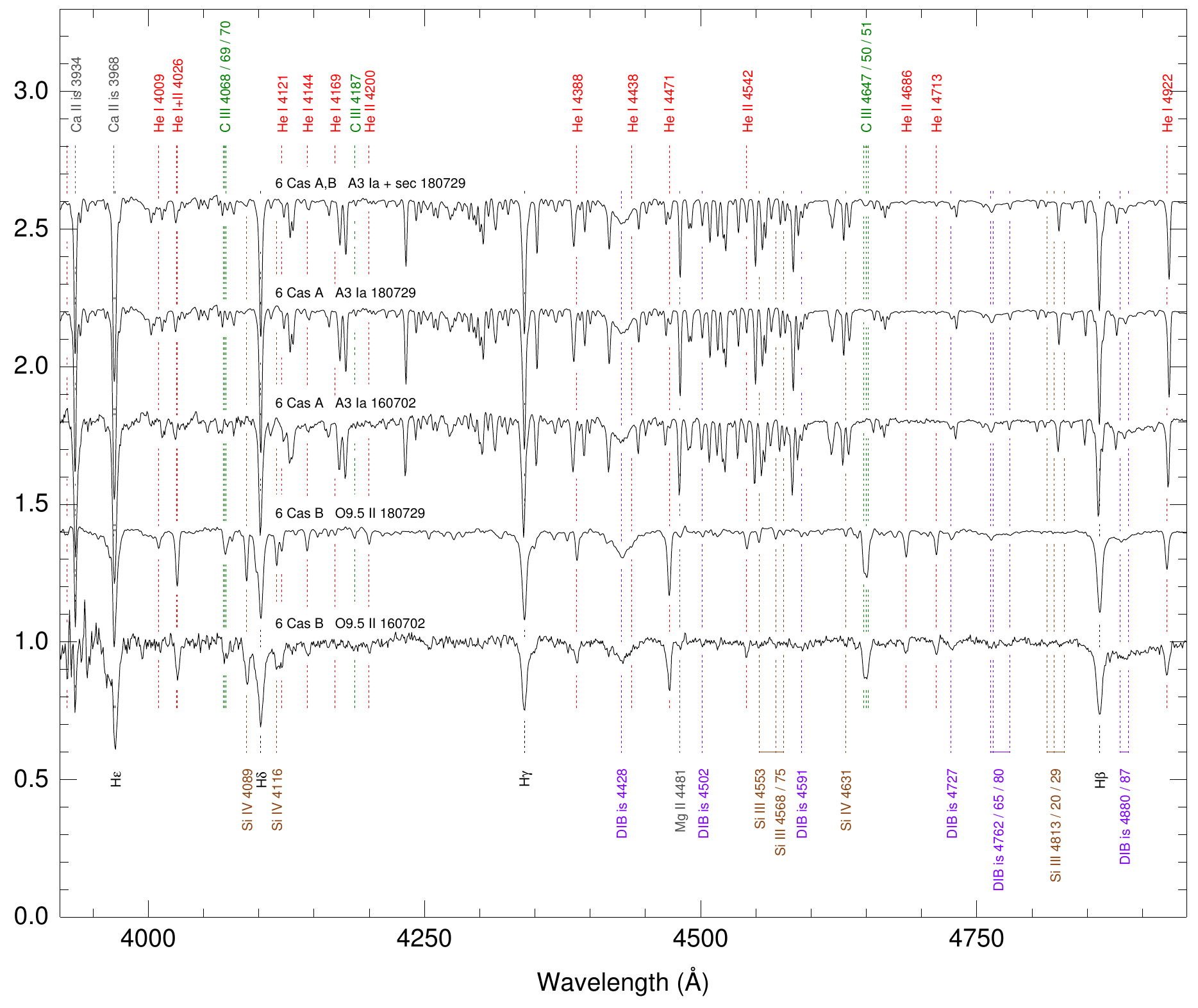}}
\caption{Rectified spectrograms for 6~Cas at the GOSSS spectral resolution of $R\sim 2500$ and on the stellar reference frame.
For each spectrogram, the name, spectral type, and evening date (YYMMDD) are shown. The 160702 data were obtained with FRODOspect@LT i.e. not with lucky spectroscopy.
The top spectrogram is the weighted combination of the two components for the 180729 epoch. Main atomic and ISM lines are indicated.}
\label{GOSSS_6_Cas}
\end{figure*}

We have also observed this system with lucky imaging (Table~\ref{AstraLuxdata}). The separations and position angles are compatible with the WDS data but the 
magnitude difference is just barely so: {\it Hipparcos} gives $\Delta H$~=~1.357~mag and we only expect a color term of $\sim0.1$~mag with $z$ photometry.
There is no appreciable motion in the 6-year span between the two epochs and there is no obvious trend when considering the positions in the WDS catalog that stretch to over a 
century ago.

We have nine epochs of this system in the high-resolution (A+B combined) \lili\ spectra. The He\,{\sc ii} lines (which originate mostly in the O star) do not
show profile changes or velocity shifts. The Si\,{\sc iii} lines (which originate mostly in the B supergiant) show velocity shifts and the He\,{\sc i} lines 
(which originate in both but mostly in the B supergiant) show profile variations and/or velocity shifts. This is consistent with the finding by \citet{Rivietal11b} that
the B-type star is an SB1 with a 117.55$\pm$0.33~d period, making FN~CMa a quadruple system (including the C component 18\arcsec\ away).

\subsection{6~Cas~A,B, the only known O star with an A supergiant companion}

\begin{table}
\caption{{\it Gaia}~DR2 astrometric data for four stars in the vicinity of 6~Cas with similar parallaxes and proper motions and their aggregate results.
         The aggregate results use external uncertainties and include the spatial covariance terms of \citet{Lindetal18b}.}
\centerline{\footnotesize
\addtolength{\tabcolsep}{-3pt}
\begin{tabular}{lr@{$\pm$}lr@{$\pm$}lr@{$\pm$}l}
{\it Gaia} DR2 ID         & \mcii{$\varpi$} & \mcii{\pmra}   & \mcii{\pmdec}  \\
                          & \mcii{(mas)}    & \mcii{(mas/a)} & \mcii{(mas/a)} \\
\hline
\num{2012942869740260096} & 0.3233&0.0319   & $-$3.600&0.046 & $-$1.781&0.043 \\
\num{2012942805332107136} & 0.3457&0.0189   & $-$3.417&0.030 & $-$1.421&0.028 \\
\num{2012942496094500096} & 0.3272&0.0257   & $-$3.442&0.039 & $-$1.387&0.035 \\
\num{2012942564813963264} & 0.3628&0.0263   & $-$3.569&0.041 & $-$1.479&0.038 \\
\hline
                          & 0.3428&0.0402   & $-$3.482&0.063 & $-$1.465&0.063 \\
\hline
\end{tabular}
\addtolength{\tabcolsep}{3pt}
}
\label{Gaia_6_Cas}
\end{table}

$\,\!$ \indent This fifth magnitude system, also known as HD~\num{223385}~A,B, was classified as A3~Ia~+~sec by \citet{Morgetal53b}, where we explicitly add ``sec'' to 
indicate that the plus sign does not refer to a hypergiant classification (a common source of confusion, note that hypergiants had not been introduced at the time). 
The B companion is currently listed in the WDS catalog
with $d$~=~1\farcs5, PA~=~194\degr, and $\Delta m$~=~2.29~mag measured in 2015. \citet{TalaGome87} were the first to suggest that the A supergiant 
had an O-type companion based on the composite UV spectrum but favored a spectroscopic, not a visual binary. The two components have separate entries in {\it Gaia}~DR2
but their parallaxes cannot be used to derive precise distances: that of A is negative and that of B has a RUWE of 5.15, indicating poor astrometric quality, likely caused
by the proximity of the A component. The {\it Gaia}~DR2 $\Delta G$ should be reliable, with a value of 2.529$\pm$0.016~mag for the A.B pair, but the $\Delta\GBP$ and 
$\Delta\GRP$ are likely not, as they are measured by aperture photometry and the two components are too close. The {\it Gaia}~DR2 positions yield a separation of 
1\farcs5016 and a position angle of 195.89\degr.

\begin{figure*}
\centerline{\includegraphics[width=\linewidth]{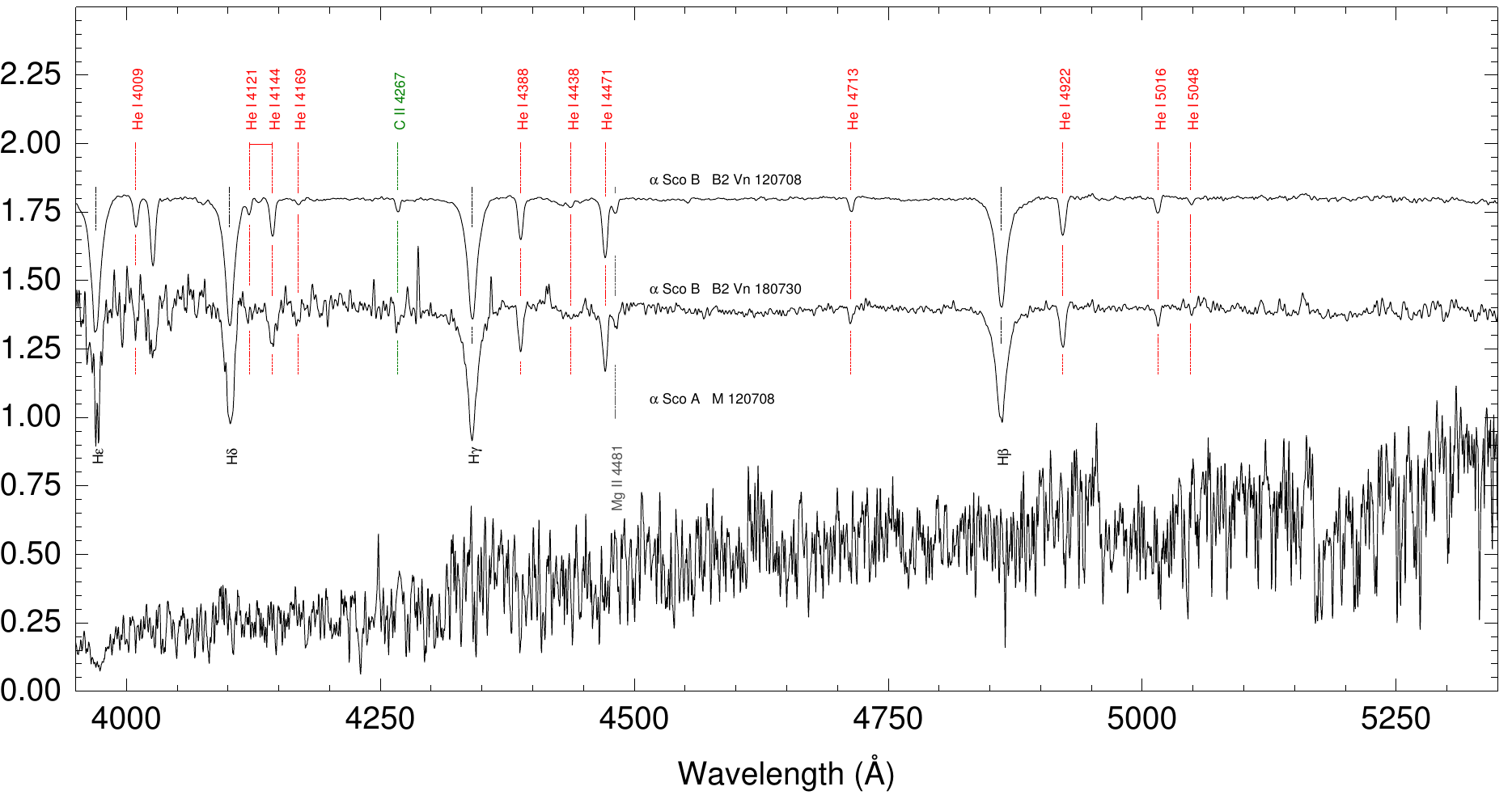}}
\centerline{\includegraphics[width=\linewidth]{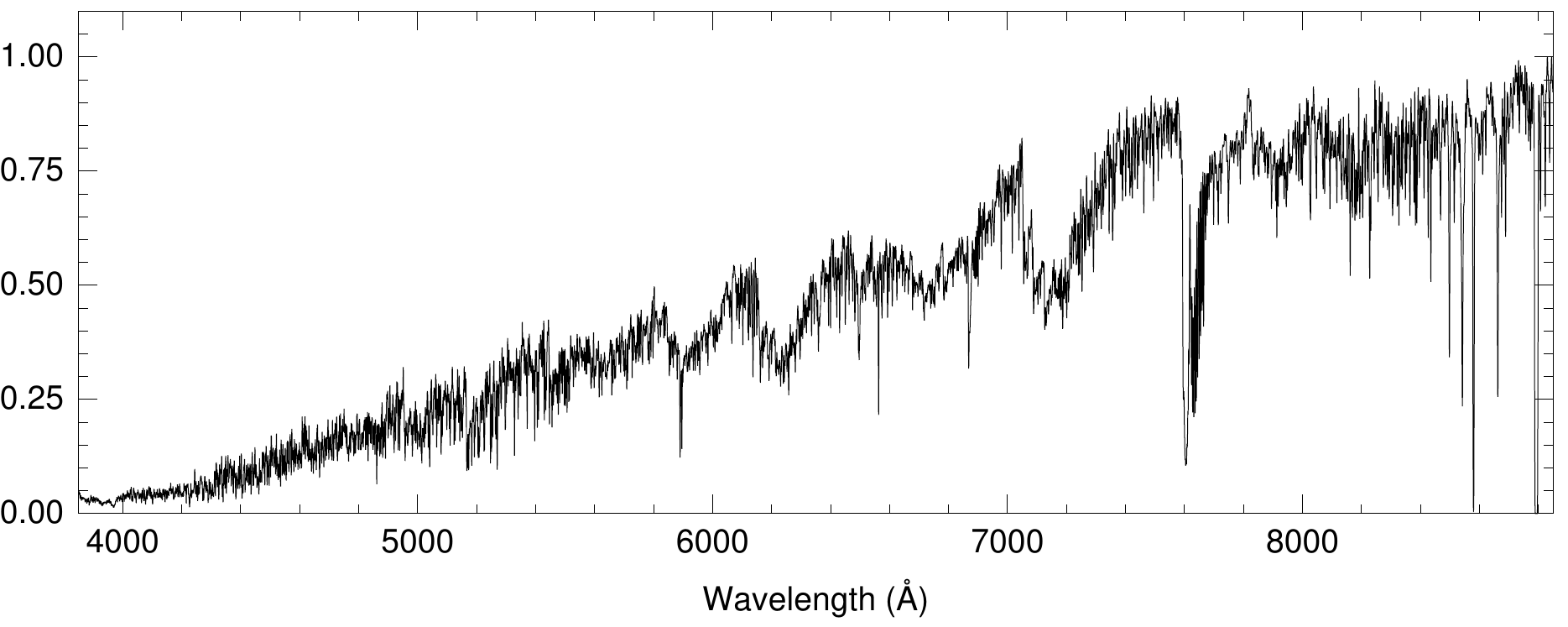}}
\caption{(Upper panel) Lucky spectrograms for $\alpha$~Sco at the GOSSS spectral resolution of $R\sim 2500$ and on the stellar reference frame.
For each spectrogram, the name, spectral type, and evening date (YYMMDD) are shown. The 120730 data were obtained with a position angle offset of 24\degr\ from the A,B line
and the 180730 data were obtained with the slit aligned with the A,B line. The spectra of the B component have been rectified while the spectrum of the A component has been 
dereddened and relative-flux calibrated. Main atomic lines are indicated. (Lower panel) $\alpha$~Sco~A spectrogram from \lili\ degraded to $R\sim 4000$, relative-flux calibrated, 
and not corrected for telluric absorption.}
\label{GOSSS_alpha_Sco}
\end{figure*}

As we cannot use the {\it Gaia}~DR2 data for 6~Cas~A,B to obtain a reliable distance to the system, we looked for nearby stars that may be associated with it. The WDS catalog lists a 
C component but it has a RUWE of 3.87, so we cannot use it. Instead, we use a variation of the method used in \citet{Maiz19}. We start with the {\it Hipparcos} proper motion for 
6~Cas and we search for {\it Gaia}~DR2 stars in the vicinity with (a) similar proper motions, (b) good RUWE, (c) $G$ magnitudes brighter than 15, (d) similar blue colors, 
and (e) self-compatible parallaxes. We found four of those, listed in Table~\ref{Gaia_6_Cas}. We combine their parallaxes using the technique described in \citet{Campetal19},
including the use of external uncertainties and a spatial covariance term \citep{Lindetal18b}, and apply the same parallax zero point and prior as before. The resulting distance 
to that group of stars is $2.78^{+0.37}_{-0.29}$~kpc, which we will use here for 6~Cas itself.

We successfully separated the two components of this system on eight occasions with lucky spectroscopy. As no significant variations in the spectral types were observed 
between epochs, we only plot in Fig.~\ref{GOSSS_6_Cas} one lucky spectroscopy epoch (including the combined spectrum) along with another epoch of the system where we were
able to obtain separate spectra with FRODOspec@LT (regular GOSSS spectroscopy). The two spectra are cleanly separated despite their very different spectral types with no sign
of cross-contamination. 6~Cas~B has an O9.5~II spectral type with no sign of any anomaly. The effect of the B component on the combined spectrum is very weak and only shows 
up in a few lines such as \SiIV{4089}, \CIIIt{4647,4650,4651}, \HeII{4686}, and \HeI{4713} located in wavelength regions without strong lines from the A supergiant. The 
FRODOspec spectrograms indicate that this system can also be resolved with regular spectroscopy but at a cost. While the two spectra for the A component in 
Fig.~\ref{GOSSS_6_Cas} are nearly identical in spectral resolution and S/N, the lucky spectrogram for the B component has a higher S/N than its regular counterpart due to 
the reduced noise introduced in the spatial disentangling process. 

We have two lucky-imaging epochs of this system (Table~\ref{AstraLuxdata}) spaced by a year and obtained prior to the average {\it Gaia}~DR2 reference epoch (2015.5). 
Combining those points with the historical data in the WDS catalog, there is little indication of a secular motion, except perhaps in the inward direction, with most observations from the 
nineteenth century and the beginning of the twentieth having separations of 1\farcs6-1\farcs7. The observed magnitude differences in our lucky images are consistent with the 
$\Delta G$ value above and the spectral types.

We have 12 \lili\ epochs covering from 2001 to 2020. The lines of the A supergiant show radial velocity variations with a peak-to-peak amplitude of $\sim$~10~km/s. The motion is 
not secular but shows variations on time scales of days, pointing towards pulsations as the likely culprit. The phenomenon was already noticed by \citet{Aydi79}.
The signature of the B component is difficult to detect in the \lili\ spectra because of the additional dilution caused by 1\farcs5 separation and the circular apertures of 
less than 3\arcsec\ diameter used by \'echelle spectrographs. In some of our epochs, however, \HeII{5412} is clearly visible.

\subsection{$\alpha$~Sco~A,B, the nearest RSG+B system}

\begin{figure*}
\centerline{\includegraphics[width=\linewidth]{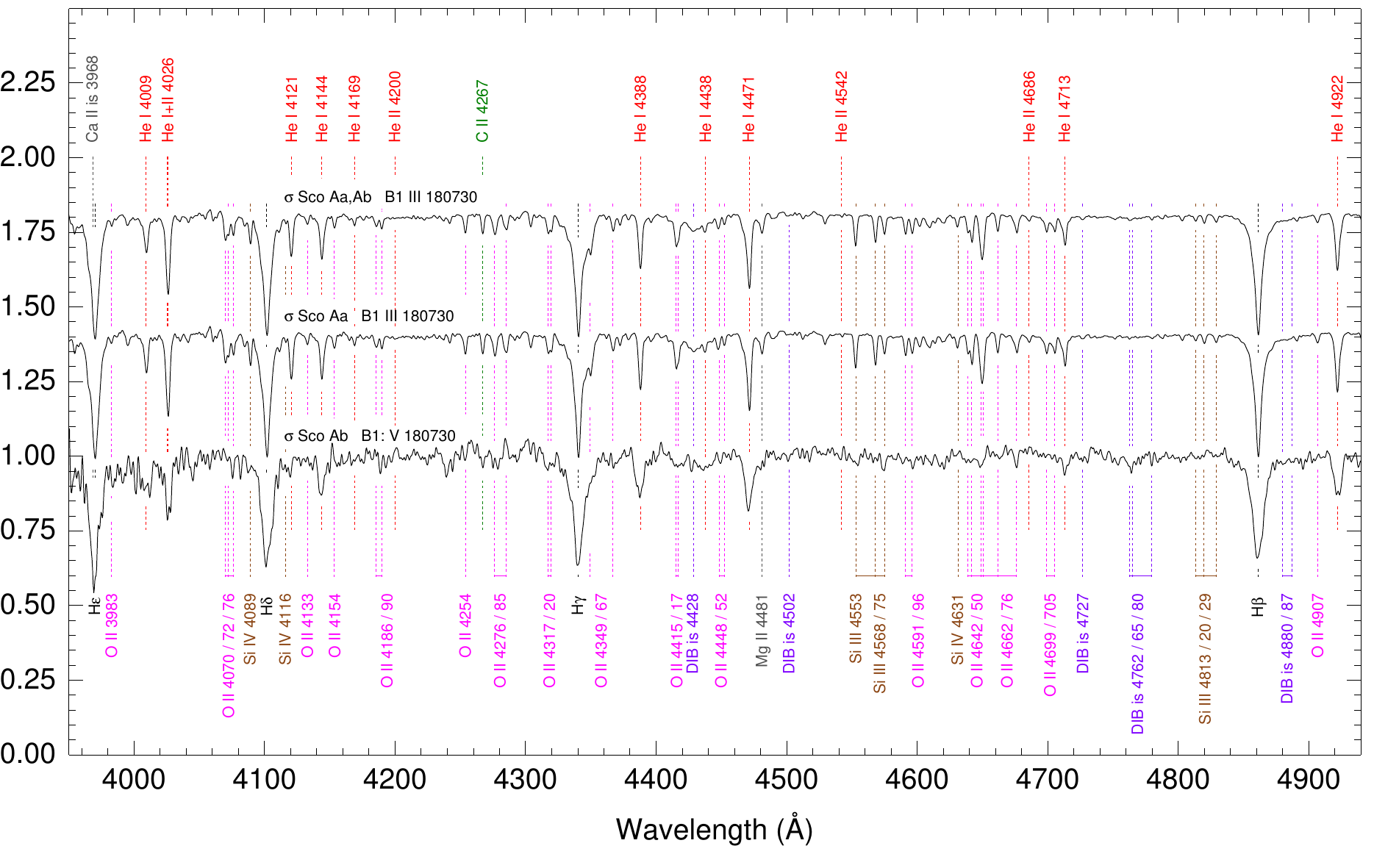}}
\caption{Rectified spectrograms for $\sigma$~Sco at the GOSSS spectral resolution of $R\sim 2500$ and on the stellar reference frame.
For each spectrogram, the name, spectral type, and evening date (YYMMDD) are shown. The top spectrogram is the weighted combination of the two components for the 180730 epoch. 
Main atomic and ISM lines are indicated.}
\label{GOSSS_sigma_Sco}
\end{figure*}

$\,\!$ \indent There are over a hundred known red supergiant stars with B-type companions in the Galaxy \citep{Pantetal20}, of which $\alpha$~Sco (Antares) is the closest one, with
a {\it Hipparcos} distance ($\alpha$~Sco is not present in {\it Gaia}~DR2) of $187^{+44}_{-30}$~pc \citep{Maizetal08a}. The companion was classified as B3 already one century ago
\citep{AdamJoy21} and subsequent authors classified it as B4~V \citep{StonStru54}, B2.5~V \citep{Garr67}, and B3~V \citep{Corb84}. The differences in classification likely do not
reflect real spectral variations but are rather a consequence of the difficulty that exists in eliminating the light from the much brighter primary less than 3\arcsec\ away. The 
B~star has not received much attention in the last decades other than the study of its associated emission nebula rich in [Fe\,{\sc ii}] lines \citep{SwinPres78,Reimetal08}. The
magnitude difference between the two components is highly dependent on the band used (see below), as expected for a pair composed of a red and a blue star (this pair is a favorite 
object among amateur astronomers who want to separate visual binaries of extreme colors and small separations, the first author still remembers his excitement when he was able to 
separate Antares through an eyepiece for the first time 34 years ago). 

We observed $\alpha$~Sco~A,B with lucky spectroscopy on four occasions and we were successful in all of them but with very different values of the S/N for the B component, an effect 
likely caused by the large magnitude difference. We measure that value in our data to be $\sim$~3.6~mag in the $B$ band but with a significant difference between 4000~\AA\ and 5000~\AA. 
Of the four epochs, the one with the best seeing (and consequently higher S/N) is the first one from 2012, when we first attempted lucky spectroscopy with ISIS@WHT. On that occasion
we centered the slit on the B component and used a position angle offset of 24\degr\ with respect to the A-B line. This reduces the effective separation slightly but it is heavily 
compensated by the effect of making the flux from the two components similar in the $B$ band, which is the most important constraint in this case (see Table~\ref{sample}). We classify the
B~component as B2~Vn, which differs from previous results in two ways: the spectral subtype is earlier and the star shows a significant line broadening likely caused by rotation.
We used the semi-automatized tool iacob-broad to measure the projected rotational velocity ($v\sin i$) from a combined Fourier transform plus goodness-of-fit technique (see 
\citealt{SimDHerr14} and references therein) using the top spectrum in Fig~\ref{GOSSS_alpha_Sco}. The absence of strong metallic lines, due to the significant line broadening, made us 
rely on He\,{\sc i} lines. This means the Stark broadening effect needs to be considered and, therefore, the amount of macroturbulent broadening cannot be adjusted. From the fitting of 
four different lines (\HeI{4387}, \HeI{4471}, \HeI{4713}, and \HeI{4922}) and the agreement between results obtained by both techniques used as consistency check (see e.g.
\citealt{Berletal20}) we obtain a $v\sin i$ for the B component of 260$\pm$30 km/s. In this way, $\alpha$~Sco~B is another example of a massive-star companion with a high rotation (see 
LS~I and STIS~I).

\citet{MorgKeen73} selected $\alpha$~Sco~A as the MK M1.5~Iab standard, although \citet{KeenMcNe89} later gave it a slightly lower luminosity at M1.5~Iab-Ib. There are many spectral 
classifications in the literature indicating earlier types, but the strength of the TiO band system in the spectrum rules them out. We have used the Mercator spectrogram from \lili\ shown in
Fig.~\ref{GOSSS_alpha_Sco} to classify $\alpha$~Sco~A following the procedures described in \citet{Dordetal18b} and their set of standards at $R=4000$. Direct comparison to HD~\num{206936}
(M2~Ia) and HD~\num{36389} (M2~Iab-Ib) shows all the temperature sensitive TiO bandheads to be essentially identical in $\alpha$~Sco~A, although $\alpha$~Ori, nominally M1-2~Ia-Iab, has 
very slightly stronger bands. The strength of the Ca\,{\sc ii} triplet and other luminosity indicators in its range agree well with those in the blue, and we consider the M1.5~Iab 
classification fully justified but noting the difficulty in distinguishing between M1.5 and M2.

We have one lucky imaging epoch of $\alpha$~Sco. The WDS catalog contains data on the system that spans a century and three quarters but the dispersion is large, likely an effect of the 
large magnitude difference between A and B. There is some evidence of a small inward and counterclockwise motion over that time span but the period must be of at least several
thousands of years.

\subsection{$\sigma$~Sco~Aa,Ab, a spectral classification and extinction standard}

\begin{figure*}
\centerline{\includegraphics[width=\linewidth]{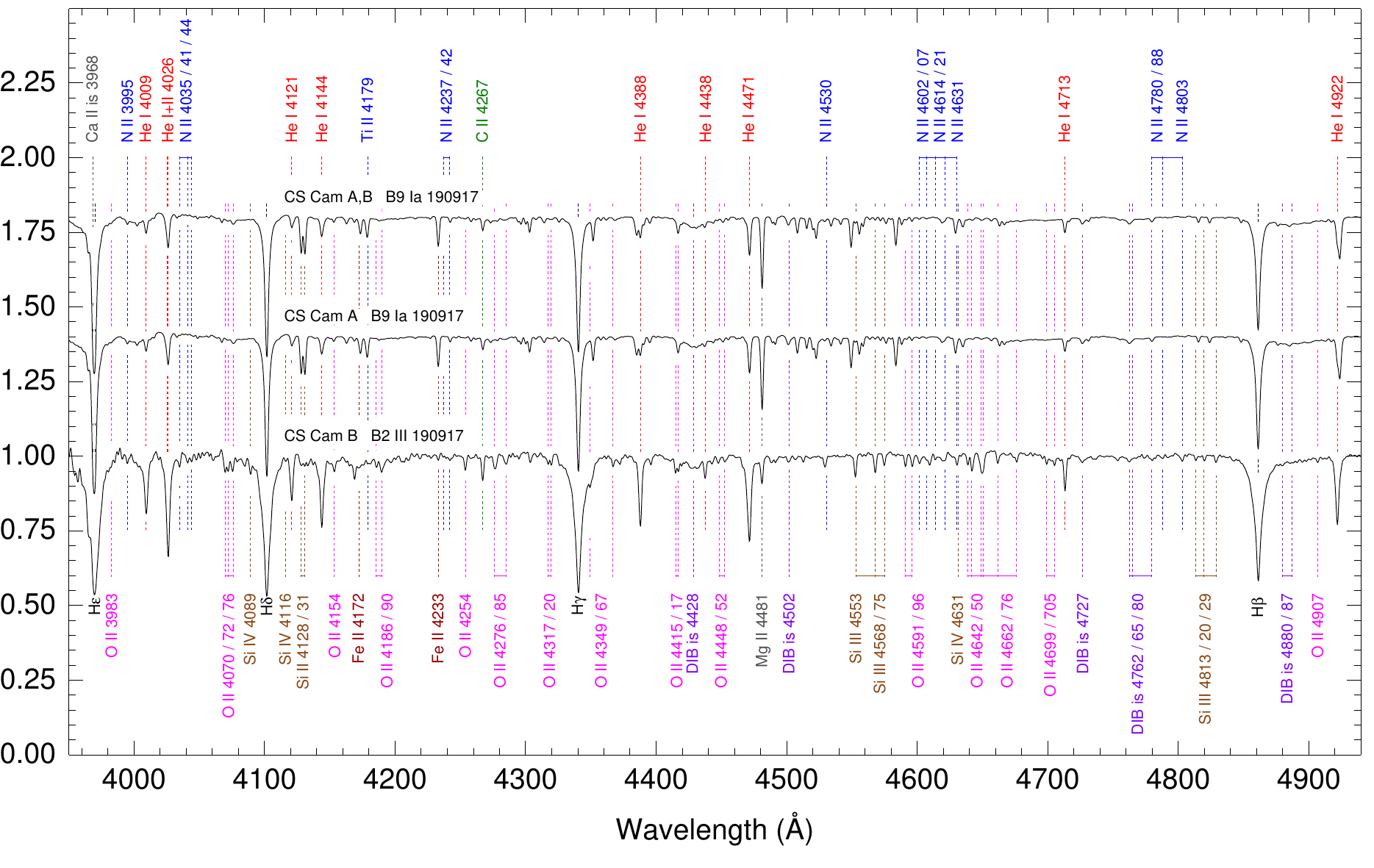}}
\caption{Rectified spectrograms for CS~Cam at the GOSSS spectral resolution of $R\sim 2500$ and on the stellar reference frame.
For each spectrogram, the name, spectral type, and evening date (YYMMDD) are shown. The top spectrogram is the weighted combination of the two components for the 190917 epoch. 
Main atomic and ISM lines are indicated.}
\label{GOSSS_CS_Cam}
\end{figure*}

$\,\!$ \indent This is a well-known object that was used as one of the spectral classification standards for B1~III by \citet{JohnMorg53}. More recently, it has also been 
used as an extinction standard as it is the prototype of $\sigma$-type sightlines \citep{Kreletal97,Camietal97,Maiz15a} in a similar way as $\zeta$~Oph is the prototype of 
$\zeta$-type sightlines. $\sigma$-type sightlines correspond to relatively lower-density regions of the ISM exposed to UV radiation while $\zeta$-type sightlines correspond to 
higher-density regions of the ISM shielded from UV radiation. This division was initially applied to diffuse interstellar bands but it has been recently discovered that a broad 
interstellar feature preferentially appears on $\sigma$-type sightlines (Ma\'{\i}z Apell\'aniz et al. 2020, accepted in MNRAS). There is no entry for this target in {\it Gaia}~DR2
but from the {\it Hipparcos}~data and the \citet{Maizetal08a} prior we derive a distance of $222^{+34}_{-26}$~pc. Alternatively, \citet{Nortetal07b} derive a distance of 
$174^{+23}_{-18}$~pc from the Aa1,Aa2 orbit (see below).

$\sigma$~Sco has long been known to be a visual multiple. A faint companion (B) has recorded positions in the WDS catalog that date back to the eighteenth century. The Aa,Ab pair was
discovered in the mid twentieth century \citep{Fins56} with a 2.2~mag difference and $d$~=~0\farcs3-0\farcs5. Finally, the Aa component is a close visual + spectroscopic
binary (Aa1,Aa2) with a 33~d eccentric orbit measured with interferometry and a 0.8~mag difference \citep{Nortetal07b}. Those authors estimate the spectral type of Aa2 to be B1 V 
but note that this is not a true spectral classification. Currently Simbad gives as a spectral classification for $\sigma$~Sco~Aa,Ab O9.5~+~B7, taken from \citet{BeavCook80}, who 
attempted the spectral separation from an unresolved (either spatially or spectroscopically) spectrum of the system without acknowledging the existence of three components in the 
system. To our knowledge, no spatially resolved spectroscopy of Aa,Ab exists in the literature.

We successfully spatially separated $\sigma$~Sco,Aa,Ab once with lucky spectroscopy (Fig.~\ref{GOSSS_sigma_Sco}). The Aa component appears as a single-lined B1~III object. It is not
surprising we do not see double lines, as the maximum velocity separation for the Aa1,Aa2 orbit is just 87~km/s \citep{Nortetal07b}, which is too low for a spectral resolution of
2500. The Ab spectrum is relatively noisy but we can constrain its spectral subtype by the absence of He\,{\sc ii} lines and the weakness of \MgII{4481} and its luminosity class by
the width of its Balmer lines: it must be a B dwarf around B1. We clearly discard the B7 dwarf classification (or even any mid-B classification) by \citet{BeavCook80}, as the Balmer
lines are too narrow for that and \MgII{4481} is weak. What about the magnitude difference with Aa? Is it compatible with the luminosity class difference at a similar spectral
subtype? It is, especially when considering that the difference between Aa1 and Ab is just 1.8~mag (accounting for the 0.8~mag between Aa1 and Aa2). Given the similar spectral types
and relatively large magnitude difference between Aa and Ab, no significant contribution from Ab is detected in the combined spectrum at the spectral resolution of 2500.

We have one lucky imaging epoch of $\sigma$~Sco and the magnitude differences we measure are compatible with those in the WDS catalog. The WDS measurements of the Aa,Ab pair
go back only to the 1970s but that is enough to reveal a significant clockwise motion with little variation in separation (possibly with a maximum around 2010) e.g. {\it Hipparcos}
measured $d$~=~0\farcs428 and PA~=~58\degr\ in 1991.25. Assuming a circular face-on orbit we obtain a period of 230-250~a. 

There are 105 high-resolution epochs of $\sigma$~Sco,Aa,Ab in \lili. Even though Aa1, Aa2, and Ab are inside the aperture, double lines are seen in many epochs but no triple lines, 
which is expected given the relatively small velocity amplitudes and how faint Ab is. In the future we plan to do spectral disentangling of the \lili\ data to confirm some of the
results in this subsection.

\begin{figure*}
\centerline{\includegraphics[width=\linewidth]{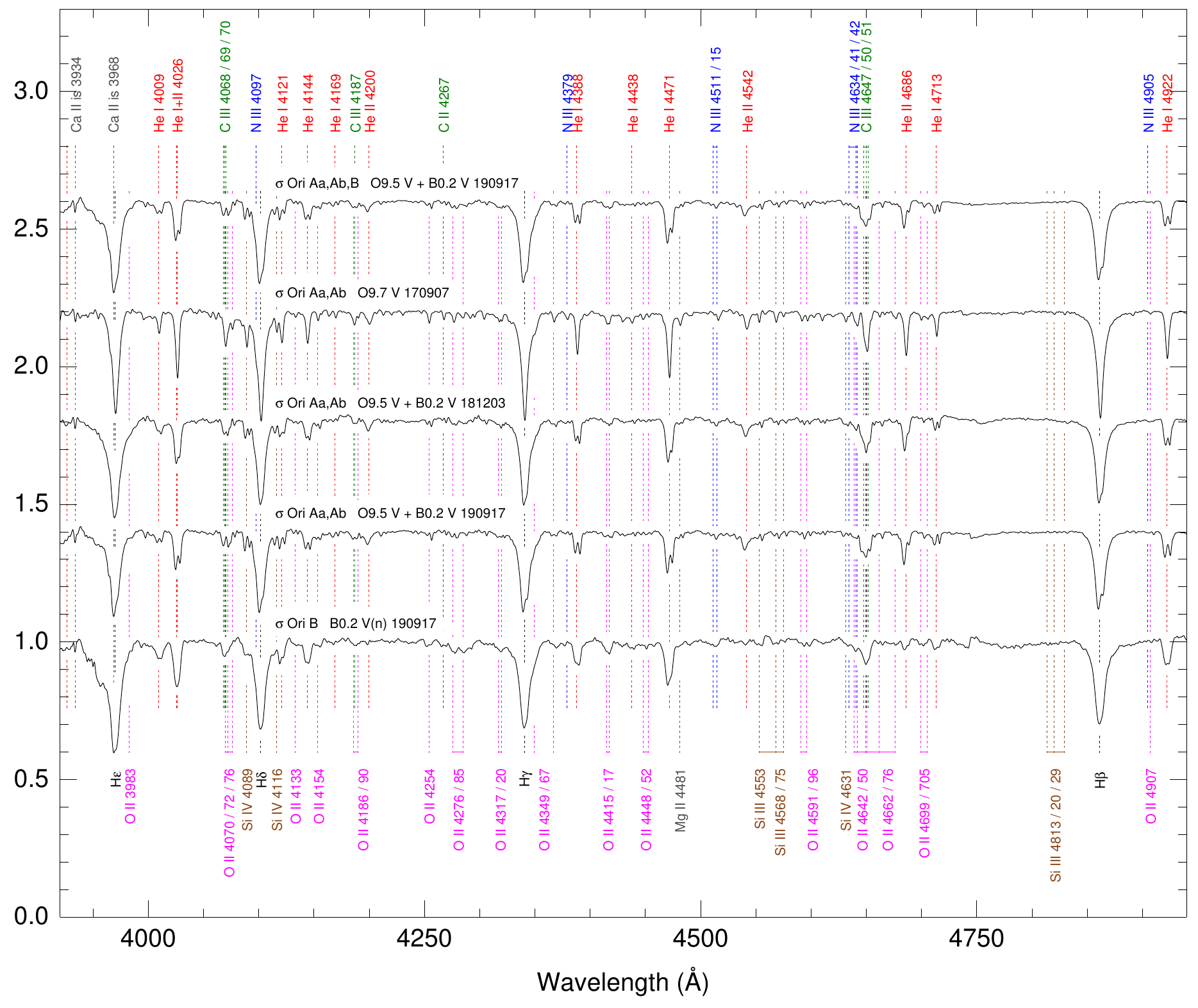}}
\caption{Rectified spectrograms for $\sigma$~Ori at the GOSSS spectral resolution of $R\sim 2500$ and on the heliocentric reference frame.
For each spectrogram, the name, spectral type, and evening date (YYMMDD) are shown. The top spectrogram is the weighted combination of the two components for the 190917 epoch. 
Main atomic and ISM lines are indicated.}
\label{GOSSS_sigma_Ori}
\end{figure*}

\subsection{CS~Cam~A,B, a pair of luminous B stars, one early and one late}

$\,\!$ \indent CS~Cam~A is a bright supergiant used by \citet{MorgRoma50} as one of their B9~Ia standards. It has a B companion located 2\farcs3 away and with a 3.6~mag difference in
the Tycho-2 bands. The companion has been mostly ignored in the literature. The {\it Gaia}~DR2 parallax for A combined with the \citet{Maizetal08a} prior and the previously used
zero point leads to a distance of $813^{+208}_{-138}$~pc. The B component also has an entry in {\it Gaia}~DR2 but its RUWE is larger than 1.4.

We successfully spatially separated CS~Cam~A,B twice with lucky spectroscopy. We show the results for one of the epochs in Fig.~\ref{GOSSS_CS_Cam}). The A component appears as a 
B9~Ia, as expected from the literature. We classify the B component as a B2~III, earlier and less luminous, which is its first ever spectral classification. In the combined spectrum
there is no sign of the secondary, other than very slight changes in some He\,{\sc i} lines that would go easily unnoticed, at least at this spectral resolution.

There are 35 \lili\ epochs of CS~Cam, where only the A component leaves a significant signal, as B is not only considerably fainter but also outside the aperture in most cases. In
the nine years covered by the observations there is no significant motion but a velocity dispersion of 2-3~km/s, a common effect in B supergiants caused by pulsations
(Sim\'on-D\'{\i}az et al., submitted to A\&A).

We have one lucky imaging epoch of CS~Cam. The magnitude differences in the $z$ and $Y$ bands are larger than in the {\it Hipparcos} $H$ band, which in turn are larger than in the 
Tycho-2 $B$ and $V$ bands, following the color differences between early- and late-type B stars. The {\it Gaia}~DR2 $G$ magnitude difference is larger but suspect, given the
corrections associated with stars that are as bright as CS~Cam~A (future data releases may change this). The WDS catalog lists pair measurements that go back 
almost two centuries with no appreciable motion detected but with a considerable scatter in the oldest measurements. Both components appear in {\it Gaia}~DR2 catalog and the
astrometric data there are in excellent agreement with our measurements. On the other hand, {\it Gaia}~DR2 detects a relative proper motion mostly in the inward direction that 
should have accumulated $\sim$0\farcs4 over the two centuries of WDS observations. One way to reconcile both results would be if the system has a near edge-on orbit that passed 
near quadrature a century or so ago. 

\begin{figure*}
\centerline{\includegraphics[width=\linewidth]{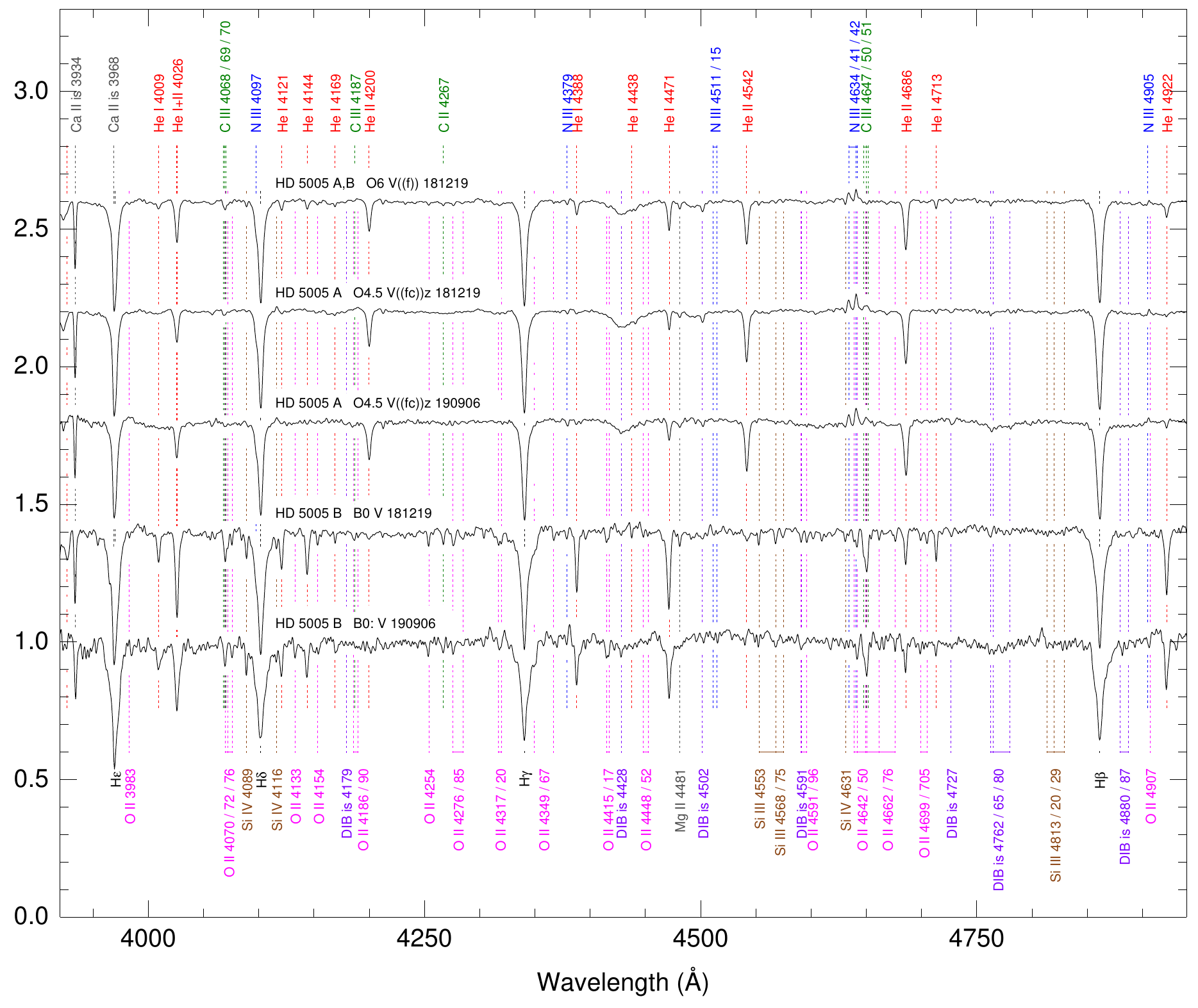}}
\caption{Rectified spectrograms for HD~5005 at the GOSSS spectral resolution of $R\sim 2500$ and on the stellar reference frame.
For each spectrogram, the name, spectral type, and evening date (YYMMDD) are shown. The top spectrogram is the weighted combination of the two components for the 181219 epoch. 
Main atomic and ISM lines are indicated.}
\label{GOSSS_HD_5005}
\end{figure*}

\subsection{$\sigma$~Ori~Aa,Ab,B, an epoch with a better velocity separation}
\label{sec_sigma_Ori}

$\,\!$ \indent This system has been already extensively described in \citet{SimDetal15a}, LS~I, and MONOS~I, to which the reader is referred for information about it. We presented 
the first spatially resolved spectra of $\sigma$~Ori~Aa,Ab and $\sigma$~Ori~B in LS~I. That observation took place on the night of 170907 (HJD~=~\num{2458004.74}), for which the 
spectroscopic orbit of Aa,Ab by \citet{SimDetal15a} predicts a velocity separation of just 15~km/s and, understandably, we were unable to kinematically separate the two components 
in the spectroscopic binary. In MONOS~I we spatially separated the Aa,Ab and B components with a new lucky spectroscopy observation obtained on the night of 181203 
(HJD~=~\num{2458456.54}). The prediction for the velocity separation at that time was much more favorable, 234~km/s, and indeed we were able to kinematically separate Aa and Ab and 
give spectral classifications for each. Here we present new spatially resolved spectra for the night of 190917 (HJD~=~\num{2458744.69}), where the predicted velocity separation was 
even larger, 288~km/s, just 1~km/s shy of the maximum value (Fig.~\ref{GOSSS_sigma_Ori}). 

The new spectral classifications are the same as the ones in MONOS~I, providing reassurance that once the velocity separation is large enough and lucky spectroscopy is performed
under good seeing conditions the result is repeatable. The spectral classification for the combined spectrum Aa,Ab,B is the same as for Aa,Ab, as the large $v\sin i$ of B and the
relative similarity between the three spectral types simply creates a dilution of the lines, a point that was already made in Fig.~5 of \citet{SimDetal15a}. The good velocity
separation between Aa and Ab allows us to make a precise measurement of the flux ratio between the two of 0.70$\pm$0.05 or about 0.4 mag, which is larger than the value in
\citet{SimDetal15a} but not by much. We also note that the latest spectrogram of $\sigma$~Ori~B could be alternatively interpreted as a partially resolved SB2 of two similar stars, 
given the asymmetry of some lines such as \HeI{4471}, which would make the $\sigma$~Ori system an SB2+SB2. However, the effect may not be real but the effect of pulsations or of 
noise in the spatial disentangling. Further observations may solve the issue, possibly lucky spectroscopy at other phases to detect a larger velocity amplitude in the alleged SB2 
or interferometry to spatially resolve it.

\begin{figure*}
\centerline{\includegraphics[width=\linewidth]{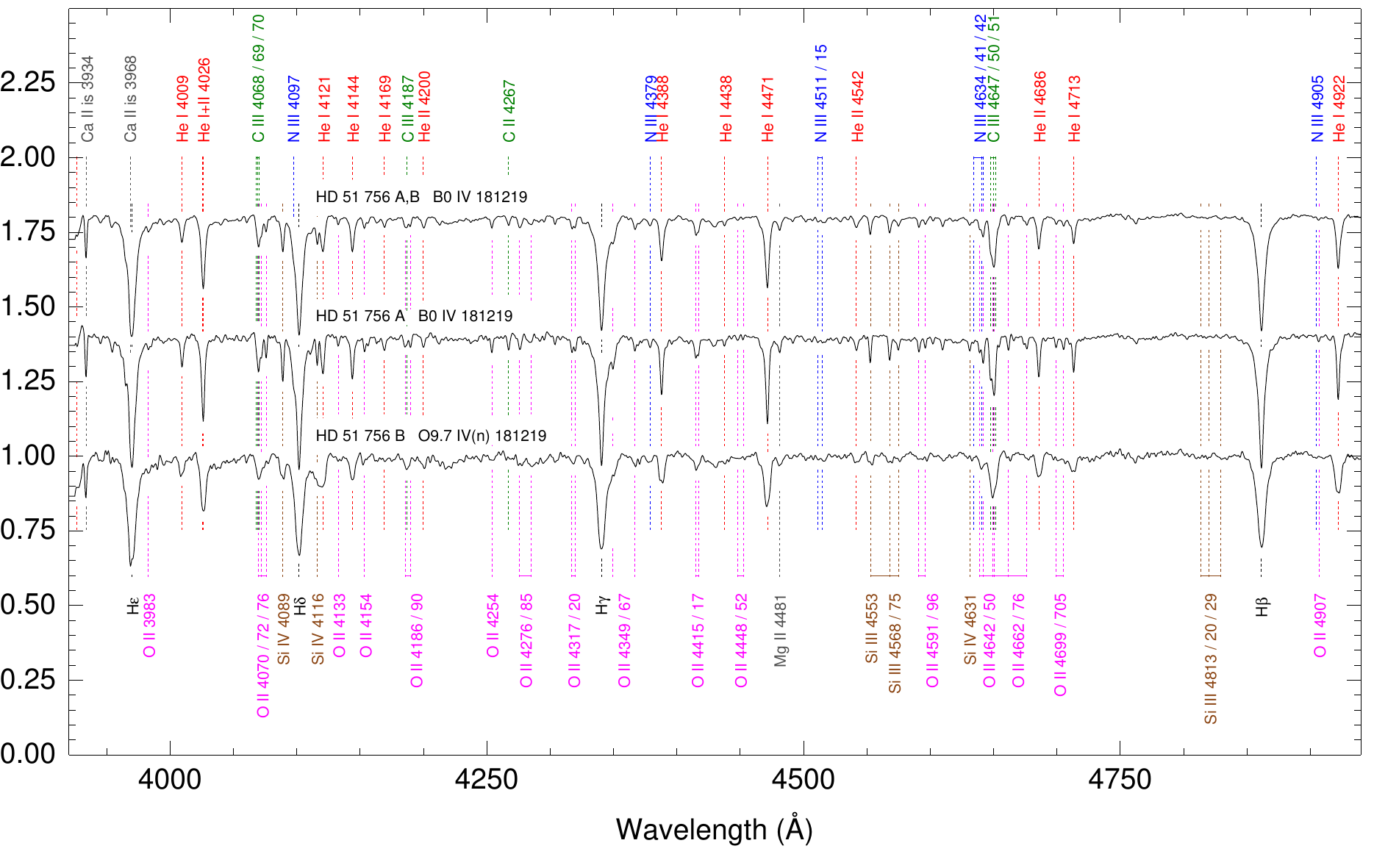}}
\caption{Rectified spectrograms for HD~\num{51756} at the GOSSS spectral resolution of $R\sim 2500$ and on the stellar reference frame.
For each spectrogram, the name, spectral type, and evening date (YYMMDD) are shown. The top spectrogram is the weighted combination of the two components for the 181219 epoch. 
Main atomic and ISM lines are indicated.}
\label{GOSSS_HD_51_756}
\end{figure*}

\subsection{HD~5005~A,B,C,D, a trapezium system with an early-type O star}

$\,\!$ \indent HD~5005 is a Trapezium system at the core of the IC~1590 cluster that ionizes the NGC~281 H\,{\sc ii} region, also known as the Pacman nebula. In GOSSS~I we classified
its four main components (A, B, C, and D) as O stars. In the case of A and B we based the spectral classifications on a spatial disentangling of the two components, which have 
$d$~=~1\farcs6 and $\Delta m$~=~1.6~mag. All four main components have {\it Gaia}~DR2 entries with good values of RUWE so we can combine them in the same way we did for 
6~Cas to obtain a distance to HD~5005 of $2.53^{+0.33}_{-0.27}$~kpc.

We have used lucky spectroscopy to spatially separate HD~5005~A,B in two epochs (Fig.~\ref{GOSSS_HD_5005}). The two epochs yield the same spectral classification for the A
component, O4.5~V((fc))z, similar to that of GOSSS~I but later by half a spectral subtype and with the z suffix that indicates that \HeII{4686} is deeper than normal stars with V
luminosity class. For the B component we get B0~V for both epochs but with an uncertain spectral subtype for the 190906 epoch due to the lower S/N generated by the spatial
disentangling. That spectral classification is later than the O9.5~V from GOSSS~I and moves HD~5005~B from the O-star to the B-star category. This is a good example of how lucky
spectroscopy allows for a cleaner separation of the spectra of close binaries compared to the traditional, long-exposure methods.

The first lucky spectroscopy epoch (181219) was obtained with the QUCAM3 setup and the second one (190906) with the EEV12 setup. The HD~5005~B QUCAM3 setup has a better S/N, as we suspected from
the comparison between the two setups, confirming that it is preferred to the EEV12 one for faint stars. Nevertheless, both setups produce valid and consistent results. 

The A,B spectrum in Fig.~\ref{GOSSS_HD_5005} is of pedagogical value. When one combines the two components (an O4.5 and a B0) we obtain an O6~V((f)) star with no major abnormalities.
For example, the \CIIIt{4647,4650,4651} triplet is in weak emission in A (hence, the c suffix) and in strong absorption in B. When both are combined (weighted by the $\Delta m$), the
feature disappears, as expected for a normal O6 dwarf. On the other hand, the \NIIIt{4634,4641,4642} triplet, also in emission in A and in absorption in B is left in emission in the
combined spectrum (as expected for an object with an ((f)) suffix) because they have similar strengths in each component and the magnitude difference allows the emission to dominate.
Therefore, it is possible for such an unresolved binary system to remain undetected in single-epoch spectroscopy with $R\sim2500$ and to make us believe that the object is a single
star of intermediate spectral type.

There are 9 \lili\ epochs of HD~5005~A,B but the data are complicated to interpret, as the separation between the two components combined with the different apertures for each
spectrograph and seeing variations make the contribution of each of the two components different in each epoch. The He\,{\sc ii} lines, dominated by A, are relatively stable in
intensity (and also in velocity, pointing towards the absence of a spectroscopic companion). On the other hand, \HeI{4922}, dominated by the B component (Fig.~\ref{GOSSS_HD_5005}), has
large artificial variations in intensity due to that effect. This is a good example of where one needs spatially resolved spectroscopy to study spectroscopic binarity.
This system is also of particular interest since is one of the cases in which quantitative spectroscopy leads to the early-O star lying closer to the ZAMS than the bulk of Galactic O-type 
stars analyzed by \citet{Holgetal20}. Knowing to what extent the B companion is affecting the derived parameters is of prime importance to ascertain the reliability of the differentiated 
position of this star in the spectroscopic HR diagram.

For HD~5005 we have three good-quality AstraLux epochs, one of them already reported in \citet{Maiz10a}, that span a period of eleven years. 
There is no appreciable motion between the A and B components in that time span.
This is consistent with the WDS data that go back a century and a half with nearly constant position angles and separations and with the relative proper motion in {\it Gaia}~DR2, 
which is essentially zero in right ascension and within two sigmas of that value in declination.


\begin{figure*}
\centerline{\includegraphics[width=\linewidth]{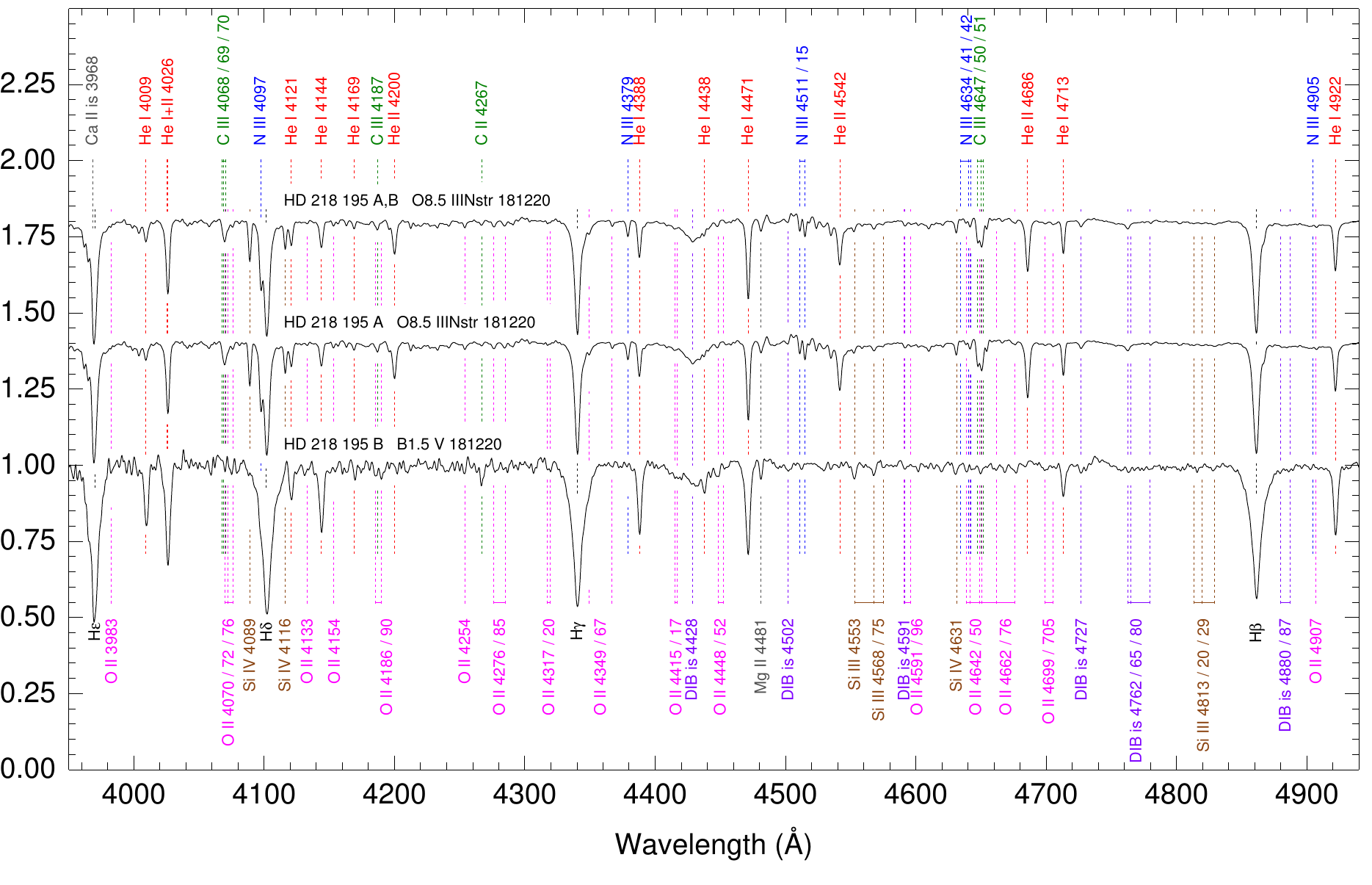}}
 \caption{Rectified spectrograms for HD~\num{218195} at the GOSSS spectral resolution of $R\sim 2500$ and on the stellar reference frame.
For each spectrogram, the name, spectral type, and evening date (YYMMDD) are shown. The top spectrogram is the weighted combination of the two components for the 181220 epoch. 
Main atomic and ISM lines are indicated.}
\label{GOSSS_HD_218_195}
\end{figure*}

\subsection{HD~\num[detect-all]{51756}~A,B, a system of two similar components with a new O star}

$\,\!$ \indent HD~\num{51756} is a bright but relatively poorly studied early-type system that has spectral classifications listed in Simbad that range in spectral subtype from B0 to B3 and in
luminosity class from IV to Ib. It is composed of a pair A,B separated by 0\farcs7 and with a small magnitude difference of just 0.3~mag. A third C component is dimmer and located 13\arcsec\
away. The A,B pair has no entries in {\it Gaia}~DR2 but the C component has a parallax of 0.5398$\pm$0.0575~mas and a good RUWE there. Using the {\it Gaia}~DR2 for the C component as we have
done previously for FN~CMa we obtain a distance of $1.79^{+0.21}_{-0.17}$~kpc.

We spatially resolve the A and B components with lucky spectroscopy (Fig.~\ref{GOSSS_HD_51_756}) and we derive classifications of B0~IV for the first (brighter) component and O9.7~IV(n) for 
the second (dimmer) component. The two spectral classifications are similar but B is a faster rotator and of an earlier subtype. This is the first time that HD~\num{51756}~B has been identified 
as an O star and, given its magnitude ($V\sim8.1$~mag), it is one of the brightest objects of that type identified recently. Note that \citet{RomLetal18} already indicated that the composite
A,B spectrum is O9.7~III and that for the composite we obtain B0~IV instead (Fig.~\ref{GOSSS_HD_51_756}), the small difference being likely caused by not matching spectral resolutions between 
standards and observations, which is a large effect around O9.7/B0 spectral subtypes due to the differences in intrinsic line width between \HeII{4542} and \SiIII{4552}. This system is similar to 
FN~CMa in that the B star in an O+B pair is capable of hiding the nature of the hotter companion in the unresolved spectrum.

There are 18 \lili\ epochs for HD~\num{51756}~A,B, where both visual components are included. The line profiles at high resolution show a clearly composite profile, with a narrow component and a
broad one, which we assign to A and B, respectively, based on our lucky spectroscopy. The narrow component shows velocity variations with a peak-to-peak amplitude of $\sim$20~km/s, suggesting
that A is an SB1. If that were the case, it would explain why A is slightly brighter than B despite being one quarter of a spectral subtype later and with the same luminosity class, as the
second star (which would be Ab if it were resolved) would contribute to the light without leaving a significant signature in the spectrum (possibly because it is a fast rotator, as B is, note that
would make this system quite similar to $\sigma$~Ori). 

We have two lucky imaging epochs for this system and our results agree with those of {\it Hipparcos} from 1991.25 in terms of separation, position angle, and
magnitude difference (there are no entries for this system in {\it Gaia}~DR2). The only slight difference is in position angle, for which there is a small counterclockwise 
movement that is also consistent with the historical WDS data that stretches back a century and a half. For a circular face-on orbit the corresponding period is around 7~ka.

\begin{figure*}
\centerline{\includegraphics[width=\linewidth]{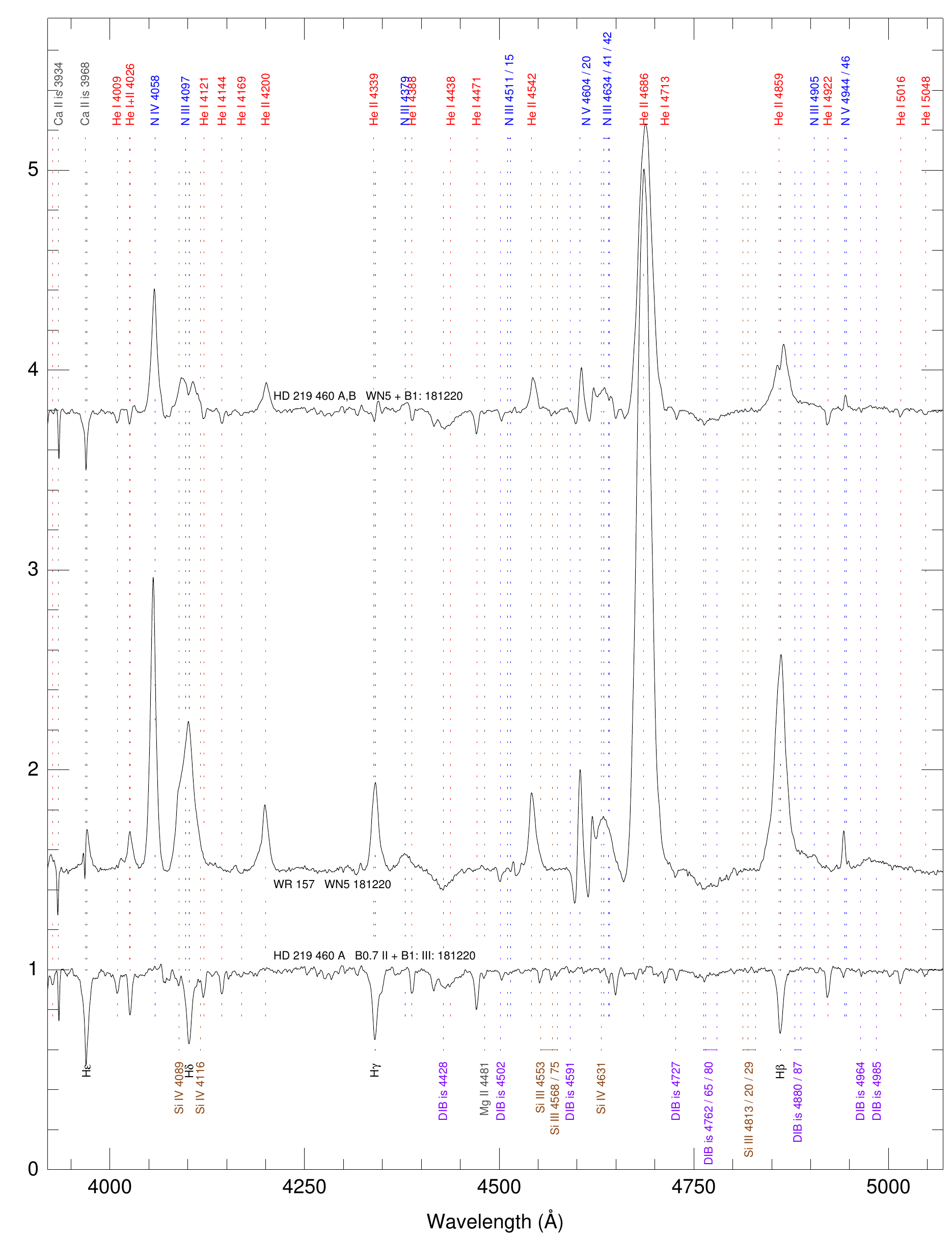}}
\caption{Rectified spectrograms for HD~\num{219460} at the GOSSS spectral resolution of $R\sim 2500$ and on the stellar reference frame.
For each spectrogram, the name, spectral type, and evening date (YYMMDD) are shown. The top spectrogram is the weighted combination of the two components for the 181220 epoch. 
Main atomic and ISM lines are indicated.}
\label{GOSSS_HD_219_460}
\end{figure*}

\begin{figure*}
\centerline{\includegraphics[width=\linewidth]{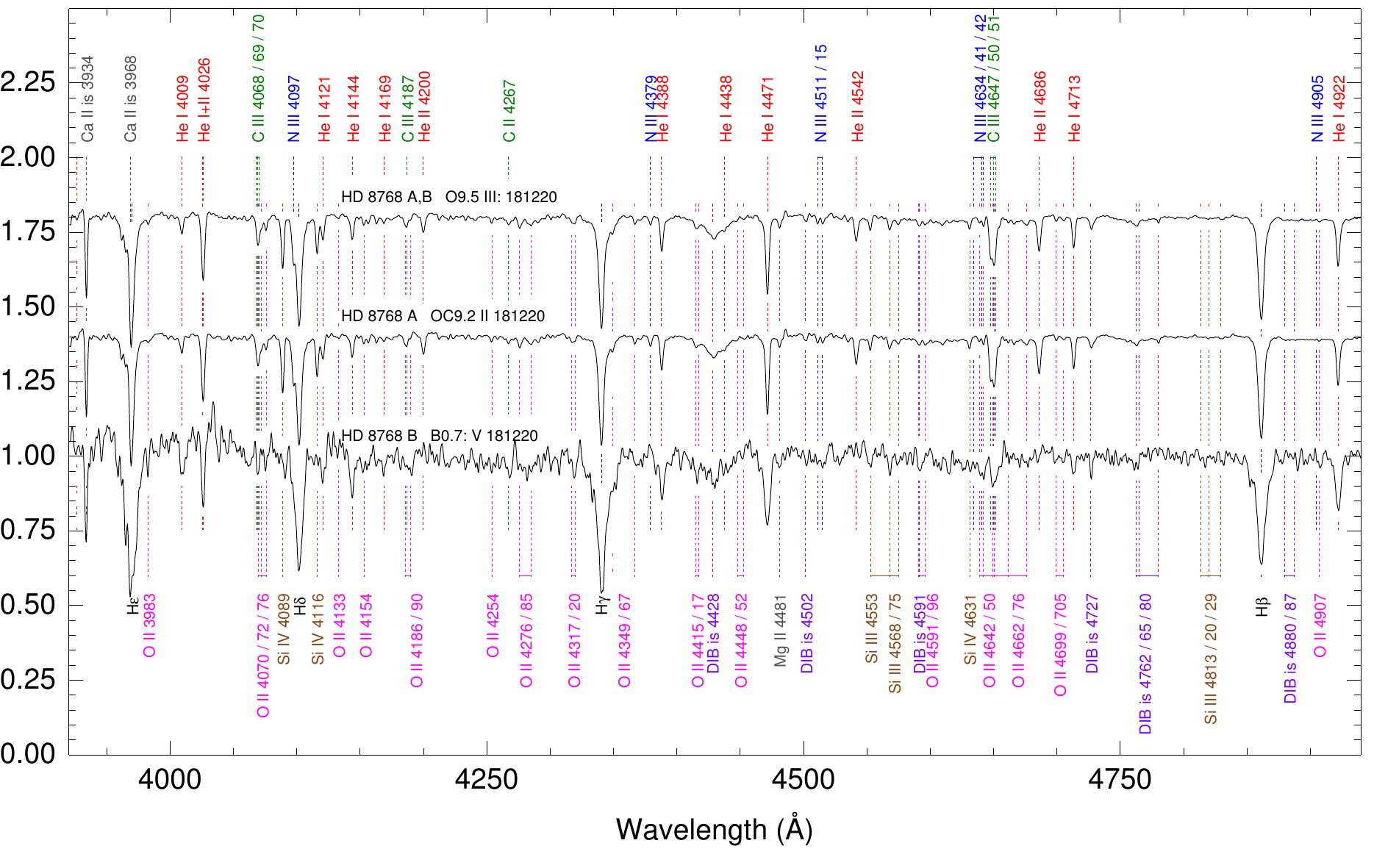}}
\caption{Rectified spectrograms for HD~\num{8768} at the GOSSS spectral resolution of $R\sim 2500$ and on the stellar reference frame.
For each spectrogram, the name, spectral type, and evening date (YYMMDD) are shown. The top spectrogram is the weighted combination of the two components for the 181220 epoch. 
Main atomic and ISM lines are indicated.}
\label{GOSSS_HD_8768}
\end{figure*}

\subsection{HD~\num[detect-all]{218195}~A,B, a system with a N-rich primary}

$\,\!$ \indent HD~\num{218195}~A is a nitrogen-enhanced O star (GOSSS~II) with a B component located at a separation of 0\farcs94 that is 2.7~mag dimmer. The WDS catalog lists two weaker 
components, C and D, farther away. The A component has an entry in {\it Gaia}~DR2 but with a RUWE of 1.67, so we cannot use its parallax to derive the distance. There is no clearly defined 
cluster around the object and the C component (Gaia~DR2~\num{2010239999586616448}) also has a high RUWE. Our last resort to obtain a {\it Gaia}~DR2 distance to the system is to assume that the D 
component (Gaia~DR2~\num{2010239999586619392}, with a RUWE of 0.94), located 11\farcs4 away, is physically bound to A,B. If that is the case the distance to the system is 
$2.29^{+0.20}_{-0.17}$~kpc.

Despite the large magnitude difference, we were able to spatially separate A and B with lucky spectroscopy and obtain a clean spectrum of the weak component. HD~\num{218195}~A maintains the 
O8.5~IIINstr classification of GOSSS~II (Fig.~\ref{GOSSS_HD_218_195}), as the B component exerts little influence on the combined spectrum at $R\sim 2500$  (but see below for the 
high-resolution case).  The B component, which was already identified as an early-B star in GOSSS~I, has a spectral type of B1.5~V. As an indication of the clean separation of the two components, 
the B spectrum has no anomalies around the wavelengths where He\,{\sc ii} lines are seen in the A spectrum. 

There are eight \lili\ epochs for HD~\num{218195}~A,B that span 14 years, with both components included in the aperture. All the strong lines look single and there are no appreciable velocity 
variations, indicating that the primary has no spectroscopic companions. However, the situation changes when one looks at weak lines where the contribution of the B component dominates despite 
the large magnitude difference, such as \CII{4267} and some O\,{\sc ii} lines. For those lines there are significant time-dependent changes and at least one epoch shows double profiles with a 
velocity separation of $\sim$70~km/s. We also note that some lines in the B spectrum in Fig.~\ref{GOSSS_HD_218_195} are slightly asymmetric. Therefore, we tentatively identify the B component 
as an SB2.

We have two lucky imaging epochs for this system, one of them already reported in \citet{Maiz10a}, and our results are in reasonable agreement with those of 
{\it Hipparcos} from 1991.25 in terms of separation, position angle, and magnitude difference. There are few historical observations of this pair in the WDS catalog but the scarce data from 
the early twentieth century point towards a small secular increase in separation.

\subsection{HD~\num[detect-all]{219460}~A,WR~157, disentangling two very different spectra}

$\,\!$ \indent HD~\num{219460} is a visual binary whose A component is of B spectral type and its B component is a Wolf-Rayet star (WR~157). The system was classified as
B1~II~+~WN4.5 by \citet{Turnetal83} based on an spatially unresolved spectrum, something that is feasible because of the very different ionic species and absorption/emission nature
of the lines present in an early-B supergiant and a Wolf-Rayet star (Fig.~\ref{GOSSS_HD_219_460}). The separation between the two components is 1\farcs37 and the magnitude difference
is small, with A being 0.4~mag brighter than B in Tycho-2~$B$. Both HD~\num{219460}~A and WR~157 have {\it Gaia}~DR2 entries but the RUWE for the latter is 1.9, prompting us to use the former to
calculate the distance. Using the same parallax zero point and prior as before we obtain $2.58^{+0.25}_{-0.21}$~kpc.

Given the small magnitude difference and the relatively large separation, this system is relatively easy to spatially separate with lucky spectroscopy. We point out, however, how
clean the disentangling is in Fig.~\ref{GOSSS_HD_219_460}: there are no signs of emission lines in the B star and the only absorption lines in the WR star are all caused by the ISM
with the only exception of the absorption components in the P-Cygni profiles for \NVd{4604,4620}. The absorption lines for the A component are asymmetric and from the narrowest
ones (e.g. \HeI{5016}) we identify the system as an SB2 caught with a velocity of $\sim$175~km/s. The primary spectroscopic component can be accurately classified as B0.7~II (quite similar to the
\citealt{Turnetal83} result) but the spectral type of the secondary spectroscopic component is more uncertain. Further epochs are necessary to see if the system can be caught with a larger
velocity difference. The B component (WR~157) is classified as WN5 according to the criteria for nitrogen lines in \citet{Smitetal96}, as our spectrum does not reach to \HeII{5412} 
or \HeI{5876}. The composite spectrum in Fig.~\ref{GOSSS_HD_219_460} shows the (diluted) combination of both visual components.

\begin{figure*}
\centerline{\includegraphics[width=\linewidth]{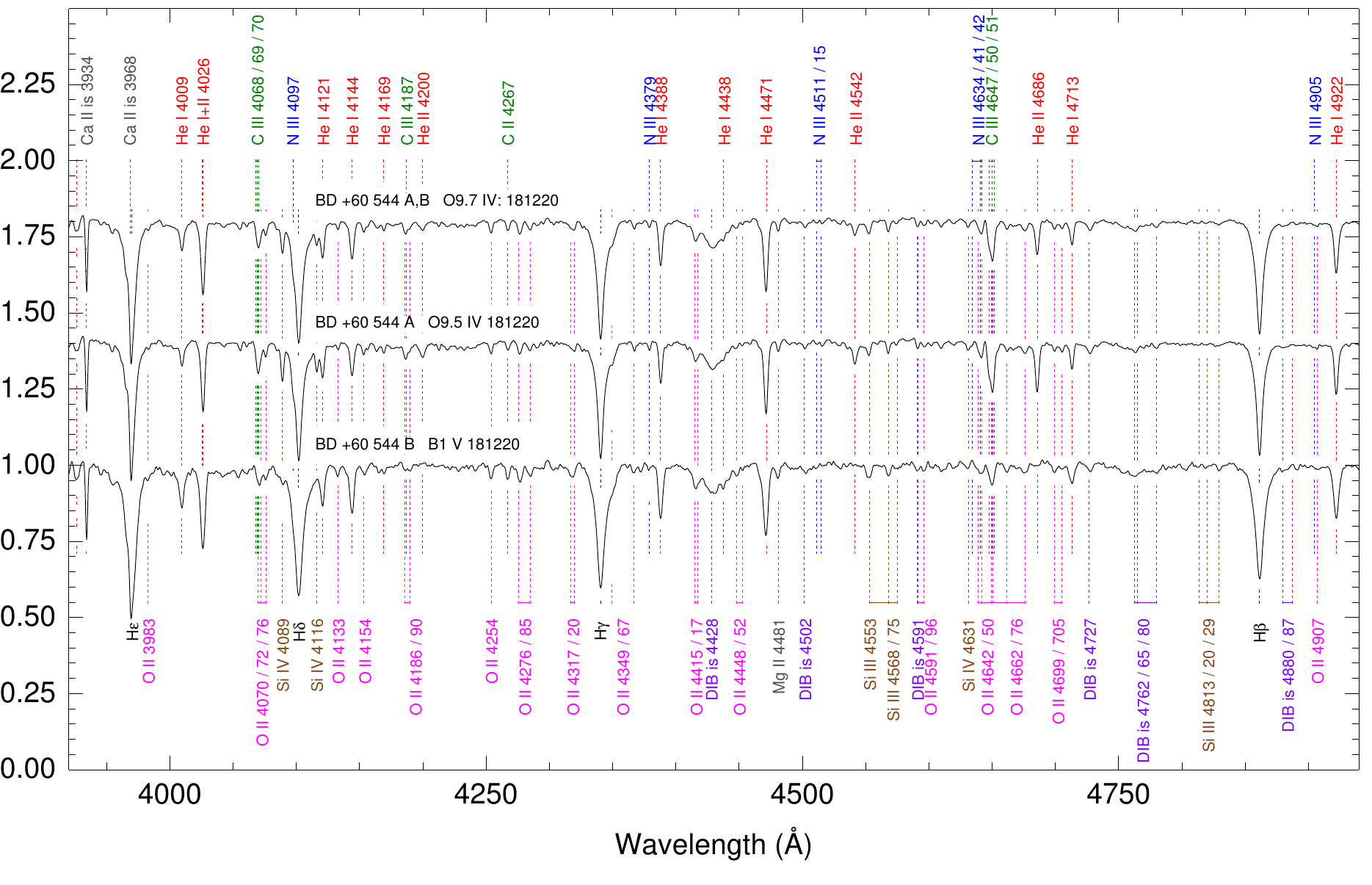}}
\caption{Rectified spectrograms for BD~$+$60~544 at the GOSSS spectral resolution of $R\sim 2500$ and on the stellar reference frame.
For each spectrogram, the name, spectral type, and evening date (YYMMDD) are shown. The top spectrogram is the weighted combination of the two components for the 181220 epoch. 
Main atomic and ISM lines are indicated.}
\label{GOSSS_BD_+60_544}
\end{figure*}

For this system we have two lucky imaging epochs. Our position angles and separations are very similar and are in excellent agreement with the {\it Gaia}~DR2 ones,
as both components are also detected there. Our magnitude differences are negative, meaning that WR~157 is brighter than HD~\num{219460}~A in our bands. Indeed, the magnitude difference
decreases from Tycho-2~$B$ to Tycho-2~$V$ to {\it Gaia}~DR2 (all positive) to $i$ to $z$ (both negative), indicating that the B component is redder than the A component. The 
{\it Gaia}~DR2 relative proper motion should lead to a counterclockwise motion of WR~157 with respect to HD~\num{219460}~A of $\sim$6\degr\ between the early twentieth century and
nowadays but no such effect is seen in the average of the two epochs around that time in the WDS catalog.

\subsection{HD~8768~A,B, an OC+B visual binary}

$\,\!$ \indent HD~8768 was classified as O9.5~IV by \citet{Morgetal55}, who were unaware of the existence of a B component with $d$~=~0\farcs66 and $\Delta m$~=~2.6~mag \citep{Maiz10a}. 
A third C component is located 5\farcs9 away but is too faint to make a significant contribution to the spectrum. This object has received little attention in
the literature compared to other O stars of similar magnitude. The pair is unresolved in {\it Gaia}~DR2 and with a negative parallax and a RUWE value of 3.2, likely as a result of the different 
contributions from each component to each observation. The C component, on the other hand, has a valid RUWE and we can use it to estimate the distance (assuming it is physically associated to the
A,B pair) with the same parallax zero point and prior as before to obtain $3.11^{+0.77}_{-0.52}$~kpc.

We spatially separated the A and B components with lucky spectroscopy (Fig.~\ref{GOSSS_HD_8768}). For A we obtain a spectral classification of OC9.2~II, as established by its weak nitrogen lines 
(e.g. \NIIId{4511,4515}) and strong carbon lines (e.g. \CIIIt{4068,4069,4070}), adding in this way a new member to the rare class of carbon-enriched O stars. The spectrogram for B is noisy as a 
consequence of the small separation and large magnitude difference, but we are able to establish that it is an early-B dwarf with a most likely subtype of B0.7. The combined spectrum shows a small 
influence of the B component in moving the spectral subtype to O9.5 and leaving an uncertain luminosity classification due to the inconsistency between different classification criteria (a problem 
common to such combinations of late-O + early-B binaries, see \citealt{SimDetal15a}). That effect explains the original classification by \citet{Morgetal55}.

We have three lucky imaging epochs of this system, one of them already reported in \citet{Maiz10a}. That data point and the {\it Hipparcos} one are the only ones in the
WDS catalog, so there is little historical information. Our data show a hint of a counterclockwise motion of B around A but the
uncertainties are large, so new epochs are needed to confirm it. The motion corresponds to a period of at least several thousand years. 

\begin{figure*}
\centerline{\includegraphics[width=\linewidth]{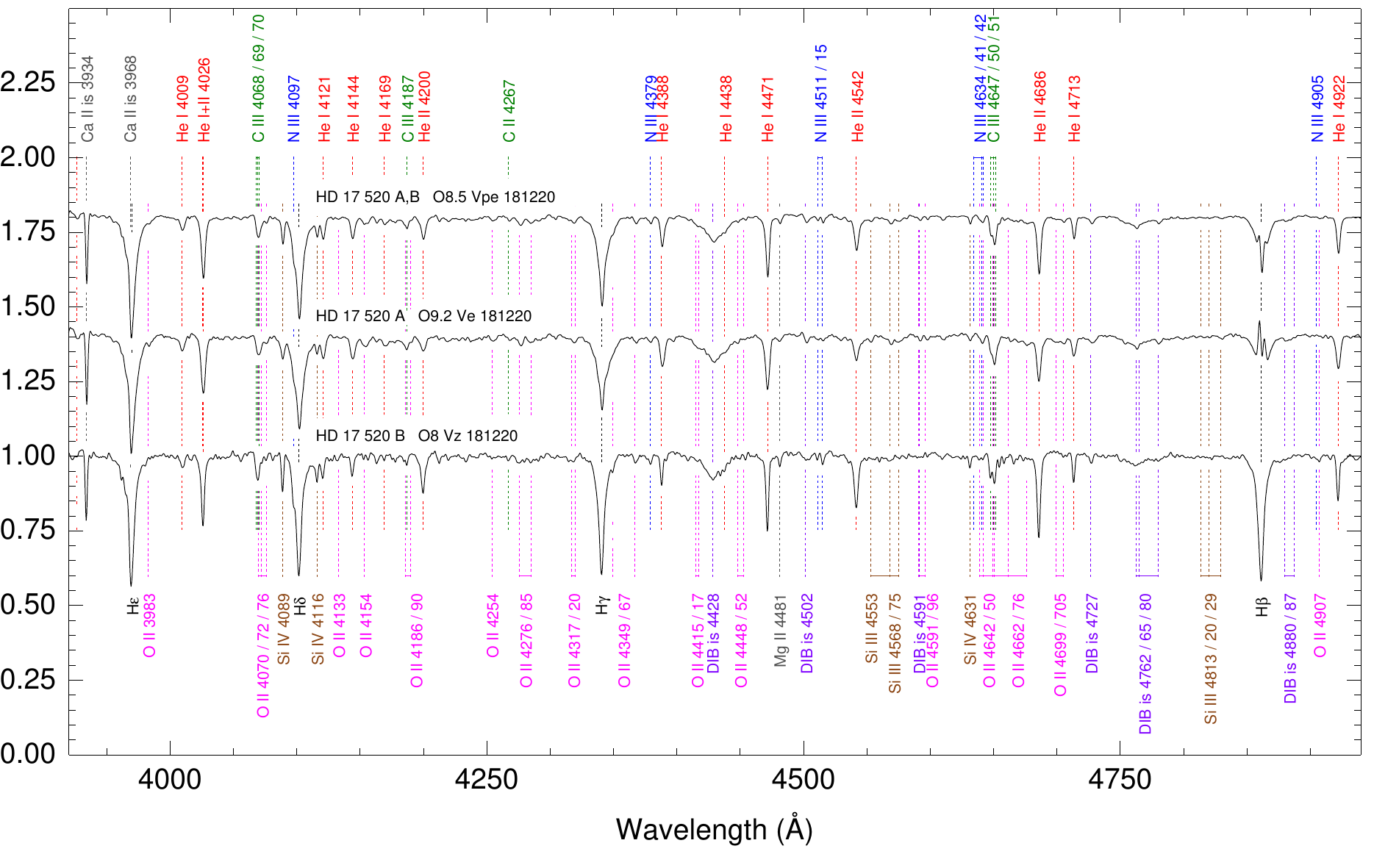}}
\caption{Rectified spectrograms for HD~\num{17520} at the GOSSS spectral resolution of $R\sim 2500$ and on the stellar reference frame.
For each spectrogram, the name, spectral type, and evening date (YYMMDD) are shown. The top spectrogram is the weighted combination of the two components for the 181220 epoch. 
Main atomic and ISM lines are indicated.}
\label{GOSSS_HD_17_520}
\end{figure*}

\subsection{BD~$+$60~544~A,B, a new O-type system}

$\,\!$ \indent This system has no previous accurate spectral classifications listed in Simbad, just two references that indicate it is of OB type \citep{Hardetal59,Ryds78}. The WDS catalog
indicates it is a pair separated by 1\farcs9 with a $\Delta m$ of 0.9~mag and a first measurement from a century ago. Both components have entries with good RUWE in {\it Gaia}~DR2 and with 
parallaxes within one sigma of each other. We combine the two with the same zero point and prior as before (and including the covariance term) to obtain a distance of $2.24^{+0.33}_{-0.25}$~kpc.

This system is relatively easy to spatially separate with lucky spectroscopy given its relatively small magnitude difference and large separation (Fig.~\ref{GOSSS_BD_+60_544}). The A component is 
classified as O9.5~IV, making this the first time it is identified as an O star. The B component, on the other hand, is classified as B1~V. The merged spectrum receives the intermediate classification
of O9.7~IV:, with the uncertainty in the luminosity class being another instance of conflicting luminosity criteria when combining a late-O star and an early-B star.

We have one lucky imaging epoch of BD~$+$60~544. Both components are also detected in {\it Gaia}~DR2 and there is good agreement with our values for the separation,
position angle, and magnitude difference.


\subsection{HD~\num[detect-all]{17520}~A,B, a system with an Oe star}

$\,\!$ \indent This system was analyzed in previous GOSSS papers and in GOSSS~III the A component was assigned a spectral type of O8~V and the B component a spectral type of O9:~Ve. The system has one
of the smallest separations in this paper (0\farcs32); its magnitude difference is 0.7~mag. Those classifications were done with regular long-slit spectroscopy and a reanalysis of the previous data has
revealed that we erroneously identified the two components by reversing the position of the slit: the one with the Oe spectrum is actually A, not B. HD~\num{17520}~A,B has a single entry in 
{\it Gaia}~DR2 but with a bad RUWE and a negative parallax. The C component, on the other hand, has a valid RUWE and we can use it to estimate the distance (assuming it is physically associated to 
A,B) with the same parallax zero point and prior as before to obtain $2.69^{+0.32}_{-0.26}$~kpc. HD~\num{17520} is in IC~1848, whose characteristics derived from {\it Gaia} data will be analyzed in more
detail in a subsequent paper of our Villafranca series of Galactic groups with OB stars (Ma{\'\i}z Apell\'aniz et al., accepted in A\&A).

Despite the small magnitude difference, we are able to cleanly separate the two components with lucky spectroscopy (Fig.~\ref{GOSSS_HD_17_520}), 
as evidenced by the opposite behavior of H$\beta$ in A and B. The A component has a
spectral classification of O9.2~Ve and the B component one of O8~Vz. The new classifications are similar to the previous GOSSS ones but with the A and B components exchanged and significantly cleaner
at the same time. The merged spectrum is classified as O8.5~V with a p suffix to indicate discrepancies between weak lines caused by the composite nature. The double-peaked emission profile in H$\beta$
characteristic of high-inclination Oe/Be stars is strong in A, absent in B, and weak in merged spectrum.

There are 16 \lili\ epochs of HD~\num{17520}~A,B and the most notorious feature is that the double-peaked emission in H$\beta$ and the even stronger one in H$\alpha$ experience not only the flux 
changes typical in Oe/Be stars but also large velocity variations with a peak-to-peak amplitude of $\sim$200~km/s. This indicates that the A component is an SB1 and explains why it is brighter than B
despite being of a later subtype. We plan to obtain additional lucky spectroscopy epochs to attempt catching A at a more favorable velocity separation and possibly identify the missing spectroscopic
component.

\begin{figure*}
\centerline{\includegraphics[width=\linewidth]{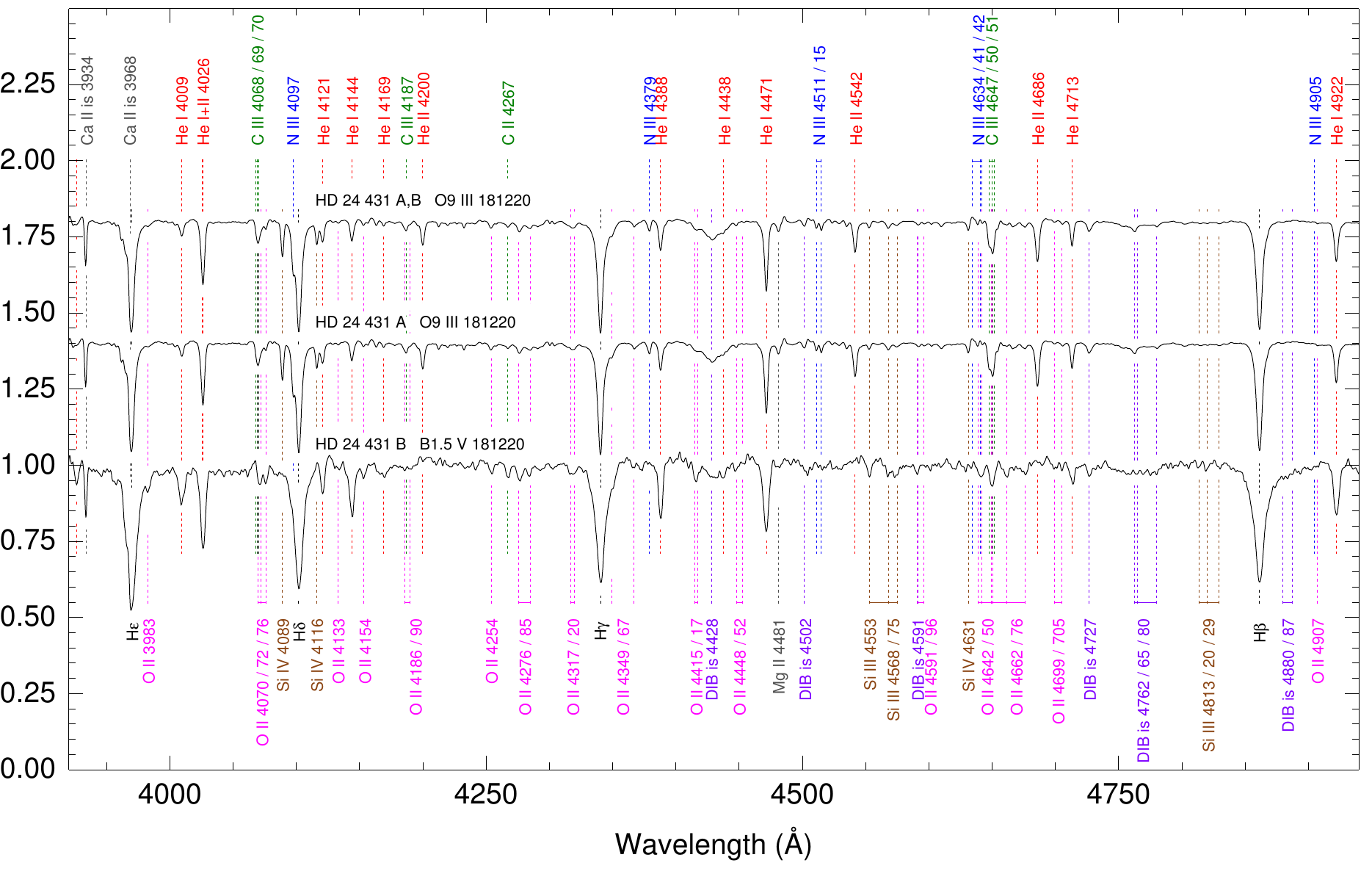}}
\caption{Rectified spectrograms for HD~\num{24431} at the GOSSS spectral resolution of $R\sim 2500$ and on the stellar reference frame.
For each spectrogram, the name, spectral type, and evening date (YYMMDD) are shown. The top spectrogram is the weighted combination of the two components for the 181220 epoch. 
Main atomic and ISM lines are indicated.}
\label{GOSSS_HD_24_431}
\end{figure*}

We have two lucky imaging epochs for this system, one of them already reported in \citet{Maiz10a}. The combination of our results and the historical WDS data indicates 
a small counterclockwise and outward motion of B with respect to A. The corresponding period would be of several thousands of years.

\subsection{HD~\num[detect-all]{24431}~A,B, a nearby isolated system}

$\,\!$ \indent HD~\num{24431} was classified as O9~III in \citet{Walb73a} and received the same classification in GOSSS~I. The WDS catalog lists a B component with $d$~=~0\farcs7 and  
$\Delta m$~=~2.9~mag. There is a single {\it Gaia}~DR2 entry for this system but it has a RUWE of 1.69, likely caused by the presence of the B component. The target is isolated and we have
been unable to find a stellar group associated with it in the {\it Gaia}~DR2 data, leaving us no alternative but to use the {\it Gaia}~DR2 parallax to estimate a distance of $812^{+61}_{-53}$~pc,
where the uncertainty is likely an underestimation due to the high RUWE. Nevertheless, we point out that {\it Hipparcos} also obtained a high value of the parallax for HD~\num{24431} (with a much
larger uncertainty) and that the spectroscopic parallax of \citet{MaizBarb18} is $\sim$1~kpc, so this system appears to be relatively nearby compared to most O stars in the sample of this paper.

We were able to separate A and B with lucky spectroscopy (Fig.~\ref{GOSSS_HD_24_431}). 
The spectral classification of A (and of the merged spectrum, where B has little contribution) is the same as the previous one from GOSSS~I.
The spectrum of B is somewhat noisy but not as much as that of e.g. HD~8768~B, which has a similar separation and magnitude difference with respect to its primary (but which is fainter in absolute
terms). Its spectral classification is B1.5~V.

We have 22 \lili\ epochs, where the light from both components is included but the contribution from A is clearly dominant. HD~\num{24431}~A appears as an SB1 in the high-resolution spectra with a
peak-to-peak amplitude of $\sim$50~km/s. To our knowledge, this is the first identification of the target as an SB1.

For this system we have two lucky imaging epochs, one of them already reported in \citet{Maiz10a}. The combination of our results and the WDS data that starts with the {\it Hipparcos} measurements 
shows a clockwise orbital motion of B with respect to A. For a face-on circular motion the estimated period is 4-5~ka.

\begin{figure*}
\centerline{\includegraphics[width=\linewidth]{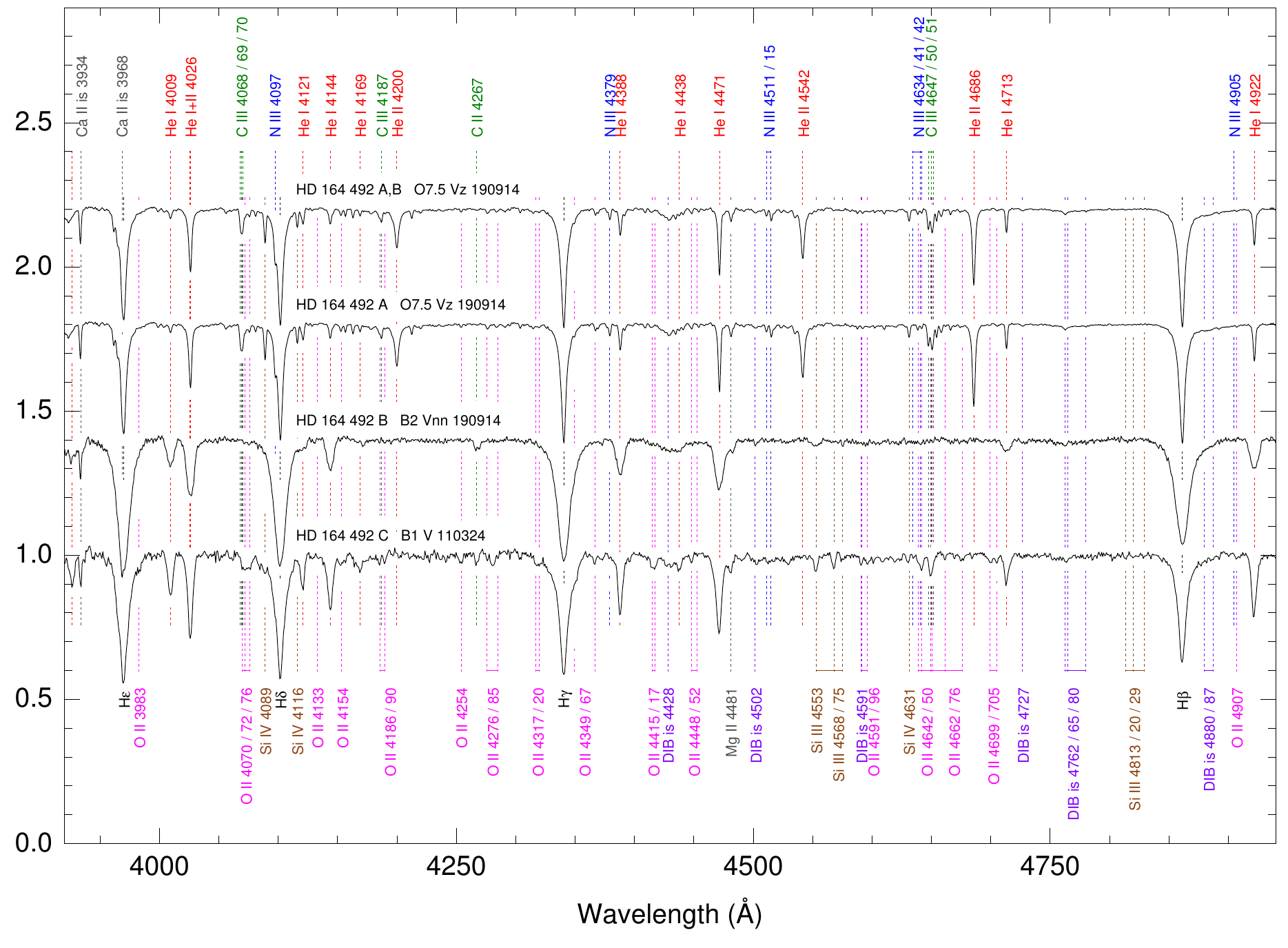}}
\caption{Rectified spectrograms for HD~\num{164492} at the GOSSS spectral resolution of $R\sim 2500$ and on the stellar reference frame.
For each spectrogram, the name, spectral type, and evening date (YYMMDD) are shown. The 110324 data were obtained with the 2.5 duPont telescope at LCO i.e. not with lucky spectroscopy.
The top spectrogram is the weighted combination of the two components for the 190914 epoch. Main atomic and ISM lines are indicated.}
\label{GOSSS_HD_164_492}
\end{figure*}

\subsection{HD~\num[detect-all]{164492}~A,B, an O star with a misclassified companion}

$\,\!$ \indent HD~\num{164492} is the high-order multiple system at the center of M20, also known as NGC~6514 or as the Trifid nebula. The A component is a well-known O7.5~Vz star (GOSSS~II) and the
WDS catalog lists another ten components, making this a borderline case between a high-order multiple system and a small cluster (considering the likelihood of additional low-mass stars hiding in the
glare of the high- and intermediate-mass ones and the existence of other components unaccounted for in the catalog, see below). A itself is listed as an Aa,Ab pair but Ab is just 25~mas away from Aa 
and is a red object with a $\Delta H$ of 3 magnitudes \citep{Sanaetal14}, so its contribution in the optical is negligible for the purposes of this paper. The second brightest component is C, which 
itself is an interesting triple system composed of a static, fast-rotating, magnetic, early-B star (C1) and a spectroscopic pair (C2+C3) of two narrow-lined early+late B stars 
\citep{Wadeetal17,Gonzetal17}. 

Simbad currently gives a spectral classification of A2~Ia for the B component that can be traced back to \citet{Gahmetal83} and that is problematic for several reasons. The first one is that it implies
the coexistence in an H\,{\sc ii}~region of a very young O7.5~Vz star and an evolved A supergiant, which under normal circumstances would require a significantly older age. The second one is that an A
supergiant should be significantly brighter in the optical and in the NIR than an O7.5~Vz star and the WDS catalog lists the A component being almost 3 magnitudes brighter than B (see also below). This
would require the A supergiant either being significantly obscured by differential extinction and/or being much farther away than the rest of the stars. The first option is excluded by the 2MASS colors
and the second one by the {\it Gaia}~DR2 parallaxes (see below). 

\begin{table}
\caption{{\it Gaia}~DR2 astrometric data for four components of the HD~\num{164492} system with compatible parallaxes and proper motions and their 
          aggregate results. The aggregate results use external uncertainties and include the spatial covariance terms of \citet{Lindetal18b}.}
\centerline{
\begin{tabular}{cr@{$\pm$}lr@{$\pm$}lr@{$\pm$}l}
comp. & \mcii{$\varpi$} & \mcii{\pmra}   & \mcii{\pmdec}  \\
      & \mcii{(mas)}    & \mcii{(mas/a)} & \mcii{(mas/a)} \\
\hline
A     & 0.6376&0.0873   &  0.426&0.122   & $-$1.765&0.097 \\ 
B     & 0.7780&0.0757   &  0.308&0.106   & $-$1.196&0.083 \\ 
D     & 0.8049&0.1071   &  1.055&0.140   & $-$2.858&0.112 \\ 
E     & 0.8793&0.0685   &  0.842&0.093   & $-$1.548&0.073 \\ 
\hline
      & 0.7841&0.0613   &  0.630&0.084   & $-$1.706&0.082 \\
\hline
\end{tabular}
}
\label{Gaia_HD_164_492}
\end{table}

To estimate the distance to HD~\num{164492} we follow a strategy similar to the one we used for 6~Cas, combining {\it Gaia}~DR2 entries with good RUWE and similar parallaxes and proper motions, but
selecting among the bright named WDS components. We exclude C due to its large RUWE (see above for its multiplicity, the likely culprit) and we keep the other four bright components 
(Table~\ref{Gaia_HD_164_492}). Applying the same parallax zero point and prior as before we obtain a distance of $1.29^{+0.12}_{-0.10}$~kpc. We note that the proper motions indicate the presence of
significant internal motions, as expected for a high-order multiple system such as this one.

\begin{figure*}
\centerline{\includegraphics[width=\linewidth]{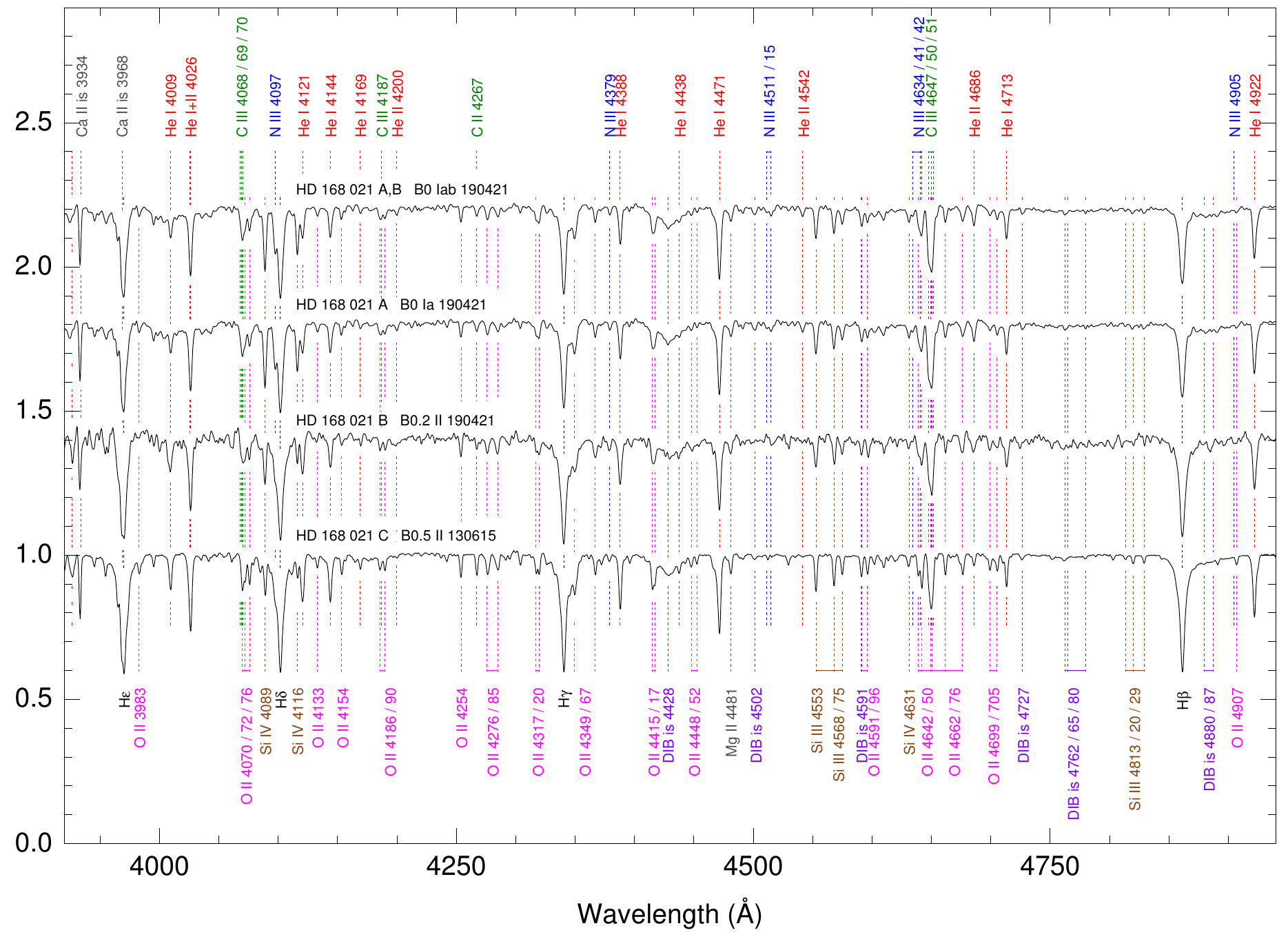}}
\caption{Rectified spectrograms for HD~\num{168021} at the GOSSS spectral resolution of $R\sim 2500$ and on the stellar reference frame.
For each spectrogram, the name, spectral type, and evening date (YYMMDD) are shown. The 130615 data were obtained with WHT but not with lucky spectroscopy.
The top spectrogram is the weighted combination of the two components for the 190421 epoch. Main atomic and ISM lines are indicated.}
\label{GOSSS_HD_168_021}
\end{figure*}

We placed the slit along the A-B direction and easily separated the two components with lucky spectroscopy (Fig.~\ref{GOSSS_HD_164_492}), as expected given the large separation (6\farcs26). The A and the 
merged A,B spectra receive the same spectral classification as in GOSSS~II (the B component is too weak to make a significant dent in the merged spectrum). The B component is a B2~Vnn star, that is, not an 
A~ supergiant but instead a mid-B star with a very large $v\sin i$ that makes it a peculiar object but not for the reason given by \citet{Gahmetal83}. Our spectral classification for HD~\num{164492}~B 
suggests possible explanations for the previous A~supergiant classification: the object could be a PMS star that was observed by \citet{Gahmetal83} while it was undergoing a shell phase or it could be a Be 
star currently in a phase with a weak disk. We also observed the C component with standard long-slit spectroscopy and assign it a spectral classification of B1~V, which must be dominated by the C2 component 
of \citet{Wadeetal17}, as the C1 component is easily hidden by its large $v\sin i$ and the C3 component is significantly fainter.

There are 12 epochs of HD~\num{164492}~A in \lili\ spanning 13 years. No changes in the spectral appearance or in velocity can be appreciated in the data. There is also one FEROS epoch of
HD~\num{164492}~B (possible given the large separation) that is consistent with our lucky spectroscopy observation. We have used that epoch to measure the $v\sin i$ of that fast rotator as we did for 
$\alpha$~Sco~B (which has a quite similar spectrum), in this case using five He\,{\sc i} lines: \HeI{4387}, \HeI{4471}, \HeI{4713}, \HeI{5876}, and \HeI{6678}). We obtain a value of $355\pm 55$~km/s. 
Finally, there are also 20 epochs of HD~\num{164492}~C in \lili\ spanning 6 years that show the complex line profiles produced by the SB3 system.

We have a single lucky imaging epoch for HD~\num{164492}. In this case we list in Table~\ref{AstraLuxdata} the measurements for three of the pairs. We did not attempt to
resolve A into Aa+Ab, as the predicted separation is too small for AstraLux. Our data for the A,B and A,C pairs agree reasonably well with the {\it Gaia}~DR2 results. For the A,H pair
our magnitude difference is compatible with that of \citet{Turnetal08}, obtained at similar wavelengths, but is significantly larger than the NIR values of \citet{Sanaetal14},
indicating that H is a redder object than A. 

%
%
%

\subsection{HD~\num[detect-all]{168021}~A,B,C, a triple system of early-B supergiants}

$\,\!$ \indent This system is a hierarchical triple composed of a close pair (A,B) separated by 0\farcs48 and with a $\Delta m$ of 1.0~mag and a more distant companion (C) 17\farcs2 away and with 
a magnitude intermediate between those of A and B, according to the WDS catalog. \citet{Morgetal53b} classified the unresolved A,B pair as B0~Ib. The C component has no precise spectral classification
in the literature. There is a single {\it Gaia}~DR2 entry for the A,B pair but it has a bad RUWE, so we cannot use it to estimate the distance to the system. HD~\num{168021}~C, on the other hand, has an
entry with a good RUWE that allows us to obtain a value of $1.62^{+0.18}_{-0.14}$~kpc.

\begin{figure*}
\centerline{\includegraphics[width=\linewidth]{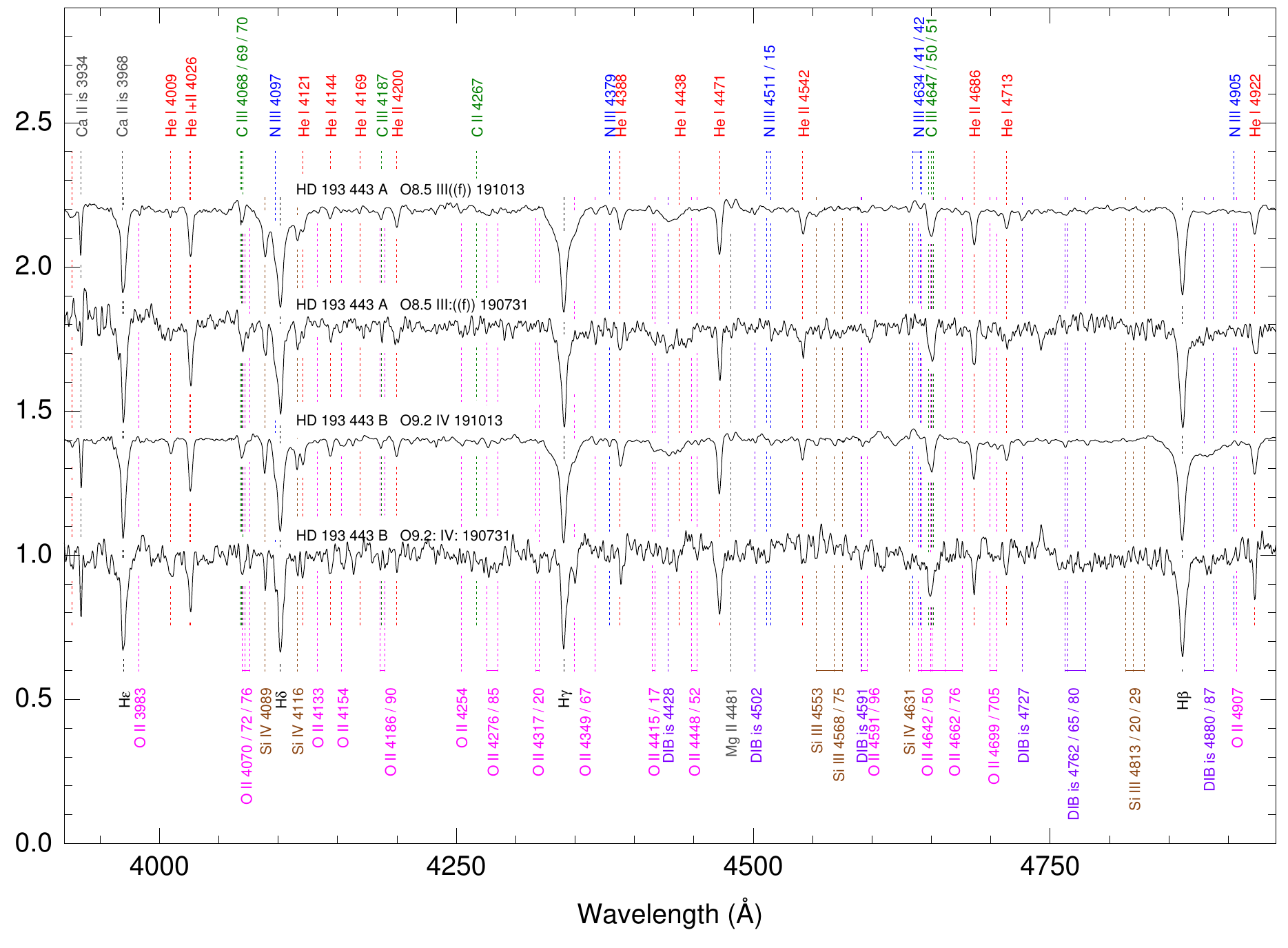}}
\caption{Rectified spectrograms for HD~\num{193443} at the GOSSS spectral resolution of $R\sim 2500$ and on the stellar reference frame.
For each spectrogram, the name, spectral type, and evening date (YYMMDD) are shown. The 191013 data were obtained with STIS@HST and the 190731 with WHT using lucky spectroscopy.
Main atomic and ISM lines are indicated.}
\label{GOSSS_HD_193_443}
\end{figure*}

We spatially resolve A and B with lucky spectroscopy (Fig.~\ref{GOSSS_HD_168_021}) 
and we find that both are relatively similar early-B supergiants. The A component is earlier and more luminous at B0~Ia while the B component is
later and less luminous at B0.2~II. The combined spectrum is classified as B0~Iab, i.e. the same spectral subtype as the \citet{Morgetal53b} classification but changing the luminosity class from Ib to
Iab. We also observed the C component with the WHT using regular long-slit spectroscopy and determined a spectral classification of B0.5~II. Therefore, HD~\num{168021} includes three similar early-type
B~supergiants.


\subsection{HD~\num[detect-all]{193443}~A,B, comparing STIS and lucky spectroscopy}

$\,\!$ \indent HD~\num{193443} is a visual double separated by 0\farcs138 and with a small magnitude difference (B is actually slightly brighter than A) that was studied in STIS~I. In that paper
spectral classifications of O8.5~III((f)) and O9.2~IV for A and B, respectively. One of the components is a spectroscopic binary \citep{Mahyetal13} and in STIS~I we favored that it is B but stated that
more observations are needed to confirm that hypothesis. There is a single entry in {\it Gaia}~DR2 for HD~\num{193443}~A,B but we cannot use it for a distance determination given its bad RUWE. The WDS
catalog lists a C component 9\farcs1 away but its RUWE is also bad. Therefore, the only way to give a {\it Gaia}~DR2 distance to the system is by its inclusion in the Cyg~OB1 association, for which
\citet{Wardetal20} give a value of $1.64\pm0.24$~kpc. We note that Simbad currently assigns a spectral classification of O9.5~V to HD~\num{193443}~C and cite \citet{Mahyetal15}. This is a case of 
mistaken identity in Simbad, as that paper is referring to the spectroscopic binary, not to a visual companion. The C component is too faint to be an O star at the same distance as the A,B pair.

Among the sample in STIS~I, HD~\num{193443}~A,B is one of the two easiest systems to separate with HST (see Fig.~4 in that paper). For that reason, we attempted to separate the pair with lucky
spectroscopy and we were able to do so on an occasion with exceptional seeing (we also attempted to separate the other ``easy'' STIS system, HD~\num{16429}~Aa,Ab but we were unsuccessful). We compare 
in Fig.~\ref{GOSSS_HD_193_443} the spectra from both setups. The spectra for each component are compatible but the lucky spectroscopy results are much noisier, illustrating the different spatial
resolving power of HST and lucky spectroscopy. This translates into uncertain spectral types derived from the ground-based data.

\begin{figure*}
\centerline{\includegraphics[width=\linewidth]{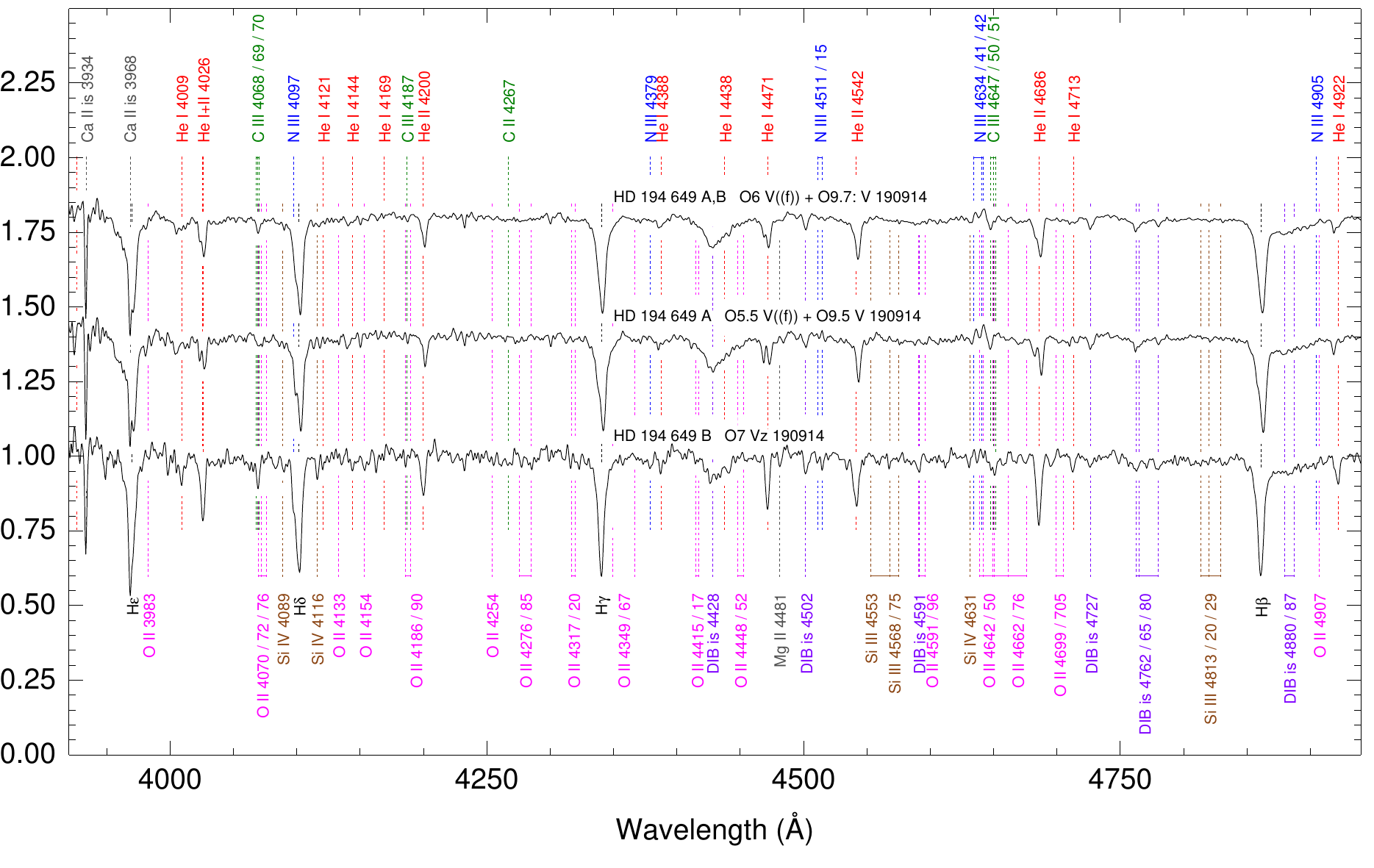}}
\caption{Rectified spectrograms for HD~\num{194649} at the GOSSS spectral resolution of $R\sim 2500$ and on the heliocentric reference frame.
For each spectrogram, the name, spectral type, and evening date (YYMMDD) are shown. 
The top spectrogram is the weighted combination of the two components for the 190914 epoch. Main atomic and ISM lines are indicated.}
\label{GOSSS_HD_194_649}
\end{figure*}

We have 29 \lili\ epochs of HD~\num{193443}~A,B, where there is a clear signal of the spectroscopic motion of different components. However, as already mentioned in STIS~I, this is an SB3 with a static
component (likely A) and two objects (likely two stars in B) moving in a spectroscopic orbit with a period of 7.467~d and a relatively low velocity amplitude ($K_1+K_2 \sim$~130~km/s,
\citealt{Mahyetal13}) which is not large enough to make the three components appear separated at any point of the orbit. In future papers we will analyse this system in more detail with the help of new
data.

\subsection{HD~\num[detect-all]{194649}~A,B, identifying the SB2 in a visual binary}

$\,\!$ \indent This system was included in MONOS~I, where we presented spatially unresolved spectroscopy and lucky imaging results. It is a pair with $\Delta m$~=~0.95~mag (with a possible
color term) and $d$~=~0\farcs40. Similarly to HD~\num{193443}, it includes a spectroscopic binary with a \num{3.39294}~d period \citep{Mahyetal13} 
but previous results were not able to spatially
separate the A and B visual components. In MONOS~I we derived spectral classifications of O6 V((f))~+~O9.7:~V with GOSSS data and of O6~IV((f))~+~O9.5~V((f)) with \lili\ data but we acknowledged the
existence of three stars contributing to the spectra and, hence, the need for further studies of the system. HD~\num{194649} has a single entry in {\it Gaia}~DR2 but without parallax or proper motions
due to the extremely high RUWE of the solution, 108.1, caused by the lack of resolution of the system into its two visual components. The system is the brightest object of a little studied stellar
cluster listed in Simbad as [KPS2012]~MWSC~3335 \citep{Kharetal13}. We have searched {\it Gaia}~DR2 to select the brightest members with good RUWE according to their proper motions and calculate the 
distance to the cluster (Table~\ref{Gaia_HD_194_649}), obtaining a result of $1.62^{+0.12}_{-0.10}$~kpc.

\begin{table}
 \caption{{\it Gaia}~DR2 astrometric data for 9 components of the [KPS2012]~MWSC~3335 cluster (not including HD~\num{194649}~A,B, for which no 
          {\it Gaia}~DR2 parallax or proper motions are available) with compatible parallaxes and proper motions and their 
          aggregate results. The aggregate results use external uncertainties and include the spatial covariance terms of \citet{Lindetal18b}.}
\centerline{\footnotesize
\addtolength{\tabcolsep}{-3pt}
\begin{tabular}{cr@{$\pm$}lr@{$\pm$}lr@{$\pm$}l}
{\it Gaia} DR2 ID         & \mcii{$\varpi$} & \mcii{\pmra}   & \mcii{\pmdec}  \\
                          & \mcii{(mas)}    & \mcii{(mas/a)} & \mcii{(mas/a)} \\
\hline
\num{2067482773947102208} & 0.6009&0.0269   & $-$3.140&0.045 & $-$4.242&0.041 \\ 
\num{2067483044525974784} & 0.6250&0.0308   & $-$3.211&0.053 & $-$4.134&0.045 \\ 
\num{2067482602148428288} & 0.5542&0.0264   & $-$3.166&0.043 & $-$4.118&0.043 \\ 
\num{2067482705227636480} & 0.5593&0.0240   & $-$3.142&0.039 & $-$4.180&0.036 \\ 
\num{2067482945745805312} & 0.5753&0.0168   & $-$3.300&0.027 & $-$4.409&0.029 \\ 
\num{2067482735291629056} & 0.6065&0.0245   & $-$3.445&0.034 & $-$3.811&0.048 \\ 
\num{2067482877026323712} & 0.5757&0.0189   & $-$2.890&0.035 & $-$4.450&0.029 \\ 
\num{2067482670867892352} & 0.5944&0.0194   & $-$3.254&0.035 & $-$4.293&0.029 \\ 
\num{2067482945745805440} & 0.5826&0.0230   & $-$3.040&0.039 & $-$4.181&0.034 \\ 
\hline
                          & 0.5852&0.0413   & $-$3.177&0.065 & $-$4.036&0.065 \\
\hline
\end{tabular}
\addtolength{\tabcolsep}{3pt}
}
\label{Gaia_HD_194_649}
\end{table}

\begin{figure*}
\centerline{\includegraphics[width=\linewidth]{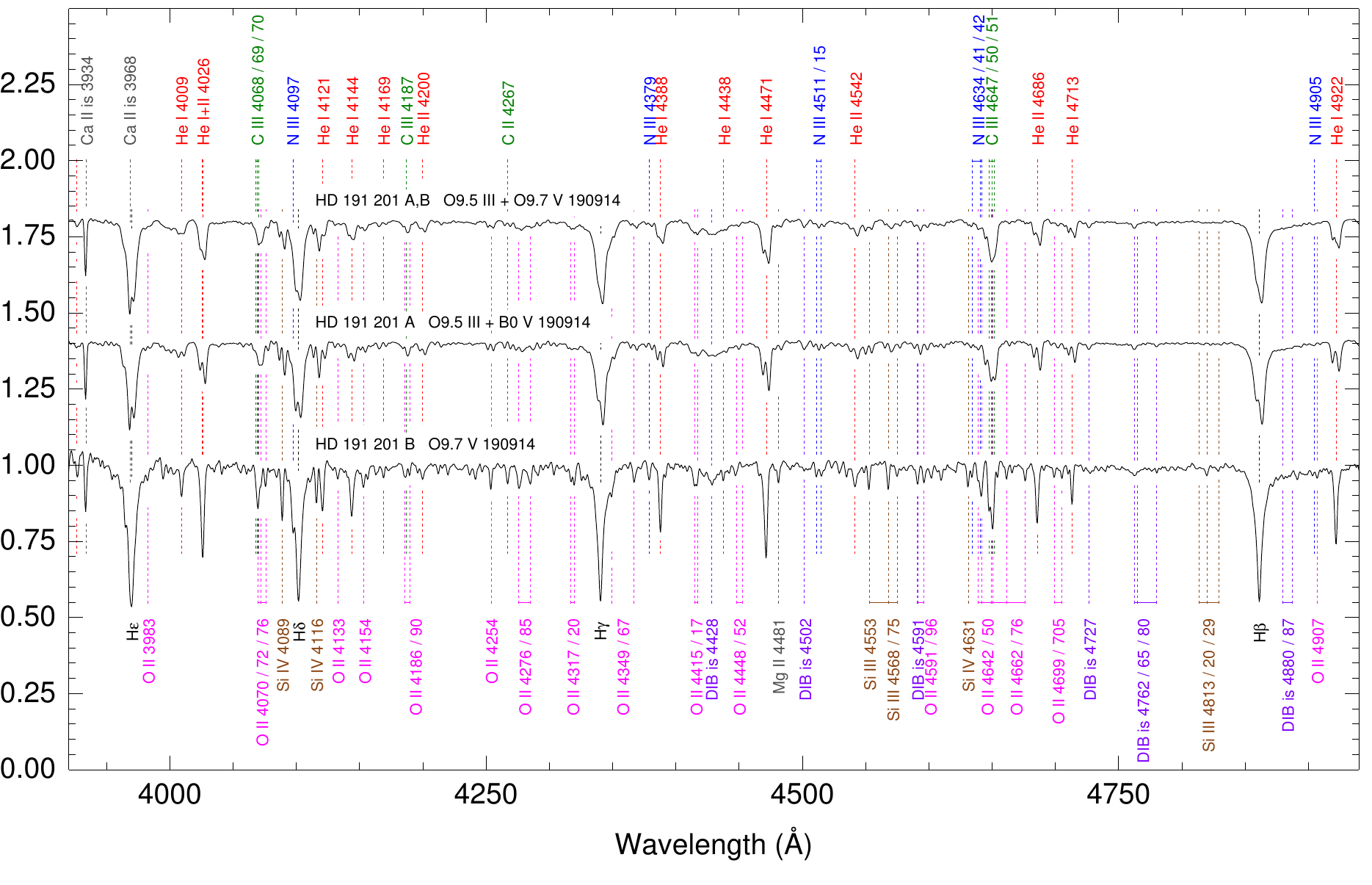}}
\caption{Rectified spectrograms for HD~\num{191201} at the GOSSS spectral resolution of $R\sim 2500$ and on the heliocentric reference frame.
For each spectrogram, the name, spectral type, and evening date (YYMMDD) are shown. 
The top spectrogram is the weighted combination of the two components for the 190914 epoch. Main atomic and ISM lines are indicated.}
\label{GOSSS_HD_191_201}
\end{figure*}

The two visual components in HD~\num{194649} are clearly resolved with lucky spectroscopy (Fig.~\ref{GOSSS_HD_194_649}). The SB2 in the system is the A component, with a spectral classification of
O5.5~V((f))~+~O9.5~V. The B component, on the other hand, is classified as O7~Vz. The merged spectrum appears as O6~V((f))~+~O9.7:~V, the same classification we obtained with GOSSS data in MONOS~I.
These results paint a coherent picture in which the late-O star (the one with the largest velocity amplitude in the SB2) is separated in velocity in the merged spectrum near quadrature (when the lucky
spectra were obtained) but the two mid-O stars (one in A and one in B) are merged into a single component in velocity. The spectral type of the primary in the merged data (O6) is an intermediate one
between the two real types (O5.5 and O7). At the same time, the velocity amplitude of the primary in the SB2 in the merged data is lower than the real value: if we measure the velocity difference 
between the two components in the merged data we obtain $\sim$275~km/s but if we do it in the separated spectrum we get $\sim$325~km/s (at $R\sim 2500$ resolution, individual uncertainties 
for components well resolved in velocity are 10-15~km/s). According to \citet{Mahyetal13}, $K_1$~=~76.3~km/s and $K_2$~=~192.1~km/s with a circular orbit, which agrees reasonably well with our 
measurement in the merged spectrum. However, that means that the true value for $K_1$ is likely to be closer to 125~km/s. The importance of that result is that it changes the mass ratio from $\sim$2.5
to $\sim$1.5. The first value is hard to reconcile with what one expects for an O6~V and an O9.5~V stars but the second value is a reasonable one for an O5.5~V and an O9.5~V stars. In the future we
plan to obtain further data and derive a new orbit within the MONOS project. Finally, the measured 0.95~mag difference between A and B is consistent with the expected value for the derived spectral
classifications.


\subsection{HD~\num[detect-all]{191201}~A,B, another visual binary with an SB2 component}

$\,\!$ \indent This system was included in MONOS~I, where we presented spatially unresolved spectroscopy and lucky imaging results. It is a pair with $\Delta m$~=~1.8~mag 
and $d$~=~0\farcs98. Similarly to HD~\num{194649}, it includes a spectroscopic binary with an \num{8.333832}~d period \citep{SticLloy01} and for this system it is clear that
it resides in the A~component. In GOSSS~I we classified the A component as O9.5~III~+~B0~IV and the B~component as O9.7~III. HD~\num{191201}~A,B has a single entry in {\it Gaia}~DR2 but with a bad RUWE
that does not allow us to use its astrometry to calculate a distance. The WDS catalog lists a C component for this multiple system that does have good-quality astrometry in {\it Gaia}~DR2. We use it to
derive a distance of $2.02^{+0.13}_{-0.11}$~kpc.

We easily resolve the two components of HD~\num{191201} with lucky spectroscopy (Fig.~\ref{GOSSS_HD_191_201}). The spectral classification for A is O9.5~III~+~B0~V, identical to that of GOSSS~I except
for the lower luminosity class of the spectroscopic secondary. The spectral classification for B is O9.7~V, downgraded in luminosity class from III with respect to GOSSS~I. The difference is likely 
caused by the lower contamination in the lucky spectroscopy data compared to the regular long-slit spectroscopy used in the original paper but there are two additional factors to consider. The first
one is that since GOSSS~I we have decided to give more weight to the Si/He criterion for the luminosity classification of late-type O stars in detriment of the He/He criterion, as the former criterion
is less sensitive to the effect of hidden binaries. The second one is that HD~191~201~B appears to be an object with a very low $v\sin i$. In any case, the change in luminosity classification yields a
much better agreement between the expected and the measured $\Delta m$ between A and B. The spectral classification for the merged spectrum is O9.5~III~+~O9.7~V and we see a similar effect as the one
described for HD~\num{194649}: the velocity difference between the two spectroscopic components is larger in the separated spectrum ($\sim$300~km/s, obtained near quadrature) than in the merged
spectrum ($\sim$225~km/s). HD~\num{191201} will also be studied in an incoming MONOS paper using \lili\ data.


\section{Summary and future work}

\begin{figure}
\centerline{\includegraphics[width=\linewidth]{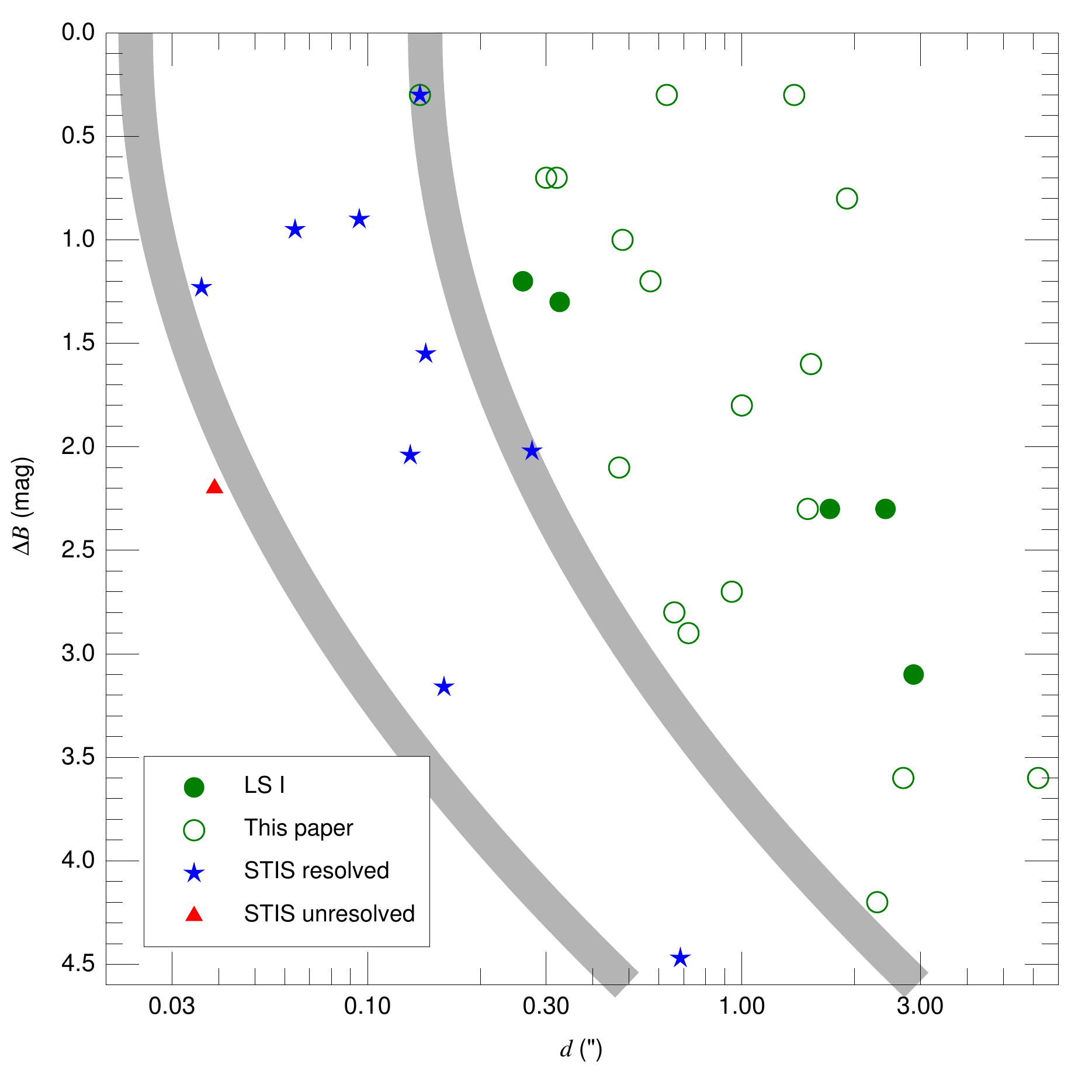}}
\caption{Binary systems resolved using lucky spectroscopy with WHT/ISIS (in LS~I and here) and HST/STIS and unresolved systems using 
HST/STIS in the $d-\Delta B$ (separation-$B$ magnitude difference) plane. The gray curves mark the tentative empirical boundaries between the regions in the plane accessible with different techniques: the 
left one is for HST/STIS and the right one for lucky spectroscopy with WHT/ISIS. This is an updated version of Fig.~4 in STIS~I.}
\label{sep_DB}
\end{figure}

$\,\!$ \indent In this paper we present spatially resolved intermediate-resolution blue-violet spectroscopy of 19 massive close binaries. The combined sample of LS~I and this paper contains 23 systems with
angular separations between 0\farcs138 and 6\farcs19 and physical plane-of-the-sky separations between 100~AU and 8000~AU. The most relevant results are:

\begin{itemize}
 \item For two stars (FN~CMa~B and 6~Cas~B) that had been previously identified as O stars based on spatially unresolved data, we obtain for the first time spatially disentangled spectra and 
       derive accurate spectral classifications. In both cases the companion is a more evolved, visually brighter supergiant, of B type in the first case and of A type in the second one. We note
       that 6~Cas is the only known Galactic case of an A supergiant + O binary.
 \item We determine that two companions to massive stars ($\alpha$~Sco~B and HD~\num{164492}~B) are fast rotators of B2 spectral type. In both cases the large $v\sin i$ likely contributed to their
       misidentification in the literature, in the first case as a B star of a later subtype and in the second one as an A supergiant. We also note the tendency, already noticed in LS~I and mentioned 
       for some further cases in STIS~I, of the visual companions of massive stars to be fast rotators of B spectral type.
 \item We have extended the technique of lucky spectroscopy to include three systems ($\alpha$~Sco~A,B, CS~Cam~A,B, and HD~\num{164492}~A,B) with extreme brightness differences (more than 3.5~mag).
 \item Using data from an occasion with exceptional seeing, we have been able to spatially separate HD~\num{193443}~A,B, a very close system ($d = 0\farcs138$) that we had also successfully separated with 
       STIS. The resulting spectra, however, are significantly noisier than the STIS equivalents, indicating that we have reached the limit of the possibilities of lucky spectroscopy (at least with our
       current setup) for objects with small magnitude differences.
 \item We have extended the range of the technique to objects much fainter than those in LS~I (down to primaries with $B=11$~mag). We have been able to do so in part due to the use of a different CCD 
       that has a better duty cycle that allows us to stay on target 98\% of the time as opposed to just 7\%.
 \item In LS~I the stars had relatively similar spectral types (mid-O to early-B). Here we have expanded the technique to demonstrate that it works with primaries of other spectral types and with systems 
       that differ greatly between primary and secondary. In this way we have observed as primaries an A~supergiant (6~Cas~A), a WR~star (HD~\num{219460}~B~=~WR~157), and an M~supergiant ($\alpha$~Sco~A).
       The secondaries in those cases are late-O to early-B stars.
 \item Three of the objects in our sample have surprising identities. HD~\num{51756}~B and BD~$+$60~544 are two bright stars identified as O-type for the first time and HD~8768~A is a new member of the 
       rare OC category. In the three cases the identification has been possible thanks to the elimination of the contamination from a nearby companion with lucky spectroscopy.
 \item For four of the visual systems in which one of the components is an SB2 ($\sigma$~Ori, HD~\num{219460}, HD~\num{194649}, and HD~\num{191201}) we identify the doubled-lined components and
       provide the spectral classifications. For HD~\num{219460}~A it is the first time the SB2 nature is noted and for HD~\num{194649} we identify the A~component as the SB2 for the first time. For 
       another seven systems (FN~CMa, $\sigma$~Sco, HD~\num{51756}, HD~\num{218195}, HD~\num{17520}, HD~\num{24431}, and HD~\num{164492}) we detect signs of spectroscopic binarity in the \lili\ data.
\end{itemize}

We plot in Figure~\ref{sep_DB} the magnitude difference as a function of separation for the systems in LS~I, STIS~I, and this paper. The boundaries for the regions accessible for the two methods are just
empirical approximations based on the existing data. In particular, the capabilities of lucky spectroscopy depend on the seeing at the moment of the observation and a star that is not spatially resolved today
may be resolved tomorrow with the same setup. Also, objects near the boundary may be spatially resolved but at the cost of a lower S/N due to the noise produced by the separation process. In that way, the
spectrum of $\sigma$~Sco~Ab in this paper is noisier than that of HD~\num{168021}, as both have similar separations but the first one has a larger $\Delta B$. Along a similar line, the lucky spectra for 
HD~\num{193443}~A,B have been separated but at the cost of a poor S/N, as the system lies at the limit of what can be done with lucky spectroscopy and the observation took place with exceptional seeing (the
spectra were easily separated with STIS with good S/N). The other object spatially resolved by STIS that lies at the limit accessible by lucky spectroscopy shown in Fig.~\ref{sep_DB} corresponds to
HD~\num{16429}~Aa,Ab. We attempted to separate that system twice with lucky spectroscopy but the conditions were not optimal and we failed. 

For the future we plan to obtain additional lucky spectroscopy along different lines:

\begin{itemize}
 \item Observe additional objects using the same configurations as in the first two papers.
 \item Obtain observations at higher spectral resolution to better study spectroscopic binaries in hierarchical systems.
 \item Extend the technique to longer wavelengths, where there are less spectral lines of interest but where lucky spectroscopy should work better and allow for the disentangling of systems with smaller
       separations and/or higher extinctions.
 \item Attempt the technique with other telescopes.
\end{itemize}

%
%
\begin{acknowledgements}
J.M.A., C.F., A.S., M.P.G., and G.H. acknowledge support from the Spanish Government Ministerio de Ciencia through grant PGC2018-\num{095049}-B-C22. 
R.H.B. acknowledges support from 
the ESAC Faculty Visitor Program.
I.N. and S.S.-D. acknowledge support from the Spanish Government Ministerio de Ciencia through grant PGC2018-\num{093741}-B-C21/22 (MICIU/AEI/FEDER, UE).
S.S.-D. also acknowledges funding from the Spanish Government Ministerio de Ciencia through grants SEV 2015-0548 and CEX2019-\num{000920}-S, and from the Canarian Agency for Research, 
Innovation and Information Society (ACIISI), of the Canary Islands Government, and the European Regional Development Fund (ERDF), under grant with reference ProID\num{2017010115}.
This paper is based on (a) lucky (and regular long-slit) spectroscopy obtained with the 4.2~m William Herschel Telescope (WHT) at the Observatorio del Roque de los Muchachos (ORM) 
on the island of La Palma, Spain; (b) lucky imaging obtained with the 2.2~m Telescope at the Centro Astron\'omico Hispano en Andaluc{\'\i}a (CAHA) in Almer{\'\i}a, Spain; (c) IFU spectroscopy 
obtained with the 2~m Liverpool Telescope (LT) at the Observatorio del Roque de los Muchachos (ORM) on the island of La Palma, Spain as part of GOSSS; (d) long-slit
spectroscopy obtained with the 2.5 duPont Telescope at the Observatorio de Las Campanas (LCO) in Chile; and (e) 
high-resolution \'echelle spectroscopy from the \lili\ project obtained with a variety of spectrographs: HERMES at the 1.2~m Mercator Telescope (MT) at the Observatorio del 
Roque de los Muchachos (ORM) on the island of La Palma, Spain; ELODIE at the 1.93~m Observatoire de Haute-Provence (OHP) Telescope, France; FEROS at the 2.2~m Telescope of the
Observatorio de La Silla in Chile; CAF\'E at the 2.2~m Centro Astron\'omico Hispano en Andaluc{\'\i}a (CAHA) Telescope, Almer{\'\i}a, Spain; FIES at the 2.5 Nordic Optical Telescope 
(NOT) at the Observatorio del Roque de los Muchachos (ORM) on the island of La Palma, Spain; and UVES at the 8.2~m Kueyen Telescope at the Observatorio Paranal in Chile.
Some of the MT and NOT data were obtained from the IACOB spectroscopic database \citep{SimDetal11a,SimDetal11c,SimDetal15b}.
This paper has made use of data from the European Space Agency (ESA) mission {\it Gaia} ({\tt https://www.cosmos.esa.int/gaia}), 
processed by the {\it Gaia} Data Processing and Analysis Consortium (DPAC, {\tt https://www.cosmos.esa.int/web/gaia/dpac/consortium}).
Funding for the DPAC has been provided by national institutions, in particular the institutions participating in the {\it Gaia} Multilateral Agreement. 
This paper has also made use of the Washington Double Star (WDS) catalog \citep{Masoetal01} and the \citet{Skif14} catalog of spectral classifications.
The authors would like to thank the personnel of the WHT, CAHA, LT, LCO, MT, La Silla, and NOT observatories for their support and hospitality throughout the years.
We dedicate this paper to our deceased colleagues, Virpi S. Niemel{\"a} and Nolan R. Walborn, who they surely would have enjoyed having access to data like the ones presented here.
\end{acknowledgements}

%
%
%
%
%
%
%

\bibliographystyle{aa}
\bibliography{general}

\begin{thebibliography}{75}
\expandafter\ifx\csname natexlab\endcsname\relax\def\natexlab#1{#1}\fi

\bibitem[{Adams \& Joy(1921)}]{AdamJoy21}
Adams, W.~S. \& Joy, A.~H. 1921, PASP, 33, 206

\bibitem[{Arias {et~al.}(2016)Arias, Walborn, Sim{\'o}n~D{\'{\i}}az, Barb{\'a},
  Ma{\'{\i}}z~Apell{\'a}niz, Sab{\'{\i}}n-Sanjuli{\'a}n, Gamen, Morrell, Sota,
  Marco, Negueruela, Le{\~a}o, Herrero, \& Alfaro}]{Ariaetal16}
Arias, J.~I., Walborn, N.~R., Sim{\'o}n~D{\'{\i}}az, S., {et~al.} 2016, AJ,
  152, 31

\bibitem[{Aydin(1979)}]{Aydi79}
Aydin, C. 1979, Ap\&SS, 64, 481

\bibitem[{Beavers \& Cook(1980)}]{BeavCook80}
Beavers, W.~I. \& Cook, D.~B. 1980, ApJS, 44, 489

\bibitem[{Berlanas {et~al.}(2020)Berlanas, Herrero, Comer{\'o}n,
  Sim{\'o}n-D{\'\i}az, Lennon, Pasquali, Ma{\'\i}z~Apell{\'a}niz, Sota, \&
  Peller{\'\i}n}]{Berletal20}
Berlanas, S.~R., Herrero, A., Comer{\'o}n, F., {et~al.} 2020, arXiv e-prints,
  arXiv:2008.09917

\bibitem[{Cami {et~al.}(1997)Cami, Sonnentrucker, Ehrenfreund, \&
  Foing}]{Camietal97}
Cami, J., Sonnentrucker, P., Ehrenfreund, P., \& Foing, B.~H. 1997, A\&A, 326,
  822

\bibitem[{Campillay {et~al.}(2019)Campillay, Arias, Barb{\'a}, Morrell, Gamen,
  \& Ma{\'{\i}}z~Apell{\'a}niz}]{Campetal19}
Campillay, A.~R., Arias, J.~I., Barb{\'a}, R.~H., {et~al.} 2019, MNRAS, 484,
  2137

\bibitem[{Corbally(1984)}]{Corb84}
Corbally, C.~J. 1984, ApJS, 55, 657

\bibitem[{Dorda {et~al.}(2018)Dorda, Negueruela, Gonz{\'a}lez-Fern{\'a}ndez, \&
  Marco}]{Dordetal18b}
Dorda, R., Negueruela, I., Gonz{\'a}lez-Fern{\'a}ndez, C., \& Marco, A. 2018,
  A\&A, 618, A137

\bibitem[{Finsen(1956)}]{Fins56}
Finsen, W.~S. 1956, Circular of the Union Observatory Johannesburg, 115, 259

\bibitem[{Gahm {et~al.}(1983)Gahm, Ahlin, \& Lindroos}]{Gahmetal83}
Gahm, G.~F., Ahlin, P., \& Lindroos, K.~P. 1983, A\&AS, 51, 143

\bibitem[{Garrison(1967)}]{Garr67}
Garrison, R.~F. 1967, ApJ, 147, 1003

\bibitem[{Garrison {et~al.}(1977)Garrison, Hiltner, \& Schild}]{Garretal77}
Garrison, R.~F., Hiltner, W.~A., \& Schild, R.~E. 1977, ApJS, 35, 111

\bibitem[{Gonz{\'a}lez {et~al.}(2017)Gonz{\'a}lez, Hubrig, Przybilla, Carroll,
  Nieva, Ilyin, J{\"a}rvinen, Morel, Sch{\"o}ller, Castro, Barb{\'a}, de~Koter,
  Schneider, Kholtygin, Butler, Veramendi, Langer, \& {BOB
  Collaboration}}]{Gonzetal17}
Gonz{\'a}lez, J.~F., Hubrig, S., Przybilla, N., {et~al.} 2017, MNRAS, 467, 437

\bibitem[{Hardorp {et~al.}(1959)Hardorp, Rohlfs, Slettebak, \&
  Stock}]{Hardetal59}
Hardorp, J., Rohlfs, K., Slettebak, A., \& Stock, J. 1959, Hamburger Sternw.
  Warner \& Swasey Obs., C01, 0

\bibitem[{Hiltner(1956)}]{Hilt56}
Hiltner, W.~A. 1956, ApJS, 2, 389

\bibitem[{Hiltner {et~al.}(1969)Hiltner, Garrison, \& Schild}]{Hiltetal69}
Hiltner, W.~A., Garrison, R.~F., \& Schild, R.~E. 1969, ApJ, 157, 313

\bibitem[{Holgado {et~al.}(2018)Holgado, Sim{\'o}n-D{\'{\i}}az, Barb{\'a},
  Puls, Herrero, Castro, Garcia, Ma{\'{\i}}z~Apell{\'a}niz, Negueruela, \&
  Sab{\'{\i}}n-Sanjuli{\'a}n}]{Holgetal18}
Holgado, G., Sim{\'o}n-D{\'{\i}}az, S., Barb{\'a}, R.~H., {et~al.} 2018, A\&A,
  613, A65

\bibitem[{Holgado {et~al.}(2020)Holgado, Sim{\'o}n-D{\'\i}az, Haemmerl{\'e},
  Lennon, Barb{\'a}, Cervi{\~n}o, Castro, Herrero, Meynet, \&
  Arias}]{Holgetal20}
Holgado, G., Sim{\'o}n-D{\'\i}az, S., Haemmerl{\'e}, L., {et~al.} 2020, A\&A,
  638, A157

\bibitem[{Johnson \& Morgan(1953)}]{JohnMorg53}
Johnson, H.~L. \& Morgan, W.~W. 1953, ApJ, 117, 313

\bibitem[{Keenan \& McNeil(1989)}]{KeenMcNe89}
Keenan, P.~C. \& McNeil, R.~C. 1989, ApJS, 71, 245

\bibitem[{Kharchenko {et~al.}(2013)Kharchenko, Piskunov, Schilbach, R{\"o}ser,
  \& Scholz}]{Kharetal13}
Kharchenko, N.~V., Piskunov, A.~E., Schilbach, E., R{\"o}ser, S., \& Scholz,
  R.-D. 2013, A\&A, 558, A53

\bibitem[{Kre{\l}owski {et~al.}(1997)Kre{\l}owski, Schmidt, \&
  Snow}]{Kreletal97}
Kre{\l}owski, J., Schmidt, M., \& Snow, T.~P. 1997, PASP, 109, 1135

\bibitem[{Law {et~al.}(2006)Law, Mackay, \& Baldwin}]{Lawetal06}
Law, N.~M., Mackay, C.~D., \& Baldwin, J.~E. 2006, A\&A, 446, 739

\bibitem[{Lesh(1968)}]{Lesh68}
Lesh, J.~R. 1968, ApJS, 17, 371

\bibitem[{Lindegren {et~al.}(2018)}]{Lindetal18b}
Lindegren, L. {et~al.} 2018, https://www.cosmos.esa.int/documents/29201/
  1770596/Lindegren\_GaiaDR2\_Astrometry\_extended.pdf

\bibitem[{Mahy {et~al.}(2013)Mahy, Rauw, De~Becker, Eenens, \&
  Flores}]{Mahyetal13}
Mahy, L., Rauw, G., De~Becker, M., Eenens, P., \& Flores, C.~A. 2013, A\&A,
  550, A27

\bibitem[{Mahy {et~al.}(2015)Mahy, Rauw, De~Becker, Eenens, \&
  Flores}]{Mahyetal15}
Mahy, L., Rauw, G., De~Becker, M., Eenens, P., \& Flores, C.~A. 2015, A\&A,
  577, A23

\bibitem[{Ma{\'{\i}}z~Apell{\'a}niz(2001)}]{Maiz01a}
Ma{\'{\i}}z~Apell{\'a}niz, J. 2001, AJ, 121, 2737

\bibitem[{Ma{\'{\i}}z~Apell{\'a}niz(2005)}]{Maiz05c}
Ma{\'{\i}}z~Apell{\'a}niz, J. 2005, in ESA Special Publication, Vol. 576, The
  Three-Dimensional Universe with Gaia, ed. C.~Turon, K.~S. O'Flaherty, \&
  M.~A.~C. Perryman, 179

\bibitem[{Ma{\'{\i}}z~Apell{\'a}niz(2010)}]{Maiz10a}
Ma{\'{\i}}z~Apell{\'a}niz, J. 2010, A\&A, 518, A1

\bibitem[{Ma{\'{\i}}z~Apell{\'a}niz(2015)}]{Maiz15a}
Ma{\'{\i}}z~Apell{\'a}niz, J. 2015, MmSAI, 86, 553

\bibitem[{Ma{\'\i}z~Apell{\'a}niz(2019)}]{Maiz19}
Ma{\'\i}z~Apell{\'a}niz, J. 2019, A\&A, 630, A119

\bibitem[{Ma{\'{\i}}z~Apell{\'a}niz {et~al.}(2008)Ma{\'{\i}}z~Apell{\'a}niz,
  Alfaro, \& Sota}]{Maizetal08a}
Ma{\'{\i}}z~Apell{\'a}niz, J., Alfaro, E.~J., \& Sota, A. 2008, arXiv:0804.2553

\bibitem[{Ma{\'{\i}}z~Apell{\'a}niz \& Barb{\'a}(2018)}]{MaizBarb18}
Ma{\'{\i}}z~Apell{\'a}niz, J. \& Barb{\'a}, R.~H. 2018, A\&A, 613, A9

\bibitem[{Ma{\'{\i}}z~Apell{\'a}niz \& Barb{\'a}(2020)}]{MaizBarb20}
Ma{\'{\i}}z~Apell{\'a}niz, J. \& Barb{\'a}, R.~H. 2020, A\&A, 636, A28 (STIS~I)

\bibitem[{Ma{\'{\i}}z~Apell{\'a}niz {et~al.}(2018)Ma{\'{\i}}z~Apell{\'a}niz,
  Barb{\'a}, Sim{\'o}n-D{\'{\i}}az, Sota, Trigueros~P{\'a}ez, Caballero, \&
  Alfaro}]{Maizetal18a}
Ma{\'{\i}}z~Apell{\'a}niz, J., Barb{\'a}, R.~H., Sim{\'o}n-D{\'{\i}}az, S.,
  {et~al.} 2018, A\&A, 615, A161 (LS~I)

\bibitem[{Ma{\'{\i}}z~Apell{\'a}niz {et~al.}(2016)Ma{\'{\i}}z~Apell{\'a}niz,
  Sota, Arias, Barb{\'a}, Walborn, Sim{\'o}n-D{\'{\i}}az, Negueruela, Marco,
  Le{\~a}o, Herrero, Gamen, \& Alfaro}]{Maizetal16}
Ma{\'{\i}}z~Apell{\'a}niz, J., Sota, A., Arias, J.~I., {et~al.} 2016, ApJS,
  224, 4 (GOSSS~III)

\bibitem[{Ma{\'{\i}}z~Apell{\'a}niz {et~al.}(2011)Ma{\'{\i}}z~Apell{\'a}niz,
  Sota, Walborn, Alfaro, Barb{\'a}, Morrell, Gamen, \& Arias}]{Maizetal11}
Ma{\'{\i}}z~Apell{\'a}niz, J., Sota, A., Walborn, N.~R., {et~al.} 2011, in HSA
  6, 467--472

\bibitem[{Ma{\'{\i}}z~Apell{\'a}niz
  {et~al.}(2019{\natexlab{a}})Ma{\'{\i}}z~Apell{\'a}niz, Trigueros~P{\'a}ez,
  Jim{\'e}nez~Mart{\'{\i}}nez, Barb{\'a}, Sim{\'o}n-D{\'{\i}}az, Pellerin,
  Negueruela, \& Souza~Le{\~a}o}]{Maizetal19a}
Ma{\'{\i}}z~Apell{\'a}niz, J., Trigueros~P{\'a}ez, E.,
  Jim{\'e}nez~Mart{\'{\i}}nez, I., {et~al.} 2019{\natexlab{a}}, in HSA 10, 420
  (\lili)

\bibitem[{Ma{\'{\i}}z~Apell{\'a}niz
  {et~al.}(2019{\natexlab{b}})Ma{\'{\i}}z~Apell{\'a}niz, Trigueros~P{\'a}ez,
  Negueruela, Barb{\'a}, Sim{\'o}n-D{\'{\i}}az, Lorenzo, Sota, Gamen,
  Fari{\~n}a, Salas, Caballero, Morrell, Pellerin, Alfaro, Herrero, Arias, \&
  Marco}]{Maizetal19b}
Ma{\'{\i}}z~Apell{\'a}niz, J., Trigueros~P{\'a}ez, E., Negueruela, I., {et~al.}
  2019{\natexlab{b}}, A\&A, 626, A20 (MONOS~I)

\bibitem[{Mason {et~al.}(1998)Mason, Gies, Hartkopf, Bagnuolo, Brummelaar, \&
  McAlister}]{Masoetal98}
Mason, B.~D., Gies, D.~R., Hartkopf, W.~I., {et~al.} 1998, AJ, 115, 821

\bibitem[{Mason {et~al.}(2001)Mason, Wycoff, Hartkopf, Douglass, \&
  Worley}]{Masoetal01}
Mason, B.~D., Wycoff, G.~L., Hartkopf, W.~I., Douglass, G.~G., \& Worley, C.~E.
  2001, AJ, 122, 3466

\bibitem[{Morgan {et~al.}(1955)Morgan, Code, \& Whitford}]{Morgetal55}
Morgan, W.~W., Code, A.~D., \& Whitford, A.~E. 1955, ApJS, 2, 41

\bibitem[{Morgan \& Keenan(1973)}]{MorgKeen73}
Morgan, W.~W. \& Keenan, P.~C. 1973, ARA\&A, 11, 29

\bibitem[{Morgan \& Roman(1950)}]{MorgRoma50}
Morgan, W.~W. \& Roman, N.~G. 1950, ApJ, 112, 362

\bibitem[{Morgan {et~al.}(1953)Morgan, Whitford, \& Code}]{Morgetal53b}
Morgan, W.~W., Whitford, A.~E., \& Code, A.~D. 1953, ApJ, 118, 318

\bibitem[{North {et~al.}(2007)North, Davis, Tuthill, Tango, \&
  Robertson}]{Nortetal07b}
North, J.~R., Davis, J., Tuthill, P.~G., Tango, W.~J., \& Robertson, J.~G.
  2007, MNRAS, 380, 1276

\bibitem[{Pantaleoni~Gonz{\'a}lez {et~al.}(2020)Pantaleoni~Gonz{\'a}lez,
  Ma{\'\i}z~Apell{\'a}niz, Barb{\'a}, \& Negueruela}]{Pantetal20}
Pantaleoni~Gonz{\'a}lez, M., Ma{\'\i}z~Apell{\'a}niz, J., Barb{\'a}, R.~H., \&
  Negueruela, I. 2020, RNAAS, 4, 12

\bibitem[{Reimers {et~al.}(2008)Reimers, Hagen, Baade, \& Braun}]{Reimetal08}
Reimers, D., Hagen, H.-J., Baade, R., \& Braun, K. 2008, A\&A, 491, 229

\bibitem[{Rivinius {et~al.}(2011)Rivinius, Stahl, {\v S}tefl, Baade, Townsend,
  \& Barrera}]{Rivietal11b}
Rivinius, T., Stahl, O., {\v S}tefl, S., {et~al.} 2011, in IAUS, Vol. 272,
  543--544

\bibitem[{Roman-Lopes {et~al.}(2018)Roman-Lopes, Rom{\'a}n-Z{\'u}{\~n}iga,
  Tapia, Chojnowski, G{\'o}mez Maqueo~Chew, Garc{\'{\i}}a-Hern{\'a}ndez,
  Borissova, Minniti, Covey, Longa-Pe{\~n}a, Fernandez-Trincado, Zamora, \&
  Nitschelm}]{RomLetal18}
Roman-Lopes, A., Rom{\'a}n-Z{\'u}{\~n}iga, C., Tapia, M., {et~al.} 2018, ApJ,
  855, 68

\bibitem[{Rydstrom(1978)}]{Ryds78}
Rydstrom, B.~A. 1978, A\&AS, 32, 25

\bibitem[{Sana {et~al.}(2014)Sana, Le~Bouquin, Lacour, Berger, Duvert, Gauchet,
  Norris, Olofsson, Pickel, Zins, Absil, de~Koter, Kratter, Schnurr, \&
  Zinnecker}]{Sanaetal14}
Sana, H., Le~Bouquin, J.-B., Lacour, S., {et~al.} 2014, ApJS, 215, 15

\bibitem[{Sim{\'o}n-D{\'{\i}}az
  {et~al.}(2015{\natexlab{a}})Sim{\'o}n-D{\'{\i}}az, Caballero, Lorenzo,
  Ma{\'{\i}}z~Apell{\'a}niz, Schneider, Negueruela, Barb{\'a}, Dorda, Marco,
  Montes, Pellerin, Sanchez-Bermudez, S{\'o}dor, \& Sota}]{SimDetal15a}
Sim{\'o}n-D{\'{\i}}az, S., Caballero, J.~A., Lorenzo, J., {et~al.}
  2015{\natexlab{a}}, ApJ, 799, 169

\bibitem[{Sim{\'o}n-D{\'{\i}}az
  {et~al.}(2011{\natexlab{a}})Sim{\'o}n-D{\'{\i}}az, Castro, Garc{\'\i}a, \&
  Herrero}]{SimDetal11c}
Sim{\'o}n-D{\'{\i}}az, S., Castro, N., Garc{\'\i}a, M., \& Herrero, A.
  2011{\natexlab{a}}, in IAUS, Vol. 272, 310--312

\bibitem[{Sim{\'o}n-D{\'{\i}}az
  {et~al.}(2011{\natexlab{b}})Sim{\'o}n-D{\'{\i}}az, Garcia, Herrero,
  Ma{\'{\i}}z~Apell{\'a}niz, \& Negueruela}]{SimDetal11a}
Sim{\'o}n-D{\'{\i}}az, S., Garcia, M., Herrero, A., Ma{\'{\i}}z~Apell{\'a}niz,
  J., \& Negueruela, I. 2011{\natexlab{b}}, in Stellar Clusters \&
  Associations: A RIA Workshop on Gaia, 255--259

\bibitem[{Sim{\'o}n-D{\'{\i}}az {et~al.}(2017)Sim{\'o}n-D{\'{\i}}az, Godart,
  Castro, Herrero, Aerts, Puls, Telting, \& Grassitelli}]{SimDetal17}
Sim{\'o}n-D{\'{\i}}az, S., Godart, M., Castro, N., {et~al.} 2017, A\&A, 597,
  A22

\bibitem[{Sim{\'o}n-D{\'{\i}}az \& Herrero(2014)}]{SimDHerr14}
Sim{\'o}n-D{\'{\i}}az, S. \& Herrero, A. 2014, A\&A, 562, A135

\bibitem[{Sim{\'o}n-D{\'{\i}}az {et~al.}(2014)Sim{\'o}n-D{\'{\i}}az, Herrero,
  Sab{\'{\i}}n-Sanjuli{\'a}n, Najarro, Garcia, Puls, Castro, \&
  Evans}]{SimDetal14}
Sim{\'o}n-D{\'{\i}}az, S., Herrero, A., Sab{\'{\i}}n-Sanjuli{\'a}n, C.,
  {et~al.} 2014, A\&A, 570, L6

\bibitem[{Sim{\'o}n-D{\'{\i}}az
  {et~al.}(2015{\natexlab{b}})Sim{\'o}n-D{\'{\i}}az, Negueruela,
  Ma{\'{\i}}z~Apell{\'a}niz, Castro, Herrero, Garcia, P{\'e}rez-Prieto, Caon,
  Alacid, Camacho, Dorda, Godart, Gonz{\'a}lez-Fern{\'a}ndez, Holgado, \&
  R{\"u}bke}]{SimDetal15b}
Sim{\'o}n-D{\'{\i}}az, S., Negueruela, I., Ma{\'{\i}}z~Apell{\'a}niz, J.,
  {et~al.} 2015{\natexlab{b}}, in HSA 8, 576--581

\bibitem[{Skiff(2014)}]{Skif14}
Skiff, B.~A. 2014, VizieR Online Data Catalog, B/mk

\bibitem[{Smith {et~al.}(1996)Smith, Shara, \& Moffat}]{Smitetal96}
Smith, L.~F., Shara, M.~M., \& Moffat, A. F.~J. 1996, MNRAS, 281, 163

\bibitem[{Sota {et~al.}(2014)Sota, Ma{\'{\i}}z~Apell{\'a}niz, Morrell,
  Barb{\'a}, Walborn, Gamen, Arias, \& Alfaro}]{Sotaetal14}
Sota, A., Ma{\'{\i}}z~Apell{\'a}niz, J., Morrell, N.~I., {et~al.} 2014, ApJS,
  211, 10 (GOSSS~II)

\bibitem[{Sota {et~al.}(2011)Sota, Ma{\'{\i}}z~Apell{\'a}niz, Walborn, Alfaro,
  Barb{\'a}, Morrell, Gamen, \& Arias}]{Sotaetal11a}
Sota, A., Ma{\'{\i}}z~Apell{\'a}niz, J., Walborn, N.~R., {et~al.} 2011, ApJS,
  193, 24 (GOSSS~I)

\bibitem[{Stickland \& Lloyd(2001)}]{SticLloy01}
Stickland, D.~J. \& Lloyd, C. 2001, The Observatory, 121, 1

\bibitem[{Stone \& Struve(1954)}]{StonStru54}
Stone, S.~N. \& Struve, O. 1954, PASP, 66, 191

\bibitem[{Swings \& Preston(1978)}]{SwinPres78}
Swings, J.~P. \& Preston, G.~W. 1978, ApJ, 220, 883

\bibitem[{Talavera \& G{\'o}mez~de Castro(1987)}]{TalaGome87}
Talavera, A. \& G{\'o}mez~de Castro, A.~I. 1987, A\&A, 181, 300

\bibitem[{Turner {et~al.}(1983)Turner, Moffat, Lamontagne, \&
  Maitzen}]{Turnetal83}
Turner, D.~G., Moffat, A.~F.~J., Lamontagne, R., \& Maitzen, H.~M. 1983, AJ,
  88, 1199

\bibitem[{Turner {et~al.}(2008)Turner, ten Brummelaar, Roberts, Mason,
  Hartkopf, \& Gies}]{Turnetal08}
Turner, N.~H., ten Brummelaar, T.~A., Roberts, L.~C., {et~al.} 2008, AJ, 136,
  554

\bibitem[{Wade {et~al.}(2017)Wade, Shultz, Sikora, Bernier, Rivinius, Alecian,
  Petit, Grunhut, \& {BinaMIcS Collaboration}}]{Wadeetal17}
Wade, G.~A., Shultz, M., Sikora, J., {et~al.} 2017, MNRAS, 465, 2517

\bibitem[{Walborn(1972)}]{Walb72}
Walborn, N.~R. 1972, AJ, 77, 312

\bibitem[{Walborn(1973)}]{Walb73a}
Walborn, N.~R. 1973, AJ, 78, 1067

\bibitem[{Ward {et~al.}(2020)Ward, Kruijssen, \& Rix}]{Wardetal20}
Ward, J.~L., Kruijssen, J.~M.~D., \& Rix, H.-W. 2020, MNRAS, 495, 663

\end{thebibliography}

\end{document}